\documentclass[%
preprint,
 amsmath,amssymb,
 aps,pre,
]{revtex4-2}

\usepackage{graphicx}
\usepackage{dcolumn}
\usepackage{bm}
\usepackage [latin1]{inputenc}

\begin{document}

\preprint{APS/123-QED}

\title{Physics of the low momentum diffusivity regime in tokamaks and its experimental applicability}

\author{Haomin Sun$^{1,*}$, Justin Ball$^{1}$, Stephan Brunner$^{1}$, Anthony Field$^{2}$, Bhavin Patel$^{2}$, Alessandro Balestri$^{1}$, Daniel Kennedy$^{2}$, Colin Roach$^{2}$, Diego Jose Cruz-Zabala$^{3}$, Fernando Puentes Del Pozo$^{3}$, Eleonora Viezzer$^{3}$, and Manuel Garcia Munoz$^{3}$}
\affiliation{$^1$Ecole Polytechnique F\'ed\'erale de Lausanne (EPFL), Swiss Plasma Center (SPC), CH-1015 Lausanne, Switzerland\\
$^2$UKAEA (United Kingdom Atomic Energy Authority), Culham Campus, Abingdon, Oxfordshire, OX14 3DB, UK\\
$^3$Department of Atomic, Molecular and Nuclear Physics, University of Seville, Seville, Spain\\
$*$ haomin.sun@epfl.ch
}


\vspace{10pt}


\date{\today}

\begin{abstract}
Strong $E\times B$ plasma flow shear is beneficial for reducing turbulent transport. However, traditional methods of driving flow shear do not scale well to large devices such as future fusion power plants. In this paper, we use a large number of nonlinear gyrokinetic simulations to study a novel approach to increase flow shear: decreasing the momentum diffusivity to make the plasma ``easier to push''. We first use an idealized circular geometry and find that one can obtain low momentum diffusivity at tight aspect ratio, low safety factor, high magnetic shear and low temperature gradient. This is the so-called Low Momentum Diffusivity (LMD) regime. To drive intrinsic momentum flux, we then tilt the flux surface, making it up-down asymmetric. In the LMD regime, this intrinsic momentum flux drives strong flow shear that can significantly reduce the heat flux and increase the critical temperature gradient. We also consider the actual experimental geometry of the MAST tokamak to illustrate that this strategy can be practical and create experimentally significant flow shear. Lastly, a preliminary prediction for the SMART tokamak is made.

\end{abstract}

\maketitle


\section{\label{sec:level1}Introduction}
The transport of energy out of tokamaks is usually dominated by turbulence \cite{DimitsShift2000}. It is therefore of great importance to find ways to mitigate turbulence in order to enable an economical fusion power plant. Sufficiently fast rotation gradients (in particular involving $E\times B$ flow shear) can significantly reduce turbulent transport \cite{Biglari1990,Stambaugh1990enhanced,JETL-GEriksson_1997,RotationMHDChu1999NF,Angioni2001IntrinsicRotation,RotationMHDWahlberg2000POP,Angioni2001IntrinsicRotation,Peeters2005LinearToroidal,schekochihin2008,deVries_2008,schekochihin2008,JETMantica2009PRL,Ida2009PRL,RotationMHDAiba2009NF,NewtonFlowShearUnderstanding2010,RotationMHDAiba2011NF,BarnesFlowShear2011,schekochihin2012,schekochihin2012,Highcock2011POP,highcock2012,schekochihin2012,ChristenFlowShear2018,ben2019,ball2019,JETJ.M.Noterdaeme_2003,JETdeVries_2006}.

In tokamaks, the toroidal symmetry of the device constrains the strong plasma flows ($v_{\zeta}=O(c_{s})$) to be toroidal, where $v_{\zeta}$ is the toroidal rotation speed and $c_s$ is the sound speed. In order to obtain strong plasma rotation, one can use external momentum injection such as Neutron Beam Injection (NBI) \cite{Groebner1990NBIrotation,Suckewer1981NBIrotation,Goumiri2016NSTXNBI} or Radio Frequency (RF) waves \cite{Hsuan1996ICHrotation,Chang1999ICRH,Chan2002RFrotation,Li_2011EASTRF,Lyu2020RFrotation} to exert a torque on the plasma. However, external injection is not expected to scale well to large devices \cite{YueqiangLiu_2004} given that the higher energy beams needed for larger devices inject less momentum per unit energy \cite{Parra_2011_PRL_Momentum_optimum}. An attractive alternative is intrinsic rotation \cite{Camenen2009TUBTrans,Camenen2010MomentumTransport}, which is rotation driven under certain conditions by turbulence in the plasma \cite{zhu2024intrinsic}. To create flow of the order of the sound speed, the symmetry properties of gyrokinetics \cite{Peeters2005LinearToroidal,ParraUpDownSym2011,Peeters_2011NFReview,Parratheory_2015} imply that one has to break the up-down symmetry of magnetic flux surface shapes \cite{ball2014,ball2018}. This method has the potential to scale well, as it does not rely on external sources.

In order to understand the underlying physics, we will start by reviewing the basics of momentum transport. In tokamaks, most of the momentum is carried by the ions due to their large mass. In steady-state, for a plasma without external momentum input, the total momentum flux must be zero ($\Pi_i=0$). Otherwise, momentum would be moving around the tokamak and the rotation profile would be changing. Therefore, the ion toroidal angular momentum flux $\Pi_i$ must cancel and can be expressed by \cite{ball2014}
\begin{equation}
\Pi_i\left(\Omega_i,\frac{d\Omega_i}{dx}\right)=0,
\label{eq_momentum00}
\end{equation}
where $x$ is the radial coordinate and $\Omega_i$ is the ion toroidal angular velocity. The momentum flux depends on many parameters, but here we explicitly note $\Omega_i$ and $d\Omega_i/dx$ to stress their importance. For relatively weak rotation and rotation shear, one can Taylor expand the momentum flux as follows
\begin{equation}
\Pi_i\left(\Omega_i,\frac{d\Omega_i}{dx}\right)\simeq\Pi_{i,int}
-n_im_iR^2_0D_{\Pi i}\frac{d\Omega_i}{dx}-n_im_iR^2_0P_{\Pi i}\Omega_i=0,
\label{eq_momentum}
\end{equation}
where $\Pi_{i,int}=\Pi_i(0,0)$ is the intrinsic momentum flux generated by up-down asymmetry and the second and third term refer to the diffusive turbulent momentum transport and the Coriolis pinch effect, respectively. The other variables defined here are: the ion density $n_i$, the ion mass $m_i$, the major radius of the tokamak $R_0$, the angular momentum diffusion coefficient $D_{\Pi i}$ and the angular momentum pinch coefficient $P_{\Pi i}$. The condition $\Pi_i=0$ implies that the intrinsic momentum flux will be balanced by the diffusive transport and pinch term \cite{Newton2006neoclassical,Wang2009PRL}. The diffusive transport corresponds to the viscous momentum transport driven by toroidal flow shear $d\Omega_i/dx$ \cite{Peeters2005LinearToroidal,Hahm2007NLtheory,Diamond2008momentumtransport,Holod2008GKSim,CassonExBshear2009,Roach_2009,Yoon_2010momentumtransportITG,ParraUpDownSym2011,Camenen_2011NFReview,Peeters_2011NFReview,Angioni_2012NFReview,Diamond_2013NFreview,ball2014}. Despite looking simple, Eq. \eqref{eq_momentum} contains rich physics. To be specific, the parameter dependence of $D_{\Pi i}$ and $P_{\Pi i}$ on tokamak geometry and other physical quantities is still poorly understood, which constitutes the main motivation of this work. For typical tokamak operation conditions, the pinch term is much smaller than the diffusive momentum transport term \cite{Peeters2007PRLpinchterm,Guttenfelder_2017,Zimmermann_2022,Zimmermann_2023}, so this paper will mainly focus on the diffusive effect. The effect of the pinch term is discussed in Appendix \ref{AppendixA}. Neglecting the pinch term, Eq. \eqref{eq_momentum} takes the simpler form
\begin{equation}
\Pi_i=\Pi_{i,int}-n_im_iR^2_0D_{\Pi i}\frac{d\Omega_i}{dx}=0.
\label{eq_momentum2}
\end{equation}
Eq. \eqref{eq_momentum2} implies that to get strong toroidal velocity shear $d\Omega_i/dx$, one can either increase the intrinsic momentum flux $\Pi_{i,int}$ or reduce the momentum diffusivity $D_{\Pi i}$. Maximizing $\Pi_{i,int}$ has been extensively addressed in previous work \cite{ball2018}, so we will concentrate on the reduction of momentum diffusivity. The strength of the ion turbulent momentum diffusivity $D_{\Pi i}$ is frequently compared against the ion turbulent heat diffusivity $D_{Q i}$ using the ion Prandtl number \cite{ball2014}
\begin{equation}
\text{Pr}_i = \frac{D_{\Pi i}}{D_{Q i}}.
\label{eq_prandtl}
\end{equation}
Minimizing this ratio of the two diffusivities is the main goal of this paper, given that we want to identify the optimal conditions for achieving strong rotation shear in order to reduce energy transport. For conventional aspect ratio tokamaks, many previous works have experimentally observed or theoretically assumed $\text{Pr}_i\sim 1$ \cite{Diamond2008momentumtransport,Diamond_2009,Tala_2011,Weisen_2012,ball2014,ball2018}. On the other hand, a lower Prandtl number has also been observed in gyrokinetic simulations \cite{Holod2008GKSim,Camenen_2011NFReview} and is beneficial because it means that a given source of momentum (external or intrinsic) will drive stronger rotation shear for a given level of turbulence and thus will be more effective at stabilizing turbulence \cite{HighcockRotationBifurcation2010,highcock2012,Ben_2019}. 

Recently, it was found that low Prandtl number can be reached at tight aspect ratio and low safety factor, suggesting that the underlying requirement is that the magnetic field line pitch is large (i.e. the field line is angled very poloidally) \cite{CassonExBshear2009,Ben_2015_intrinsic,Ben_2019}. However, due to limitations in computational resources, previous works did not carry out fine scans of these two parameters to more accurately characterize the boundary of the low momentum diffusivity regime (which we will define as $\text{Pr}_i=0.5$, see Sec. \ref{circularresults}), nor did it identify the optimal parameter set with the lowest Prandtl number. 
The calculation of toroidal angular momentum flux \cite{ParraUpDownSym2011} was often simplified in many previous works. Some previous works were only able to include the parallel component of the toroidal angular momentum flux \cite{Holod2008GKSim,CassonExBshear2009}, while others \cite{Ben_2015_intrinsic,Ben_2019} considered toroidal angular momentum flux generated by parallel flow shear instead of the physical toroidal flow shear that is present in tokamaks \cite{AbelGyrokineticsDeriv2012}. Additionally, a self-consistent study of the efficiency of intrinsic flow shear in stabilizing turbulence has not been performed. On the experimental side, recent results have shown that a Prandtl number below one can be obtained in practice \cite{deVries_2008,McDermott_2011,BuchholzPrandtl2015,Guttenfelder_2017,Hornsby_2018,Zimmermann_2022,Zimmermann_2023}. However, the experimental momentum transport data is often hard to interpret because the three terms in Eq. \eqref{eq_momentum} are difficult to separate, especially for tight aspect ratio spherical tokamaks \cite{Zimmermann_2022} for which the Prandtl number is often low \cite{Meyer_2009,Kaye_2021_ST_Review}. A comprehensive theoretical study would therefore be very valuable to provide a complete picture of toroidal angular momentum diffusivity over an extended parameter space, and how it can be combined with rotation drive mechanisms to stabilize turbulence.

In this paper, we comprehensively study the Low Momentum Diffusivity (LMD) regime by carrying out a large number of Non-Linear (NL) local gyrokinetic simulations using the GENE code. We start from a tight aspect ratio circular geometry and find that the LMD regime is reached at tight aspect ratio, low safety factor, normal to high magnetic shear, and low temperature gradient. We then consider a magnetic geometry with a tilted elliptical poloidal cross section to generate intrinsic momentum flux. Since the steady state operation of a tokamak implies zero total momentum flux, $\Pi_i=0$, we scan the flow shear $d\Omega_i/dx$ to find the value at which the total momentum flux is zero. This corresponds to the steady state flow shear value driven by intrinsic momentum flux. In this way, for the first time, we self-consistently determine with local gyrokinetic simulations the flow shear generated by up-down asymmetry and at the same time quantify the reduction of the heat flux resulting from associated turbulence stabilization. Lastly, to verify the experimental relevance of LMD regime, we also perform simulations considering the actual experimental parameters of the MAST tokamak. By comparing simulations considering the experimental flux surfaces and the artificially tilted experimental flux surfaces, we show that intrinsic momentum flux can drive significant flow shear under experimentally realistic conditions. A prediction for the SMART tokamak is also made, which can be verified in future experiments. Note that we have assumed a negligible pinch term in these studies. However, we show in Appendix \ref{AppendixA} that considering the pinch term makes the intrinsic flow shear even stronger. This work therefore not only provides the first comprehensive numerical study of low toroidal angular momentum diffusivity, but also demonstrates a new way to create fast rotation in future fusion power plants.

This paper is organized as follows. An introduction to the simulation model implemented in GENE is provided in Sec. \ref{Model}. This is followed by the presentation and analysis of the simulation results for a circular geometry and a tilted ellipse geometry in Sec. \ref{circulargeo} and \ref{tiltellipsegeo}. The verification of the experimental applicability considering MAST equilibria is shown in Sec. \ref{MASTgeo}. A prediction for the SMART tokamak is given in Sec. \ref{SMARTgeo}. Finally, conclusions are drawn in Sec. \ref{conclusion}.


\section{Simulation Model}\label{Model}
Turbulence in tokamaks is weakly collisional and highly anisotropic, as it is very elongated along the magnetic field but remains very narrow across it. To correctly model such turbulence, a gyrokinetic model is necessary. The gyrokinetic model takes an average over the gyromotion of particles, effectively reducing phase space from six to five dimensions and furthermore removing the fast cyclotron time scale. This significantly decreases computational cost. GENE is a well-developed and thoroughly benchmarked gyrokinetic code \cite{JenkoGENE2000,GoerlerGENE2011}. In this paper, we exclusively use the flux tube (local) version of GENE and we always use the Miller representation \cite{Millergeometry1998} to analytically parameterize the magnetic flux surface geometry. GENE considers $(x,y,z)$ as spatial coordinates and the $(v_{||},\mu)$ velocity space coordinates, where $(x,y,z)$ are the radial coordinate, binormal coordinate and straight field line poloidal angle $\chi$, respectively and $(v_{||},\mu)$ are the parallel velocity and magnetic moment $\mu=m_sv^2_{\perp}/2B$. Here $m_s$ is the particle mass, $v_{\perp}$ is the perpendicular velocity and $B$ is the magnetic field strength. In the Fourier representation used by GENE, the spatial coordinates become $(k_x,k_y,z)$, where $k_x$ and $k_y$ are radial and binormal wave numbers, respectively. It is important to note that we consider toroidal flow shear in our simulations instead of just parallel flow shear \cite{mcmillan2019} as in previous work. Furthermore, instead of just the parallel momentum flux \cite{CassonExBshear2009,BarnesFlowShear2011}, we compute the exact toroidal angular momentum flux according to the following relation \cite{Sugammafluxderiv1998,ParraUpDownSym2011,ball2016a} 
\begin{multline}\label{momentumflux}
\Pi_s(k_x,k_y,z)=-\frac{2\pi i}{m_s} \left\langle C_y k_y B \phi(k_x,k_y,z,t) 
\int dv_{||}d\mu h_s(-k_x,-k_y,z,t)\right. \\
\left. \times \left[\frac{I}{B}v_{||}J_0(k_{\perp}\rho_s)+\frac{i}{\Omega_s}\frac{dx}{d\psi}\frac{\mu k^x}{m_s}\frac{2J_1(k_{\perp}\rho_s)}{k_{\perp}\rho_s}\right] \right\rangle_{\Delta t}.
\end{multline}
The second term in the equation is often small for typical parameters, but becomes important in the LMD regime. Here, $h_s=\delta f_s+Z_se\phi F_{Ms}/T_s$ is the non-adiabatic contribution to the perturbed distribution function of species $s$, where $F_{Ms}$ is the background Maxwellian distribution, $\phi$ is the electrostatic potential associated to the fluctuations, $e$ is the elementary charge, $T_s$ is the temperature of species $s$, and $Z_s$ is the charge number of the species $s$ being considered. $\langle ...\rangle_{\Delta t}=\Delta t^{-1}\int_{\Delta t}dt(...)$ represents a time average over a sufficiently long time window $\Delta t$ in the saturated quasi-steady state, $k^x=k_x|\Vec{\nabla}x|^2+k_y\Vec{\nabla}x\cdot\Vec{\nabla}y$, $\psi$ is the poloidal magnetic flux, $C_y=(1/B_{ref}) d\psi/dx$ is a geometrical coefficient calculated by GENE based on flux surface information (where $B_{ref}$ is a reference value of the magnetic field), $\Omega_s=Z_seB/m_s$ is the particle gyro-frequency, $I=RB_{\zeta}$ is the toroidal field flux function and $J_0$ and $J_1$ are the zeroth and first order Bessel functions of the first kind. Employing the full toroidal flow shear and the exact toroidal angular momentum flux allows us to accurately conduct the first comprehensive study of low toroidal angular momentum diffusivity with a large number of NL gyrokinetic simulations.

\section{Results using circular geometry}\label{circulargeo}
\subsection{Simulation parameters}\label{circularparameter}
In order to develop a general picture of the parameter dependence of the Prandtl number, we start with concentric circular flux surfaces. To drive non-zero toroidal angular momentum flux, a toroidal flow shear $d\Omega_i/dx=-0.12(q/r_0)c_s/R_0$ is considered, composed of an $E\times B$ flow shear $\omega_{\perp}=-(r_0/q)d\Omega_{i}/d x=0.12c_s/R_0$ and the parallel flow shear $\omega_{||}$ such that the total flow shear on each flux surface is purely toroidal. Here $r_0$ is the minor radial location of the flux tube center and $c_s=\sqrt{T_e/m_i}$ is the sound speed. Since $\omega_{||}$ is calculated in GENE based on the value of $\omega_{\perp}$ such that they together give pure toroidal flow shear, they are not independent. For convenience, we use $\omega_{\perp}$ to quantify the strength of toroidal flow shear in this paper. We perform a large parameter scan with tight aspect ratio $\epsilon=r_0/R_0=0.36$ over a three-dimensional (3D) parameter space covering safety factor values $1.05\leq q\leq 4.55$ with increments $\Delta q=0.5$, magnetic shear $\hat{s}\in\bigl\{0.1, 0.4, 0.6, 0.8, 1.0, 1.2, 1.4, 1.6\bigl\}$ and temperature gradient $4.96\leq R_0/L_{Ti}\leq 12.96$ with $\Delta (R_0/L_{Ti})=1$. In total, 576 NL gyrokinetic simulations are performed. The other parameters are set to be the same as the well-known Cyclone Base Case parameters \cite{DimitsShift2000}, assuming an adiabatic electron response, equal ion and electron temperatures $T_e=T_i$, and a density gradient of $R_0/L_n=2.22$. The grid parameters that we used for simulations are shown in Table \ref{tabletightpara}. The dominant instability in these simulations is the toroidal Ion Temperature Gradient (ITG) instability. Convergence checks have been performed for the simulations with parameters in the corners of the considered 3D parameter space. Using the numerical parameters shown in Table \ref{tabletightpara}, the deviation of the fluxes from the fully converged results is within $15\%$.

\begin{table*}
\caption{\label{tabletightpara} The nominal numerical GENE parameters for tight aspect ratio ($\epsilon=0.36$) NL simulations with adiabatic electrons and flow shear $\omega_{\perp}R_0/c_s=0.12$, where $\Delta k_y\rho_i=0.05$. The numbers $(n_{k_x},n_{k_y},n_z,n_{v_{||}},n_{\mu})$ refer to the number of $k_x$, $k_y$ mode being considered, and number of grid points along $z$, $v_{||}$ and $\mu$ directions, respectively.}
\begin{ruledtabular}
\begin{tabular}{ccccccc}
$\Delta k_x$&$k_y\rho_i$&$z$&$v_{||}/\sqrt{2T_i/m_i}$&$\sqrt{\mu/(T_i/B)}$&$t/(R_0/c_s)$&$(n_{k_x},n_{k_y},n_z,n_{v_{||}},n_{\mu})$\\ \hline
$0.2\pi\Delta k_y$&$[0.05,3.2]$&$[-\pi,\pi)$&$[-3,3]$&$[0,3]$&[0,1000]&$(192,64,64,32,9)$\\
\end{tabular}
\end{ruledtabular}
\end{table*}



\subsection{Simulation results of the 3D parameter scan}\label{circularresults}
\begin{figure}
    \centering
    \includegraphics[width=0.77\textwidth]{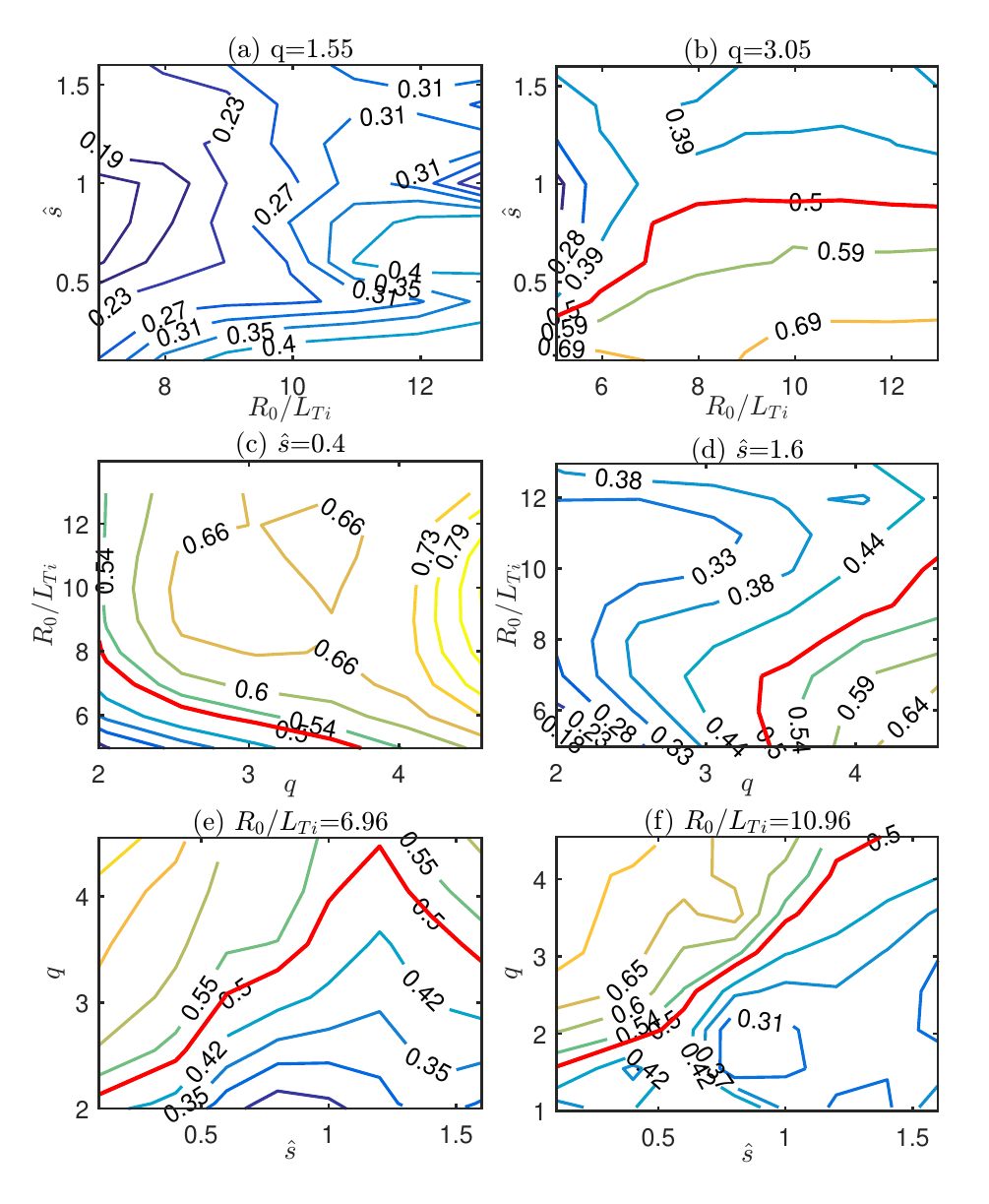}
    \caption{Contour plots of Prandtl number fixing one of the three parameters $(q,\hat{s},R_0/L_{Ti})$: (a) $q=1.55$, (b) $q=3.05$, (c) $\hat{s}=0.4$, (d) $\hat{s}=1.6$, (e) $R_0/L_{Ti}=6.96$, and (f) $R_0/L_{Ti}=10.96$. The red lines in the plots denote the contour $\text{Pr}_i=0.5$, which we define as the boundary of LMD regime.}
    \label{fig2contour}
\end{figure}

The large scan allows us to obtain a full picture of the dependence of the Prandtl number on various parameters. In simulations, the normalized Prandtl number is calculated by the following expression
\begin{equation}\label{eq_Prandtlnumber}
\text{Pr}_i=\frac{\hat{\Pi}_i}{\hat{Q}_i}\frac{R_0}{L_{Ti}}\frac{\epsilon}{q\omega_{\perp}}\frac{c_s}{R_0},
\end{equation}
where $\hat{\Pi}_i$ and $\hat{Q}_i$ are the toroidal angular momentum flux and heat flux in gyroBohm units $c^2_s m_i n_i R_0 (\rho_i/R_0)^2$ and $c_s n_i T_i(\rho_i/R_0)^2$, where $\rho_i$ is the ion gyroradius. Figure \ref{fig2contour} shows some example contour plots of Prandtl number holding one of our three parameters $(q,\hat{s},R_0/L_{Ti})$ constant. Note that only simulations with nonlinearly unstable turbulence are shown in the figure, since for stable conditions, we cannot define the Prandtl number. The red lines in the plots denote $\text{Pr}_i=0.5$, which we choose to define as the boundary of the LMD regime. As we can see from the figure, the dependence of the Prandtl number on $q$, $\hat{s}$, and $R_0/L_T$ is non-trivial. Comparing Fig. \ref{fig2contour} (a) and (b), we see that a low safety factor is favorable for a low Prandtl number. In contrast to previous work \cite{Zimmermann_2022}, the Prandtl number depends strongly on magnetic shear $\hat{s}$, where variations by a factor of $2$ can be observed for $0<\hat{s}<1.5$. Figure \ref{fig2contour} (c) and (d) indicate that a low temperature gradient typically reduces Prandtl number, especially for the cases with low $\hat{s}$ (see Fig. \ref{fig2contour} (e)), but this effect is weaker compared to the effects of $\hat{s}$ and $q$. Figure \ref{fig2contour} (e) and (f) show more clearly that a high magnetic shear ($\hat{s}=1.6$) leads to a reduced Prandtl number compared to a lower magnetic shear $\hat{s}=0.4$. On the other hand, Fig. \ref{fig2contour} (a), (b) and (c) also indicate that the decrease of $\text{Pr}_i$ with $\hat{s}$ is not monotonic. The optimum magnetic shear exists around $\hat{s}\simeq 1$.

\begin{figure}
    \centering
    \includegraphics[width=0.85\textwidth]{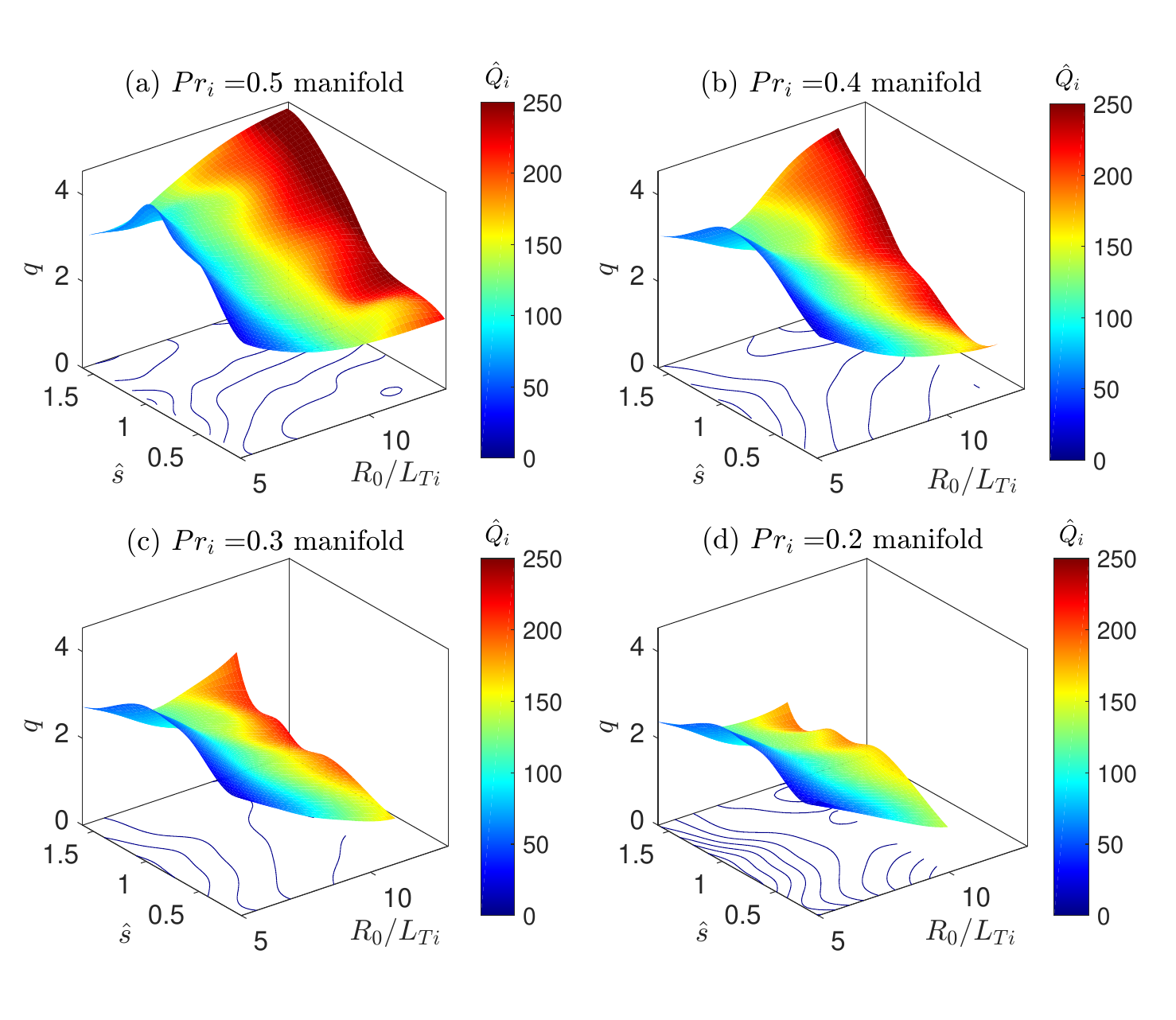}
    \caption{The constant Prandtl number manifolds for (a) $\text{Pr}_i=0.5$, (b) $\text{Pr}_i=0.4$, (c) $\text{Pr}_i=0.3$ and (d) $\text{Pr}_i=0.2$. The manifolds are obtained from linear interpolation of the 3D matrix data $\text{Pr}_i(q,\hat{s},R_0/L_{Ti})$. The height of the manifolds and the contour lines below denote the $q$ value at which the corresponding Prandtl number is reached for a given $(\hat{s},R_0/L_{Ti})$. The color map on the manifolds denotes the ion heat flux $\hat{Q}_i$ (in gyroBohm units).}
    \label{fig1manifold}
\end{figure}

Figure \ref{fig1manifold} further plots some constant Prandtl number manifolds in the 3D parameter space $(q,\hat{s},R_0/L_{Ti})$. In these plots, only data corresponding to nonlinearly unstable cases are shown. The parameter regimes below the manifolds shown in Fig. \ref{fig1manifold} have Prandtl numbers smaller than the considered manifold value denoted in the titles. The parameter regime below the manifold in Fig. \ref{fig1manifold} (a) is therefore what we define as the LMD regime, i.e., where $\text{Pr}_i<0.5$. Similarly, we see from plots in Fig. \ref{fig1manifold} that $\hat{s}\simeq 1$ favors low $\text{Pr}_i$. Further looking at Fig. \ref{fig1manifold} (b)-(d), we also see that a low temperature gradient is in most cases beneficial for reducing Prandtl numbers. On the other hand, on these manifolds, the ion heat flux $\hat{Q}_i$ increases nearly proportionally with $R_0/L_{Ti}$, indicating that $q$ and $\hat{s}$ do not strongly affect the heat flux in the LMD regime. In order to efficiently reduce the heat flux in experiments, one needs to create a strong rotation. Here, we propose to get it by combining the LMD regime with some drive of rotation (e.g. intrinsic rotation from up-down asymmetry as shown in Section \ref{tiltellipsegeo}). 

\begin{figure*}
    \centering
    \includegraphics[width=0.87\textwidth]{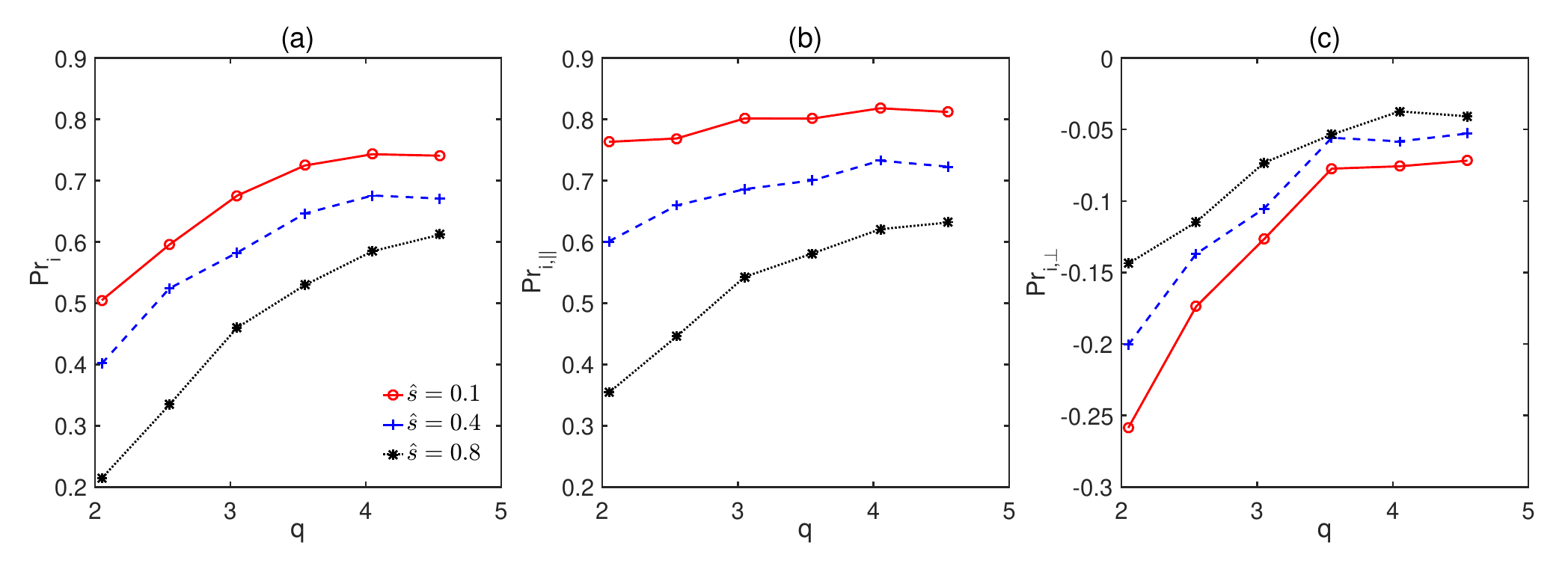}
    \caption{The (a) full Prandtl number, (b) Prandtl number only considering the parallel contribution to the toroidal angular momentum flux, and (c) Prandtl number only considering the perpendicular contribution to the toroidal angular momentum flux for $R_0/L_{Ti}=6.96$ as well as $\hat{s}=0.1$ (red solid), $\hat{s}=0.4$ (blue dashed), and $\hat{s}=0.8$ (black dotted).}
    \label{fig3paraperppri}
\end{figure*}

We would like to emphasize the importance of including the perpendicular contribution to the toroidal angular momentum flux. This is demonstrated by Fig. \ref{fig3paraperppri}, which shows the parallel $\text{Pr}_{i,||}$ and perpendicular $\text{Pr}_{i,\perp}$ Prandtl numbers. These are defined by replacing $\hat{\Pi}_i$ in Eq. \eqref{eq_Prandtlnumber} with either $\hat{\Pi}_{i,||}$ or $\hat{\Pi}_{i,\perp}$ (whose explicit expressions are given in Appendix A of Ref. \cite{Sun2024NF}). As we can see, the perpendicular contribution is not negligible, especially in the LMD regime. Neglecting the perpendicular contribution to the toroidal angular momentum flux results in an overestimate of the Prandtl number.

\subsection{Effect of aspect ratio on the Prandtl number}\label{circulardis}
\begin{figure*}
    \centering
    \includegraphics[width=0.87\textwidth]{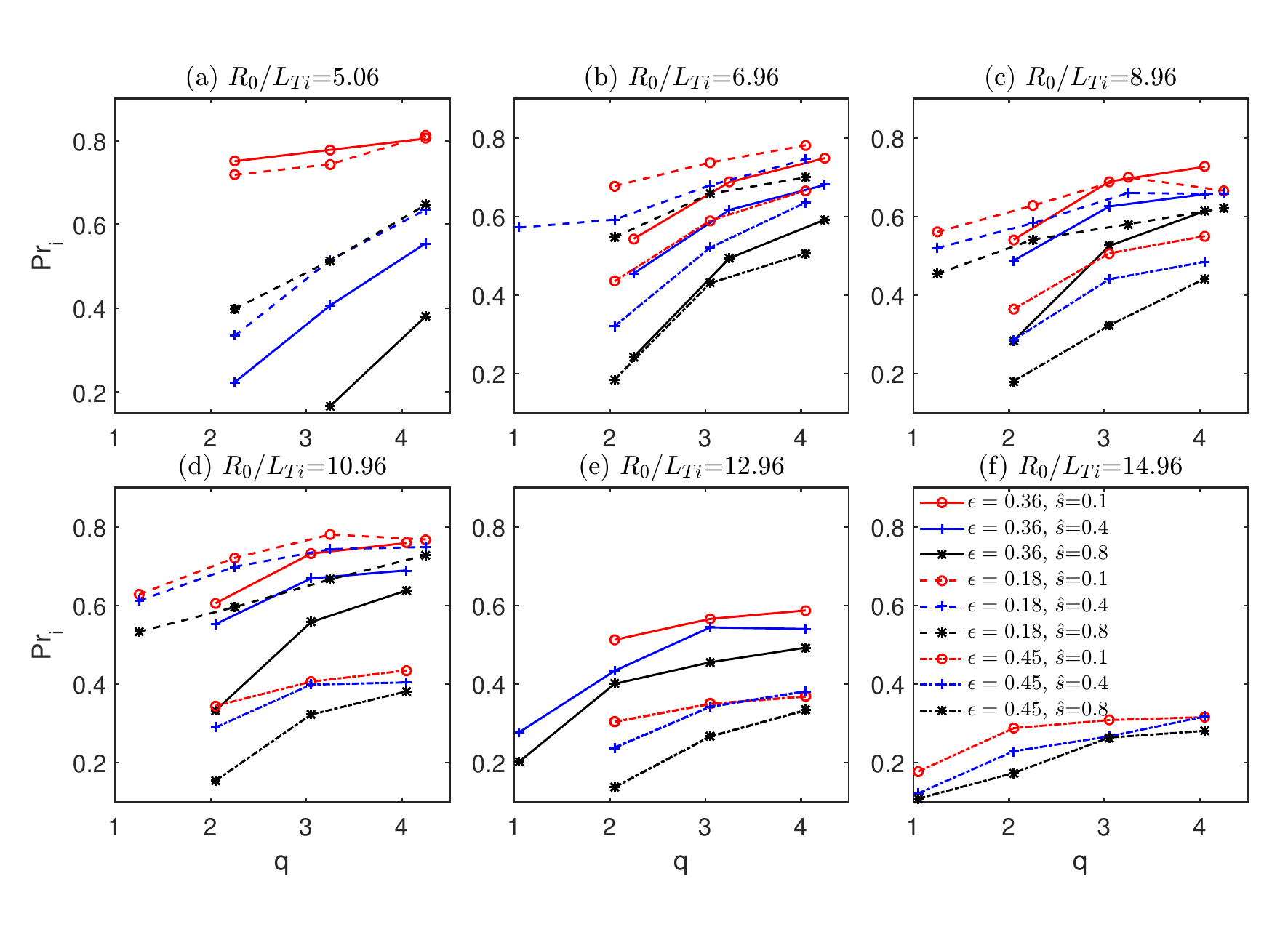}
    \caption{The Prandtl number as a function of $q$ from NL simulations with adiabatic electrons for $\epsilon=0.36$ (solid), $\epsilon=0.18$ (dashed), and $\epsilon=0.45$ (dash-dotted) at $\hat{s}=0.1$ (red), $\hat{s}=0.4$ (blue), and $\hat{s}=0.8$ (black) for (a) $R_0/L_{Ti}=5.06$, (b) $R_0/L_{Ti}=6.96$, (c) $R_0/L_{Ti}=8.96$, (d) $R_0/L_{Ti}=10.96$, (e) $R_0/L_{Ti}=12.96$, and (f) $R_0/L_{Ti}=14.96$.}
    \label{fig3-1aspectratioeffect}
\end{figure*}

Besides the large 3D parameter scan over $(q,\hat{s},R_0/L_{Ti})$, we also consider several other important effects on the Prandtl number. Figure \ref{fig3-1aspectratioeffect} illustrates the effect of the aspect ratio $\epsilon=r_0/R_0$ on the Prandtl number. Once again, only the simulation data corresponding to nonlinearly unstable turbulence is shown. Comparing the solid, dashed and dotted lines for $\epsilon=0.36, 0.18, 0.45$, respectively, we see that the Prandtl numbers for the cases with tight aspect ratio are smaller than those with a larger aspect ratio. This difference becomes more evident for the cases with larger magnetic shear. Therefore, one concludes that tight aspect ratio is beneficial for obtaining a lower Prandtl number. 
\subsection{Effect of kinetic electrons on the Prandtl number}\label{kineticelectroneffect}
\begin{figure}
    \centering
    \includegraphics[width=0.85\textwidth]{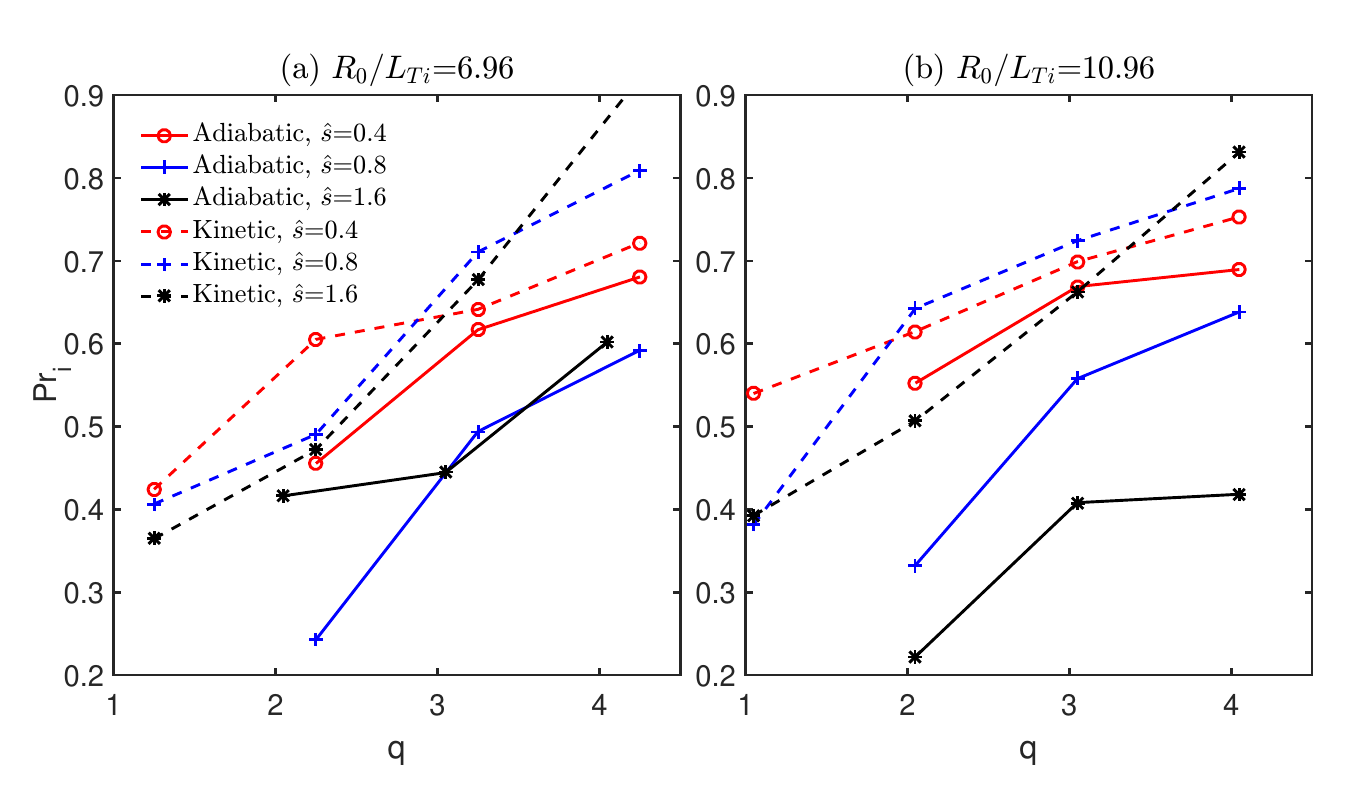}
    \caption{The Prandtl number as function of $q$ from NL simulations with adiabatic electrons (solid) and kinetic electrons (dashed) at $\hat{s}=0.4$ (red), $\hat{s}=0.8$ (blue), and $\hat{s}=1.6$ (black) for (a) $R_0/L_{Ti}=6.96$ and (b) $R_0/L_{Ti}=10.96$.}
    \label{fig3-2kineticeffect}
\end{figure}

Figure \ref{fig3-2kineticeffect} shows the effect of kinetic electrons on the Prandtl number, where we set the density and temperature gradient of the electrons to be the same as the ions. We have checked that the dominant instability remains ITG for all cases considered here. We can see that accounting for kinetic response of electrons increases the Prandtl number when the other input parameters are held fixed. In this comparison, it is however important to note that the cases with kinetic electrons are further away from marginal stability as they have a lower instability threshold. Since the Prandtl number tends to increase with the instability drive $R_0/L_{Ti}$, it makes sense to also expect a higher Prandtl number for the cases with kinetic electrons. Most importantly, one should notice that one can still obtain a low Prandtl number (below 0.4) for certain parameters, and the general trends defining the LMD regime (low $q$, high $\hat{s}$ and low $R_0/L_{Ti}$) that we obtained using adiabatic electrons remain the same with kinetic electrons. Therefore, our simulations with adiabatic electrons correctly capture the parameter dependence of the Prandtl number, which is an important conclusion of this section.

\subsection{Effect of the type of turbulent drive on the Prandtl number}\label{TEMeffect}
\begin{figure*}
    \centering
    \includegraphics[width=0.87\textwidth]{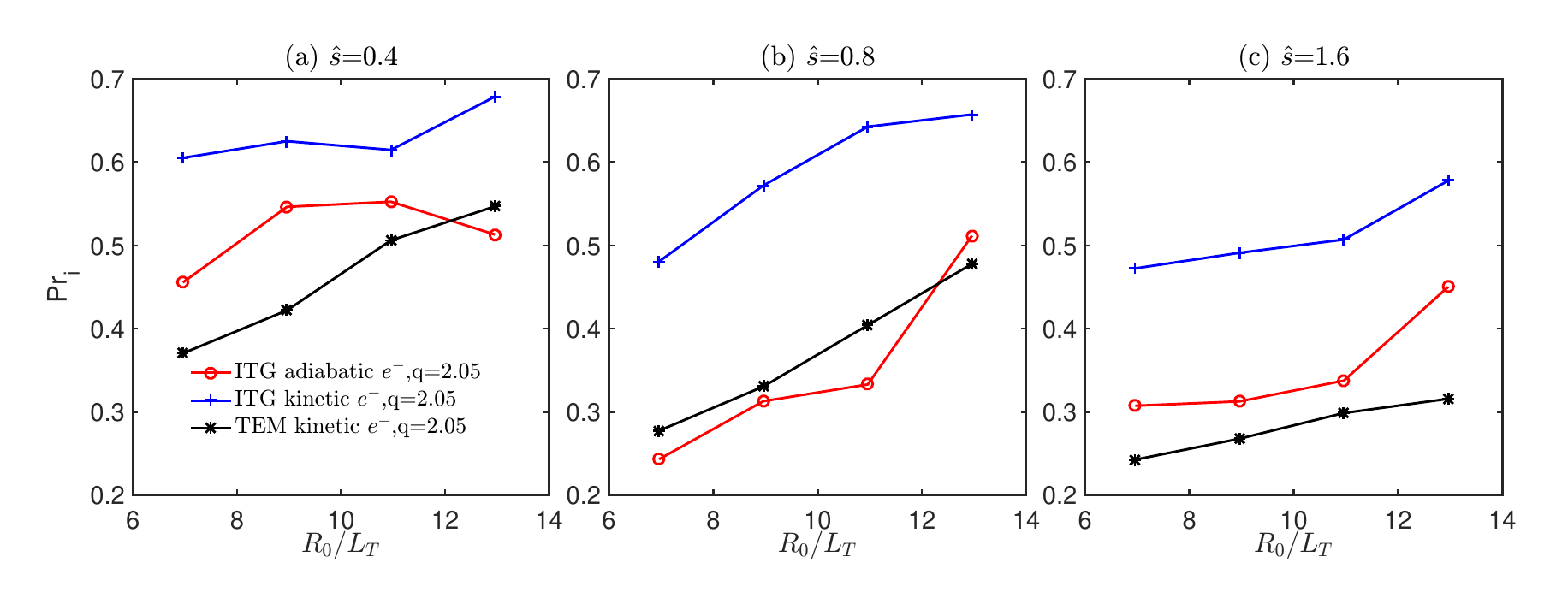}
    \caption{The Prandtl number as a function of $R_0/L_T$ computed for ITG driven turbulence with adiabatic electrons (red), kinetic electrons (blue) and for pure TEM driven turbulence (black) at $q=2.05$ for (a) $\hat{s}=0.4$, (b) $\hat{s}=0.8$, and (c) $\hat{s}=1.6$. Note that for the ITG cases $R_0/L_T$ stands for $R_0/L_{Ti}$, while for the TEM cases, $R_0/L_T$ denotes $R_0/L_{Te}$.}
    \label{fig3-3TEMeffect}
\end{figure*}

Finally, in Fig. \ref{fig3-3TEMeffect}, we study the effect of the type of turbulent drive on the Prandtl number. Here we consider some pure Trapped Electron Mode (TEM) driven cases, where we set $R_0/L_{Ti}=0$, $R_0/L_n=2.22$, $\epsilon=0.36$, and vary $R_0/L_{Te}$ and $\hat{s}$ at a constant $q=2.05$. Kinetic electrons must be considered for these pure TEM cases. For these TEM cases, the electron heat flux $\hat{Q}_e$ dominates over the ion heat flux $\hat{Q}_i$, so we change the Prandtl number definition from Eq. \eqref{eq_Prandtlnumber} to
\begin{equation}\label{eq_PrandtlnumberTEM}
\text{Pr}_i=\frac{\hat{\Pi}_i}{\hat{Q}_e}\frac{R_0}{L_{Te}}\frac{\epsilon}{q\omega_{\perp}}\frac{c_s}{R_0}.
\end{equation}
From Fig. \ref{fig3-3TEMeffect}, we see that the pure TEM cases have similar Prandtl numbers as the ITG dominated cases with adiabatic electrons. Therefore, we see that the instability type does not affect the Prandtl number significantly. The more significant factor once again appears to be how far the system is away from marginal stability: lower Prandtl numbers are achieved when the system is close to marginal stability.

\section{Results using tilted ellipse geometry}\label{tiltellipsegeo}
\subsection{Simulation parameters}
Next, we will consider the intrinsic flow shear driven by up-down asymmetric flux surfaces. Such cases are practically important for predicting the self-consistent rotation gradient that can be achieved in experiments. Up-down asymmetry creates an intrinsic toroidal momentum flux, which will give rise to a rotation gradient that will quickly grow and drive a diffusive momentum flux. In steady state and in the absence of external momentum sources, these two fluxes must cancel \cite{ball2014,ball2016a,ball2018}. Otherwise, the finite momentum flux would cause the rotation profile to change in time. The rotation gradient expected in experiment is therefore the one that achieves $\Pi_i=0$. In order to determine the self-consistent effect of this flow shear on the heat flux, we scan the flow shear $\omega_{\perp}$ to find the value at which the toroidal angular momentum flux $\Pi_i$ goes to zero and then evaluate the corresponding value of the heat flux. By doing so, we self-consistently predict how much flow shear will be created by up-down asymmetry as well as the corresponding steady state heat flux. 

\begin{figure}
    \centering
    \includegraphics[width=0.85\textwidth]{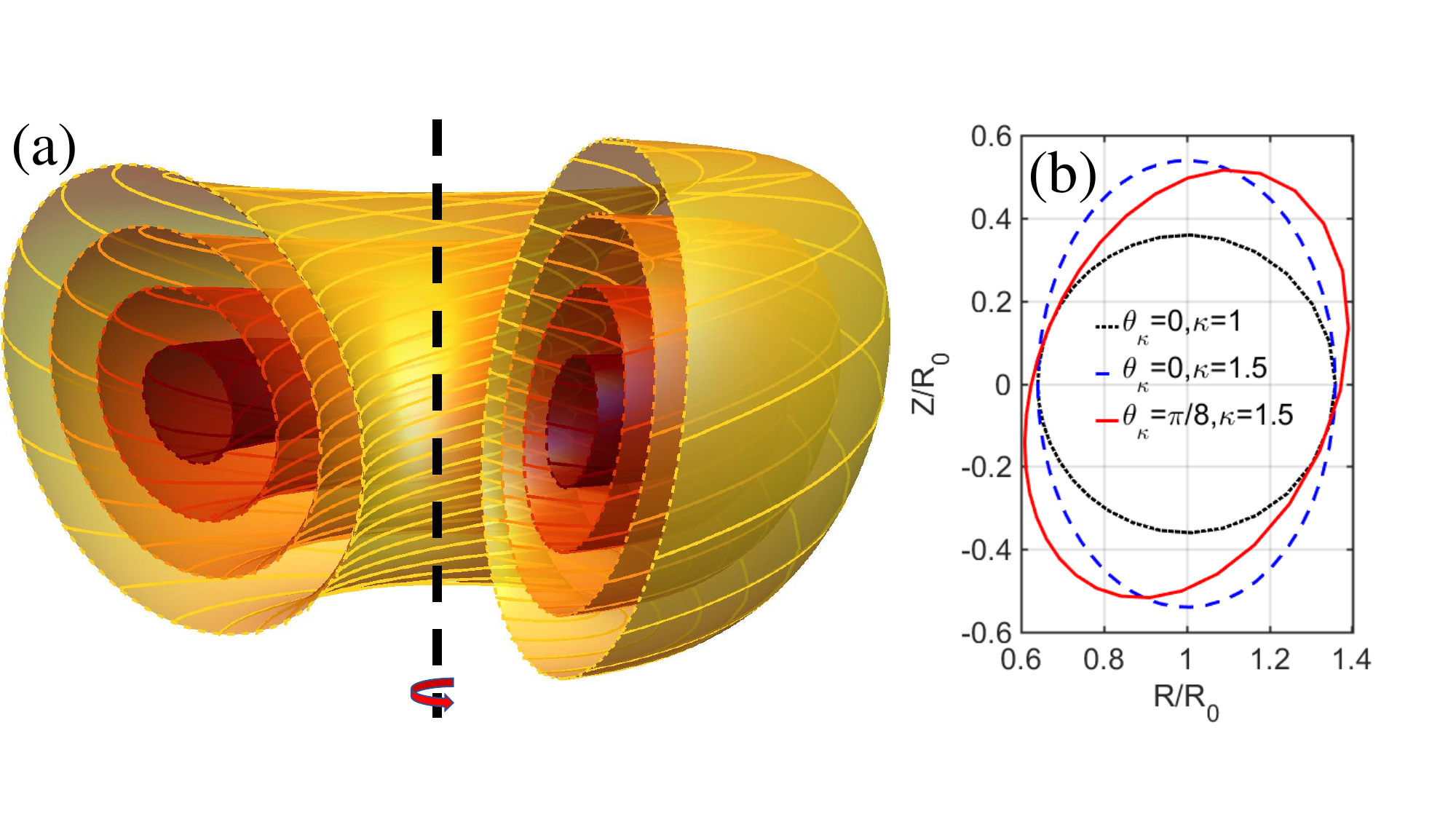}
    \caption{Subplot (a) shows an illustration of up-down asymmetric magnetic flux surfaces, shown with the axis of toroidal symmetry (denoted by the black dashed line). Subplot (b) shows the poloidal cross-sectional shape of the flux surfaces of circular (black dotted), elongated (blue dashed), and tilted elliptical (red solid) flux surfaces. The circular and tilted elliptical geometries are considered in Section \ref{circulargeo} and \ref{tiltellipsegeo}, respectively.}
    \label{fig3updownasyfluxsurface}
\end{figure}

To do so, we take the up-down asymmetric flux surface, illustrated by Fig. \ref{fig3updownasyfluxsurface} (a) and parameterized by Miller geometry. It has an elongation of $\kappa=1.5$, aspect ratio of $\epsilon=0.36$ and tilt angle of $\theta_{\kappa}=\pi/8$ (see also Fig. \ref{fig3updownasyfluxsurface} (b)). Based on previous work \cite{ball2018}, this tilt angle creates the strongest intrinsic momentum flux. Adiabatic electrons are assumed. We again scan over a 3D parameter space of $q=[1.05 - 5.05]$ with $\Delta q=1$, $\hat{s}=\bigl\{0.1, 0.4, 0.8, 1.2, 1.6\bigl\}$ and various values of $R_0/L_{Ti}$. The values of $R_0/L_{Ti}$ are set differently for each $(q,\hat{s})$ to make sure they are above the critical gradient needed to drive turbulence. For each $(q,\hat{s},R_0/L_{Ti})$ set, we further scan the flow shear $\omega_{\perp}$ to find the value for which $\hat{\Pi}_i=0$. Carrying out this flow shear scan with NL simulations would be very computationally expensive, so we used a Quasi-Linear (QL) model \cite{Sun2024NF} to estimate this flow shear value. The QL model combines the contributions from different linear eigenmodes with different $k_y$ mode numbers and ballooning angles $\chi_0$. This model predicts the ratio of momentum flux to heat flux as a function of $\omega_{\perp}$ and is able to find the $\omega_{\perp}$ for which $\hat{\Pi}_i=0$, namely, $\omega^{\Pi_i=0}_{\perp}$. We then perform a single NL simulation at this flow shear to determine the heat flux reduction. For each point in the 3D parameter set $(q,\hat{s},R_0/L_{Ti})$, several other NL simulations are performed at $\omega_{\perp}=\bigl\{0, -0.1, -0.3\bigl\} c_s/R_0$ to calculate the Prandtl number. Therefore, four NL simulations at $\omega_{\perp}=\bigl\{0, -0.1, -0.3, \omega^{\Pi_i=0}_{\perp}\bigl\} c_s/R_0$ are performed for each point in the 3D parameter set $(q,\hat{s},R_0/L_{Ti})$. With the help of the QL model, we reduce the number of required NL gyrokinetic simulations from 1000 to around 400. Note that the QL model is only used to obtain the steady state value of flow shear. The grid parameters are similar to those shown in Table \ref{tabletightpara}, except $n_{\mu}$ is increased from $9$ to $16$. Convergence checks have been performed for the simulations at the corners of the considered parameter domain. 

\subsection{Simulation results}
\begin{figure}
    \centering
    \includegraphics[width=0.83\textwidth]{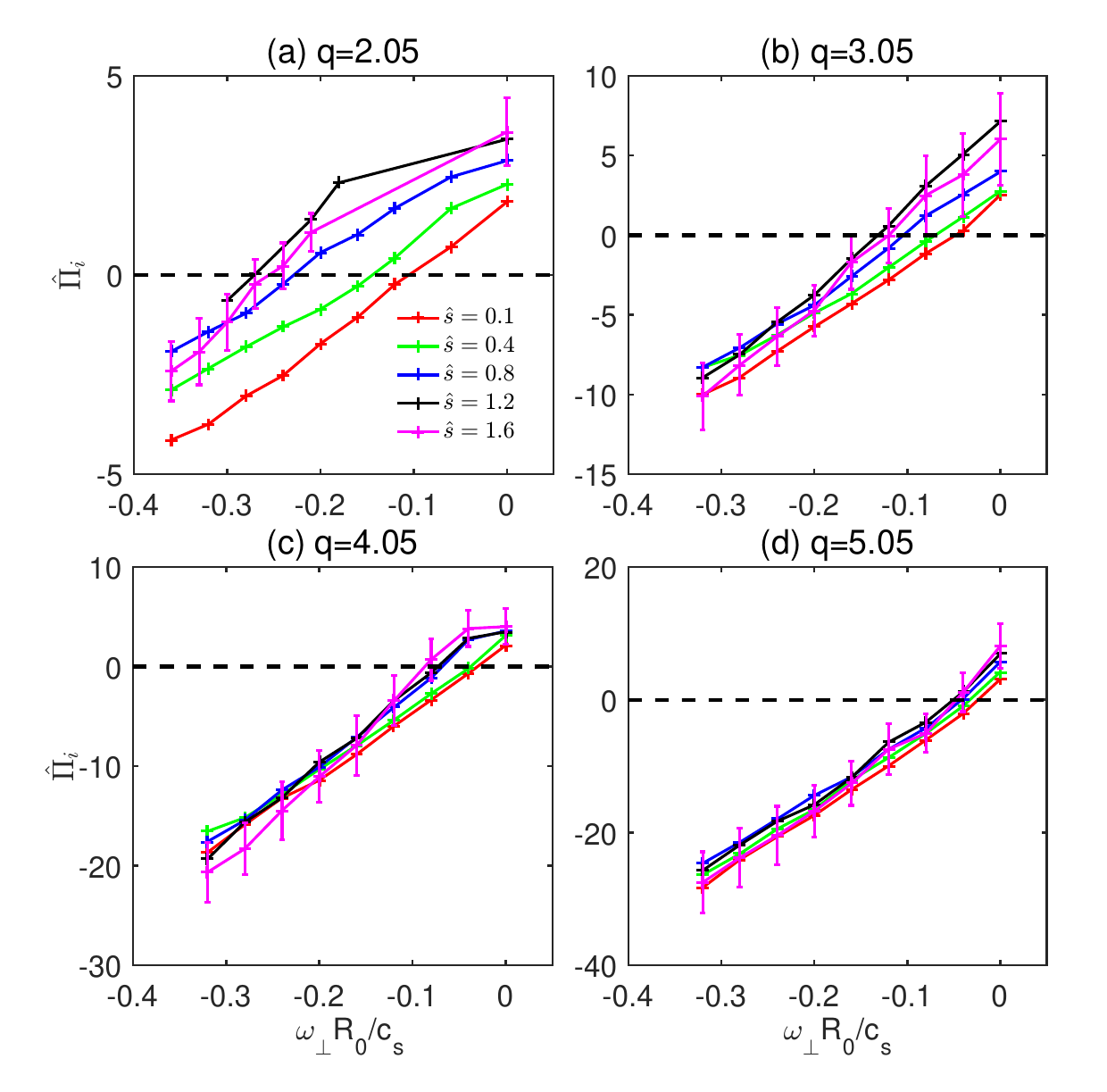}
    \caption{The ion toroidal angular momentum flux $\hat{\Pi}_i$ as a function of flow shear $\omega_{\perp}$ for NL simulations for $\hat{s}=0.1$ (red), $\hat{s}=0.4$ (green), $\hat{s}=0.8$ (blue), $\hat{s}=1.2$ (black) and $\hat{s}=1.6$ (magenta) as well as (a) $q=2.05$, (b) $q=3.05$, (c) $q=4.05$ and (d) $q=5.05$. The other physical parameters are $\epsilon=0.36$, $R_0/L_{Ti}=10.96$, $\kappa=1.5$ and $\theta_{\kappa}=\pi/8$.}
    \label{fig4flowshearmom}
\end{figure}

\begin{figure}
    \centering
    \includegraphics[width=0.83\textwidth]{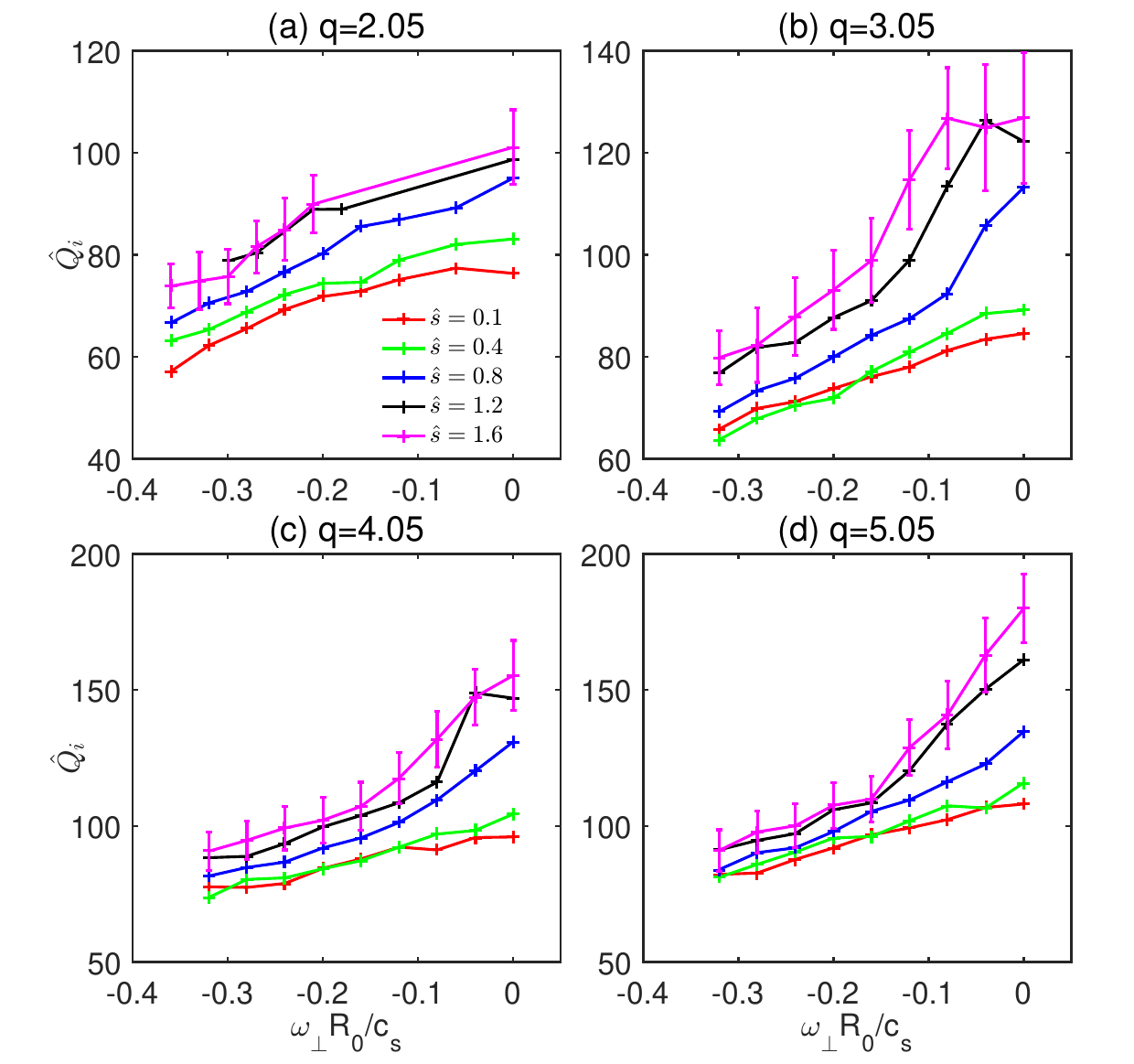}
    \caption{Same as Fig. \ref{fig4flowshearmom} but for the ion heat flux $\hat{Q}_i$.}
    \label{fig4flowshearheat}
\end{figure}

\begin{figure}
    \centering
    \includegraphics[width=0.83\textwidth]{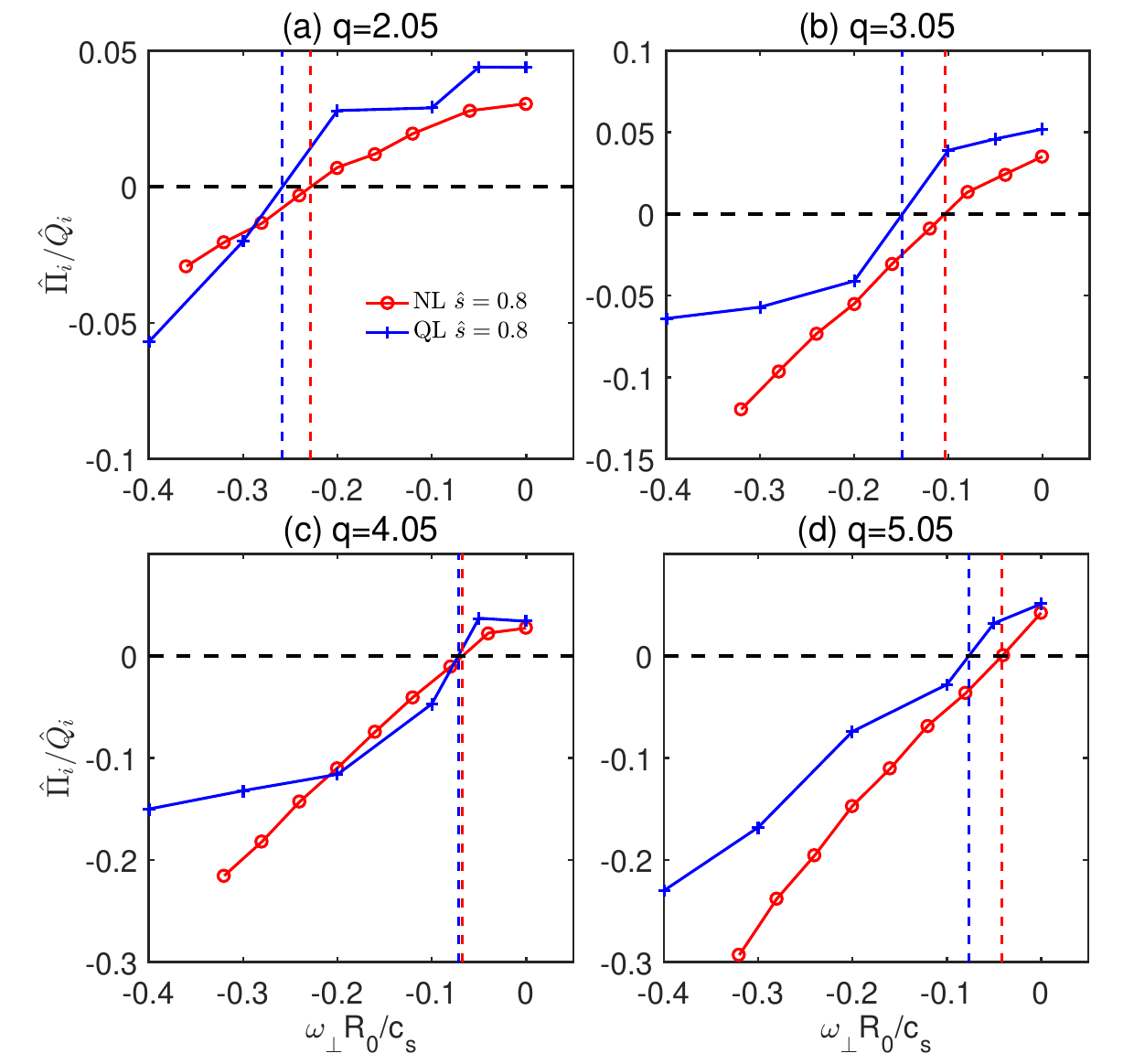}
    \caption{Ratio of ion momentum to heat flux $\hat{\Pi}_i/\hat{Q}_i$ as a function of flow shear $\omega_{\perp}$ obtained with NL simulations (red) and the QL model (blue) for (a) $q=2.05$, (b) $q=3.05$, (c) $q=4.05$ and (d) $q=5.05$. We take $\hat{s}=0.8$ cases, with all the other physical parameters the same as in Fig. \ref{fig4flowshearmom}.}
    \label{fig4flowshearQLNLcompare}
\end{figure}

To show how one finds the $\omega_{\perp}$ value at which $\hat{\Pi}_i=0$, Figures \ref{fig4flowshearmom}, \ref{fig4flowshearheat} and \ref{fig4flowshearQLNLcompare} present some NL and QL simulations at $R_0/L_{Ti}=10.96$. In Figures \ref{fig4flowshearmom} and \ref{fig4flowshearheat}, we plot the NL momentum flux $\hat{\Pi}_i$ and the NL heat flux $\hat{Q_i}$ as a function of $\omega_{\perp}$ for different $\hat{s}$ and $q$ values. At zero flow shear, the momentum flux $\hat{\Pi}_i$ is positive, indicating that the up-down asymmetric flux surface creates a positive intrinsic momentum flux. As we change $\omega_{\perp}$ to increasingly negative values, $\hat{\Pi}_i$ gradually drops to zero and then to negative values. The value of $\omega_{\perp}$ at which $\hat{\Pi}_i=0$ occurs (denoted by the black dashed lines in Fig. \ref{fig4flowshearmom}) indicates the flow shear value driven by the intrinsic momentum flux. As expected from the results in Section \ref{circulargeo}, we see that one can get stronger flow shear at low $q$ and high $\hat{s}$. As the magnitude of $\omega_{\perp}$ is increased, Fig. \ref{fig4flowshearheat} shows that the heat flux is reduced as expected as well \cite{CassonExBshear2009,WaltzFlowShear1998,Biglari1990}. We notice from Fig. \ref{fig4flowshearmom} that the dependence of the momentum flux on flow shear is nearly linear, while the heat flux shown in Fig. \ref{fig4flowshearheat} is not linear. This interesting observation will be discussed in Section \ref{tiltellipsediscussion}. Figure \ref{fig4flowshearQLNLcompare} plots the momentum to heat flux ratio $\hat{\Pi}_i/\hat{Q}_i$ as a function of $\omega_{\perp}$ for $\hat{s}=0.8$ and compares results obtained with both the QL model and NL simulations. Despite some differences between corresponding curves, a good match is observed for the $\omega_{\perp}$ values at which $\hat{\Pi}_i$ goes to zero (denoted by the vertical dashed lines). Therefore, for the cases in the 3D parameter space $(q,\hat{s},R_0/L_{Ti})$ with $R_0/L_{Ti}\neq10.96$, the steady state flow shear values are estimated with the QL model, using the same approach as given in Ref. \cite{Sun2024NF}.

One can still calculate the Prandtl number for the up-down asymmetric cases even though the momentum flux is no longer purely diffusive. Based on Eq. \eqref{eq_momentum2} and $\omega_{\perp}=-(r_0/q)d\Omega_i/dx$, the momentum diffusivity can be written as 
\begin{equation}
    D_{\Pi i}=\frac{\Delta \Pi_i}{\Delta\omega_{\perp}n_i m_i}\frac{\epsilon}{q c_s},
\end{equation}
where $\Delta \Pi_i$ is the difference in $\Pi_i$ for two different cases with different $\omega_{\perp}$ (that differs by $\Delta\omega_{\perp}$). The expression for heat diffusivity is 
\begin{equation}
    D_{Q i}=\frac{Q_i}{n_i dT_i/dx}.
\end{equation}
Therefore, the Prandtl number, defined by $\text{Pr}_i=D_{\Pi i}/D_{Q i}$, becomes
\begin{equation}\label{eq_Prandtlnumber220}
    \text{Pr}_i=\frac{\Delta\hat{\Pi}_i}{\hat{Q}_i}\frac{R_0}{L_{Ti}}\frac{\epsilon}{q\Delta\omega_{\perp}}\frac{c_s}{R_0}.
\end{equation}
Although this gives the correct expression of Prandtl number, one has to remember that it comes from Eq. \eqref{eq_momentum2}, which is simply a Taylor expansion with respect to $d\Omega_i/dx$. This expression can be expected to hold for small values of the flow shear, but not necessarily when it is large. In particular, Fig. \ref{fig4flowshearmom} (a) and (c) show that linear extrapolating the two right-most data points does not give a good estimate of the zero crossing, which is what we are most interested in. Therefore, the Prandtl number expression that we use is
\begin{equation}\label{eq_Prandtlnumber22}
\overline{\text{Pr}}_i=\Delta\left(\frac{\hat{\Pi}_i}{\hat{Q}_i}\right)\frac{R_0}{L_{Ti}}\frac{\epsilon}{q\Delta\omega_{\perp}}\frac{c_s}{R_0}.
\end{equation}
Here $\Delta(\hat{\Pi}_i/\hat{Q}_i)$ is the difference in the ratio $\hat{\Pi}_i/\hat{Q}_i$ for two cases with different values of flow shear, with a difference of $\Delta\omega_{\perp}$ and $\overline{\text{Pr}}_i$ denotes that the Prandtl number is different from Eq. \eqref{eq_Prandtlnumber220}. There are three reasons why we use Eq. \eqref{eq_Prandtlnumber22}: First, when taking two cases with non-zero $\omega_{\perp}$ (e.g. we use $\omega_{\perp}=-0.1,-0.3 c_s/R_0$), we find a good linear relation between $\hat{\Pi}_i/\hat{Q}_i$ and $\omega_{\perp}$ (see Fig. \ref{fig4flowshearQLNLcompare}). Therefore, in this way, a good prediction to the zero-crossing point $\omega^{\Pi_i=0}_{\perp}$ can be found.
Second, the ratio $\hat{\Pi}_i/\hat{Q}_i$ can be more easily estimated using QL models (compared to estimating $\hat{\Pi}_i$ alone). Third, if one was to use Eq. \eqref{eq_Prandtlnumber220}, it is not obvious how to evaluate $\hat{Q}_i$. A natural choice would be from a simulation with $\omega_{\perp}=0$, but this would require an additional simulation. We note that Eqs. \eqref{eq_Prandtlnumber220} and \eqref{eq_Prandtlnumber22} are closely related through the formula $\Delta(\hat{\Pi}_i/\hat{Q}_i)=\Delta\hat{\Pi}_i/\hat{Q}_i-\hat{\Pi}_i/\hat{Q}^2_i\Delta \hat{Q}_i$. 
For almost all cases in our 3D parameter set, the second term $-\hat{\Pi}_i/\hat{Q}^2_i\Delta \hat{Q}_i$ is small compared to the first term, thereby justifying the usage of Prandtl number expression Eq. \eqref{eq_Prandtlnumber22}. 

\begin{figure*}
    \centering
    \includegraphics[width=0.87\textwidth]{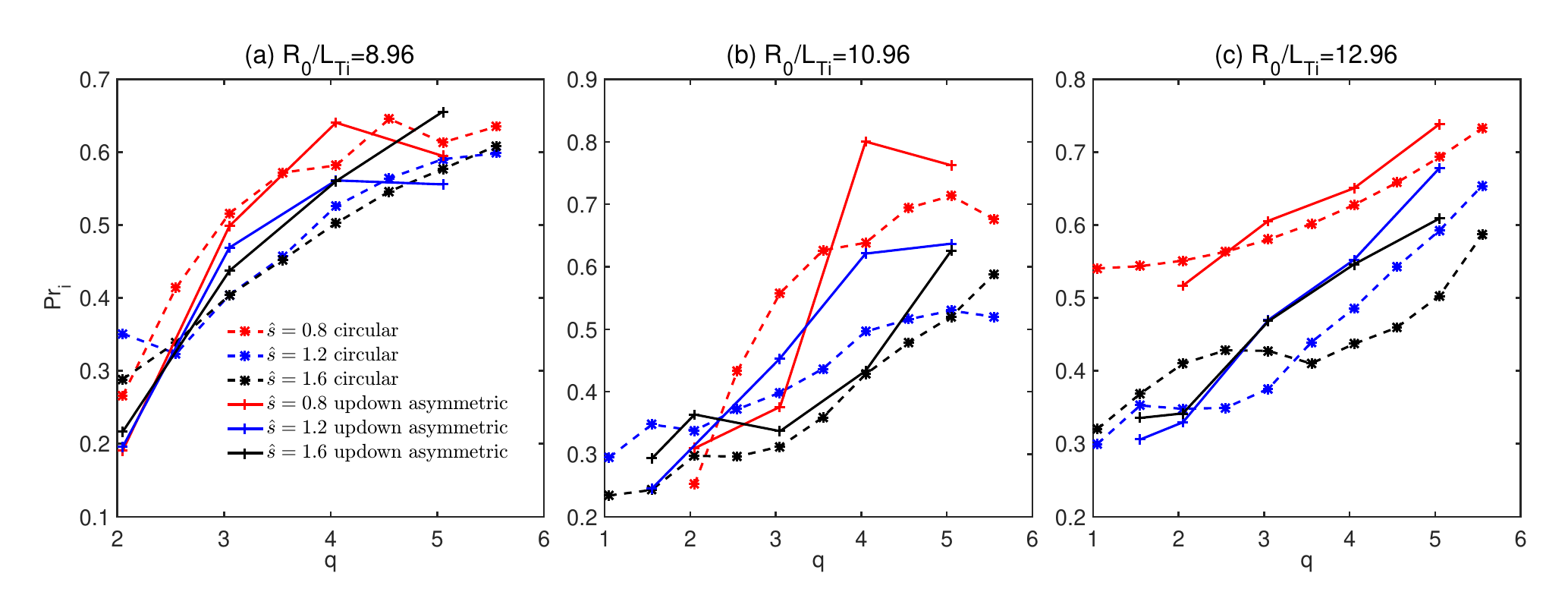}
    \caption{The ion Prandtl number $\text{Pr}_i$ as a function of the safety factor $q$ calculated by NL simulations with circular flux surfaces (dashed) and up-down asymmetric (full line) for $\hat{s}=0.8$ (red), $\hat{s}=1.2$ (blue), and $\hat{s}=1.6$ (black) as well as (a) $R_0/L_{Ti}=8.96$, (b) $R_0/L_{Ti}=10.96$ and (c) $R_0/L_{Ti}=12.96$.}
    \label{fig5Prandtlcompare}
\end{figure*}

Therefore, to calculate the Prandtl number for the up-down asymmetric cases, we considered two flow shear values $\omega_{\perp}R_0/c_s=-0.1$ and $-0.3$ for each point in the 3D parameter set $(q,\hat{s},R_0/L_{Ti})$. This is because the Prandtl number remains proportional to the slope of the red lines in Fig. \ref{fig4flowshearQLNLcompare}. The precise choice of flow shear values is not too important since, in the NL simulations, $\hat{\Pi}_i/\hat{Q}_i$ presents a nearly linear dependence on $\omega_{\perp}$ over a wide range of values away from $\omega_{\perp}=0$ (see Fig. \ref{fig4flowshearQLNLcompare}). Figure \ref{fig5Prandtlcompare} compares the Prandtl number between cases with circular geometry and up-down asymmetric cases, with all the other input parameters being the same. As we can see, the Prandtl numbers remain quite similar. This indicates that the Prandtl number is relatively invariant with respect to elongation and tilt of the magnetic flux surface. One can therefore assume that the general parameter dependence shown in the low momentum diffusivity manifolds in Section \ref{circulargeo} still holds true in the case of up-down asymmetric flux surfaces. 

\begin{figure*}
    \centering
    \includegraphics[width=1.05\textwidth]{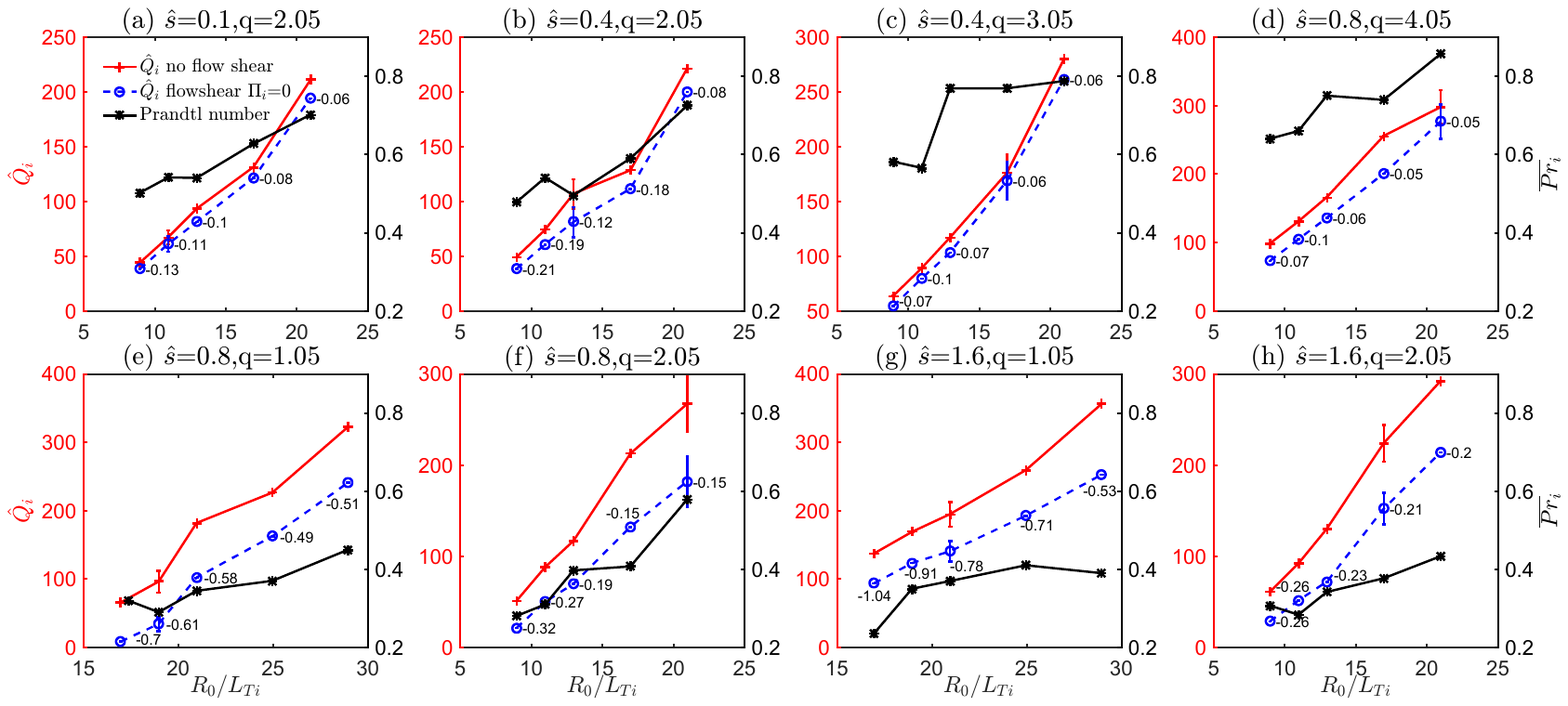}
    \caption{The heat flux (left vertical axis) accounting for the intrinsic rotation shear $\omega_{\perp}$ driven by up-down asymmetry (blue, determined by enforcing $\hat{\Pi}_i=0$), compared to the cases without any flow shear (red) for (a) $\hat{s}=0.1, q=2.05$, (b) $\hat{s}=0.4, q=2.05$, (c) $\hat{s}=0.4, q=3.05$, (d) $\hat{s}=0.8, q=4.05$, (e) $\hat{s}=0.8, q=1.05$, (f) $\hat{s}=0.8, q=2.05$, (g) $\hat{s}=1.6, q=1.05$, and (h) $\hat{s}=1.6, q=2.05$. The flow shear values $\omega_{\perp}R_0/c_s$ required to ensure $\hat{\Pi}_i=0$ are given by the number next to each blue point. The Prandtl numbers (black, right vertical axis) are also shown. The errorbars calculated by a rolling average over the time traces are shown for some example cases.}
    \label{fig6heatreduction}
\end{figure*}

Figure \ref{fig6heatreduction} shows the ion heat flux $\hat{Q}_i$ as a function of $R_0/L_{Ti}$ for up-down asymmetric flux surfaces, using both the intrinsic rotation shear $\omega_{\perp}$ such that $\hat{\Pi}_i=0$ (denoted by blue dashed lines) and $\omega_{\perp}=0$ (denoted by red solid lines). Comparing these two results enables us to estimate the impact of the self-consistent flow shear on the heat flux. Subplots (a)-(d) show four parameter sets that are not in the LMD regime, while subplots (e)-(h) show four parameter sets in the LMD regime. In the LMD regime, one can generate strong flow shear, which causes a significant heat flux reduction. If a case is very close to marginal stability, a nearly complete quench of turbulence can be observed (e.g. subplot (e)). As a result, by extrapolating the red curves and blue curves, the critical gradients increased by up-down asymmetry are $24\%$ for subplot (e) and $20\%$ for subplot (f). The percentages are calculated by $((R_0/L_{Ti})^{\Pi_i=0}_{crit}-(R_0/L_{Ti})^{\omega_{\perp}=0}_{crit})/(R_0/L_{Ti})^{\omega_{\perp}=0}_{crit}$, where $(R_0/L_{Ti})^{\Pi_i=0}_{crit}$ and $(R_0/L_{Ti})^{\omega_{\perp}=0}_{crit}$ are critical temperature gradients of the blue and red curves, respectively. On the other hand, both the flow shear created by up-down asymmetry and the heat flux reduction are relatively weak outside of the LMD regime. From this figure, it can be clearly seen how a lower Prandtl number enables flow shear stabilization of the turbulence. A lower Prandtl number facilitates larger flow shear, and this flow shear in turn suppresses the turbulence and reduces the heat flux. 

\begin{figure}
    \centering
    \includegraphics[width=0.8\textwidth]{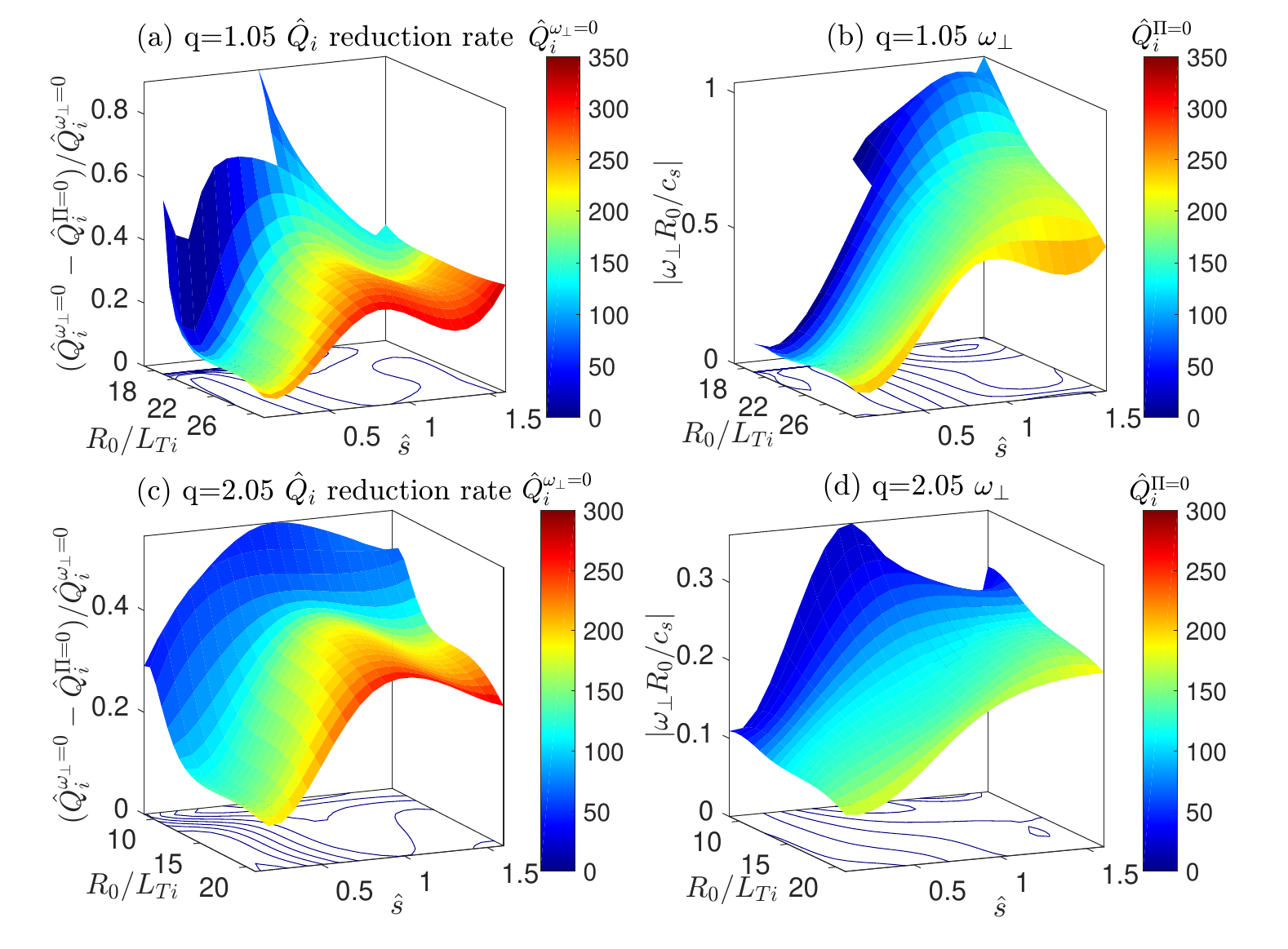}
    \caption{The fractional heat flux reduction as a function of $\hat{s}$ and $R_0/L_{Ti}$ for (a) $q=1.05$ and (c) $q=2.05$, and the corresponding value of $\omega_{\perp}$ that ensures $\hat{\Pi}_i=0$ for (b) $q=1.05$ and (d) $q=2.05$. The color maps on the surfaces in (a) and (c) denote the heat flux without flow shear $\hat{Q}^{\omega_{\perp}=0}_i$, while the color maps on the surfaces in (b) and (d) denote the heat flux at zero momentum flux $\hat{Q}^{\Pi_i=0}_i$.}
    \label{fig7heatreduction}
\end{figure}

The fractional heat flux reduction and the flow shear are summarized in Fig. \ref{fig7heatreduction}. The fractional heat flux reduction is defined as $(\hat{Q}^{\omega_{\perp}=0}_{i}-\hat{Q}^{\Pi=0}_{i})/\hat{Q}^{\omega_{\perp}=0}_{i}$, where $\hat{Q}^{\omega_{\perp}=0}_{i}$ and $\hat{Q}^{\Pi=0}_{i}$ are the heat flux without flow shear and at zero momentum flux, respectively. We can see that a larger fractional reduction occurs near marginal stability, given that the flow shear is primarily modifying the critical gradient but not the stiffness of the heat flux curve versus $R_0/L_{Ti}$. Additionally, the fractional reduction tends to peak at around $\hat{s}=0.8$. The gap area without data points in (a) and (b) at low $\hat{s}$ and low $R_0/L_{Ti}$ denotes a stable region without turbulence for which one cannot get a meaningful fractional heat flux reduction. Subplots (b) and (d) represent the flow shear generated by the intrinsic momentum flux at steady state. Studying the flow shear, one sees that it generally increases with $\hat{s}$. This is both because the intrinsic momentum flux typically increases with the increase of $\hat{s}$, as shown in Fig. \ref{fig4flowshearmom}  \cite{Diamond_2013NFreview,Parratheory_2015}, but also because of a lower Prandtl number.  On the other hand, we see that getting closer to marginal stability does not significantly increase $\omega_{\perp}$. 

\subsection{Discussion of results}\label{tiltellipsediscussion}
\begin{figure}
    \centering
    \includegraphics[width=0.92\textwidth]{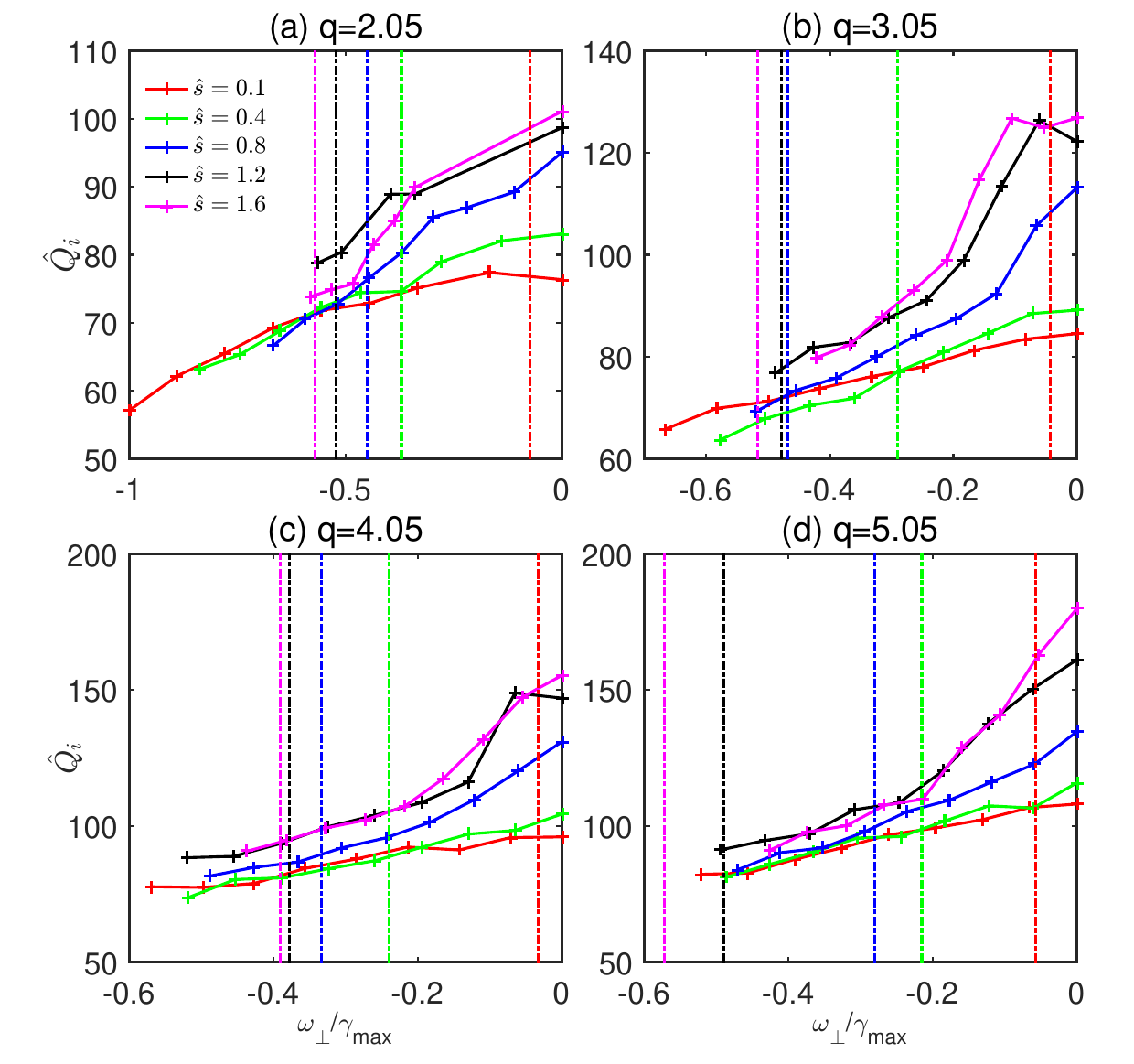}
    \caption{Ion heat flux $\hat{Q}_i$ as a function of $\omega_{\perp}/\gamma_{max}$ for different $\hat{s}$ at $R_0/L_{Ti}=10.96$ and (a) $q=2.05$, (b) $q=3.05$, (c) $q=4.05$, and (d) $q=5.05$, illustrating the suppression of heat flux by flow shear. The vertical dashed lines indicate the flow shear values $\omega_{\perp}=\hat{s}z_{avg}/\tau_{NL}$ for the corresponding solid curves with the same color. Here, $\gamma_{max}$ is the maximum linear growth rate for $\omega_{\perp}=0$ and thus represents a different normalization for the different $\hat{s}$ curves.}
    \label{figmeetingpoint}
\end{figure}

In this section, we discuss the physics behind the suppression of $\hat{Q}_i$ by flow shear. The effect of the pinch term in the momentum transport equation is neglected here but is discussed in Appendix \ref{AppendixA}. As shown in Figs. \ref{fig4flowshearmom} and \ref{fig4flowshearheat}, the variation of momentum flux is nearly linear with flow shear near the region where the curves pass through $\hat{\Pi}_i=0$, while the variation of heat flux with flow shear is not linear. An interesting observation is that the flow shear is more effective at reducing $\hat{Q}_i$ at higher $\hat{s}$, comparing the different results shown in Fig. \ref{fig4flowshearheat}. When the flow shear is zero, the heat flux is higher for larger $\hat{s}$, which is a well-known result. This is because $\hat{s}\simeq 1$ better aligns the linear modes with the curvature drive responsible for toroidal ITG turbulence \cite{Antonsen_toroidalITG_1996,Dimits2001ITG, Camenen2016popITG}. On the other hand, as shown by Fig. \ref{figmeetingpoint}, when normalizing the flow shear for each curve by the corresponding maximum linear growth rate at $\omega_{\perp}=0$ (denoted by $\gamma_{max}$), we see that the curves at different $\hat{s}$ values converge onto one another at high flow shear. A similar phenomenon has also been observed in a previous work \cite{CassonExBshear2009}. This behavior, as well as the flow shear values at which the different curves meet each other in Fig. \ref{fig4flowshearheat}, can be understood using the ``double shearing'' coordinate system \cite{NewtonFlowShearUnderstanding2010,ball2019} given by 
\begin{equation}\label{doublecoord1}
    K_x=k_x-k_y\omega_{\perp}t-k_y \hat{s}z
\end{equation}
and
\begin{equation}\label{doublecoord2}
    Z=z+\omega_{\perp}t/\hat{s}, 
\end{equation}
where $K_x$ and $Z$ are the radial wavenumber and parallel direction of the double shearing coordinate system, respectively. This coordinate system accounts for the fact that magnetic shear causes an eddy to be radially sheared as it moves along the field line in an analogous way to how flow shear affects eddies in time. In gyrokinetic equation, $K_x$ is the only way that $\hat{s}$ plays a role regulating the turbulence. Therefore, when the last term in Eq. \eqref{doublecoord1} becomes negligible as one increases the flow shear, the turbulence is no longer affected by the magnetic shear and the different curves should coincide \cite{ball2019,Sun2024NF}. We can evaluate the flow shear value at which the third term $k_y \hat{s}z$ becomes unimportant by equating second and third terms in Eq. \eqref{doublecoord1}, leading to the following estimate
\begin{equation}\label{zbbalance}
    \omega_{\perp}\tau_{NL}\sim \hat{s}z_{avg}.
\end{equation}
Here $\tau_{NL}$ is the nonlinear decorrelation time, which is evaluated as the time interval $\Delta t$ such that the auto-correlation of the electrostatic potential associated to the fluctuations
\begin{equation}\label{eq_NLCdecorr}
C(\Delta t)=\langle\phi_{NZ}(x,y,z=0,t)\phi_{NZ}(x,y,z=0,t+\Delta t)\rangle_{x,y,t}/\langle|\phi_{NZ}|^2\rangle_{x,y,t}
\end{equation}
drops to $1/e$. The subscript ``NZ'' denotes the non-zonal component and the bracket average is $\langle...\rangle_{x,y,t}=1/(L_xL_yT)\int^{L_x/2}_{-L_x/2}dx\int^{L_y/2}_{-L_y/2}dy\int^{t_0+T}_{t_0}dt(...)$, where $t_0$ is the starting time of the quasi-steady state, $L_x$ and $L_y$ are the box sizes in $x$ and $y$ direction, and $T=100R_0/c_s$. The fluctuation weighted average value of $z$ in Eq. \eqref{zbbalance} is estimated by
\begin{equation}\label{eq_zavg}
    z_{avg}=\sqrt{\frac{\int dz |\phi(k_x=0,k_y=k_{y,max},z)|^2 J(z) z^2}{\int dz |\phi(k_x=0,k_y=k_{y,max},z)|^2 J(z)}},
\end{equation}
where $k_{y,max}$ is the $k_y$ with the maximum linear growth rate, $J(z)$ is the Jacobian, and $\phi$ comes from the same NL simulations as used to calculate $\tau_{NL}$. We can thus use Eq. \eqref{zbbalance} to estimate the value of $\omega_{\perp}=\hat{s}z_{avg}/\tau_{NL}$ at which the magnetic shear term starts to become negligible compared to the flow shear term. When this is the case, the curves for different $\hat{s}$ in Fig. \ref{figmeetingpoint} should meet each other. The critical value of $\omega_{\perp}$ is shown by the different vertical dashed lines in Fig. \ref{figmeetingpoint}, which show reasonable agreement. We can see that one can obtain a reasonable estimate for the $\omega_{\perp}$ value at which the flow shear suppression dominates.

\section{Simulation results using a MAST experimental equilibrium}\label{MASTgeo}
\begin{figure*}
    \centering
    \includegraphics[width=0.94\textwidth]{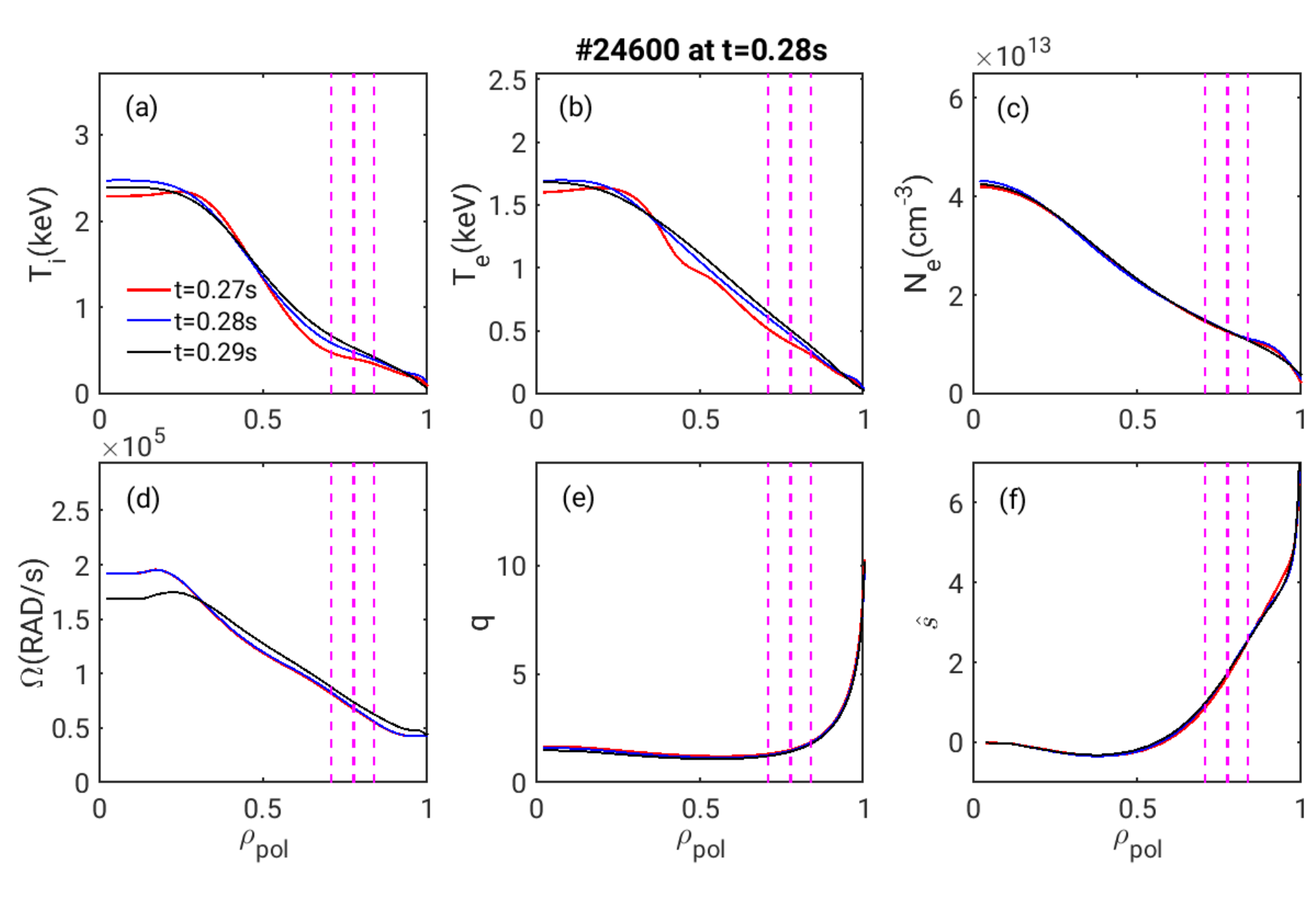}
    \caption{Profiles of (a) ion temperature, (b) electron temperature, (c) electron density, (d) toroidal rotation angular velocity, (e) safety factor, and (f) magnetic shear for MAST shot 24600 at $t=\bigl\{0.27s, 0.28s, 0.29s\bigl\}$. The pink vertical dashed lines denote the three radial locations that we consider for detailed analysis.}
    \label{fig8MASTkinetic}
\end{figure*}

Based on the results in Sections \ref{circulargeo} and \ref{tiltellipsegeo}, a combination of low $q$, tight aspect ratio, normal to high $\hat{s}$ and low $R_0/L_T$ ensures low values of Prandtl number. Furthermore, combining low Prandtl number with up-down asymmetry can create strong intrinsic flow shear. One might be concerned that the combination of low $q$ and high $\hat{s}$ may not be easily achieved in experiments. In this section, we show that this combination is in fact possible in real spherical tokamaks and would create strong flow shear if the magnetic flux surfaces were tilted. To do so, we consider experimental equilibria achieved in the Mega Ampere Spherical Tokamak (MAST) \cite{MAST_Lloyd_2011,Guttenfelder_2017,Harrison_2019} and perform a study similar to that in Section \ref{tiltellipsegeo}. We use the kinetic profiles of MAST shot number 24600 at $t=0.28s$, shown in Fig. \ref{fig8MASTkinetic}, where we define the radial coordinate $\rho_{pol}=\psi^{1/2}_n$ using the normalized poloidal flux $\psi_n$. Shot 24600 has 3MW of on-axis co-NBI injection starting from $t=0.2s$, creating strong flow shear and driving momentum flux \cite{Field_2011NF}. The kinetic profiles are obtained from running the interpretative transport code TRANSP \cite{TRANSPoriginal}. This particular shot was chosen because it has a low $q$ profile over a large radial range. The outer flux surfaces have a tight aspect ratio and a relatively low value of $q$. Around $t=0.28s$, the system reaches a quasi-steady state with relatively steady kinetic profiles and no significant MHD activity \cite{Field_2011NF}. For use in gyrokinetic simulations, the kinetic profiles are averaged over $t=0.27-0.29s$ to reduce the numerical uncertainty from the TRANSP calculations. This uncertainty is reflected in the three different instantaneous time traces shown in Fig. \ref{fig8MASTkinetic}. We consider three radial locations $\psi_n=\bigl\{0.5, 0.6, 0.7\bigl\}$, corresponding to $\rho_{pol}=\bigl\{0.71, 0.77, 0.84\bigl\}$ and perform detailed analysis using GENE. We used the Pyrokinetic code \cite{Patel_pyrokinetics_2022} to obtain GENE input files from TRANSP calculations. These three locations were chosen because they have turbulence dominantly driven by the ITG instability. We also tried simulating other radial locations, but they were either dominated by Micro-Tearing Modes (MTM) or Electron Temperature Gradient (ETG) modes, which were much more computationally challenging. The physical parameters of the three radial locations are given in Table \ref{table11}.

\begin{figure}
    \centering
    \includegraphics[width=0.8\textwidth]{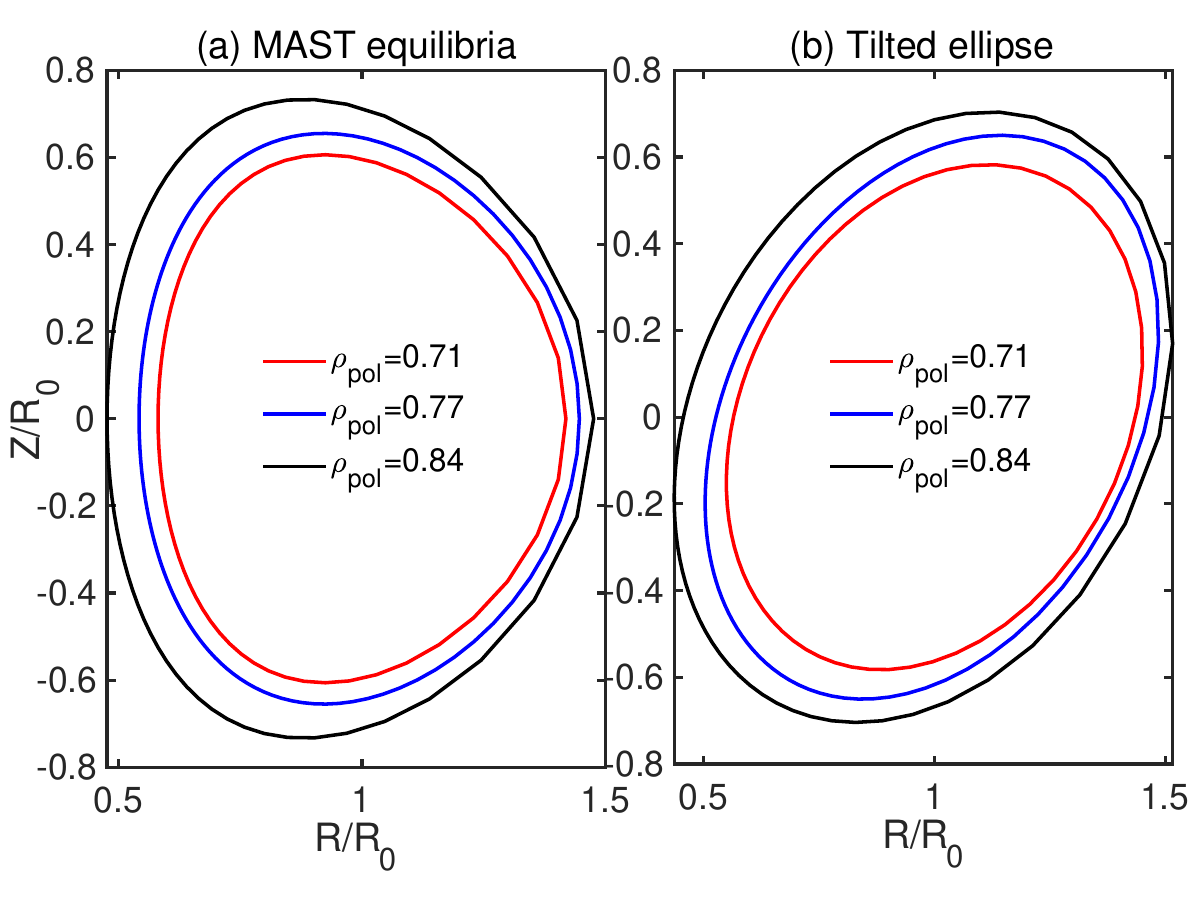}
    \caption{The magnetic flux surface shapes at $\rho_{pol}=\bigl\{0.71, 0.77, 0.84\bigl\}$ for (a) MAST shot 24600 at time $t=0.28s$ and (b) an artificial tilted elliptical geometry without triangularity, but with an elongation tilt angle of $\theta_{\kappa}=\pi/8$.}
    \label{fig9MASTgeo}
\end{figure}

\begin{figure}
    \centering
    \includegraphics[width=0.85\textwidth]{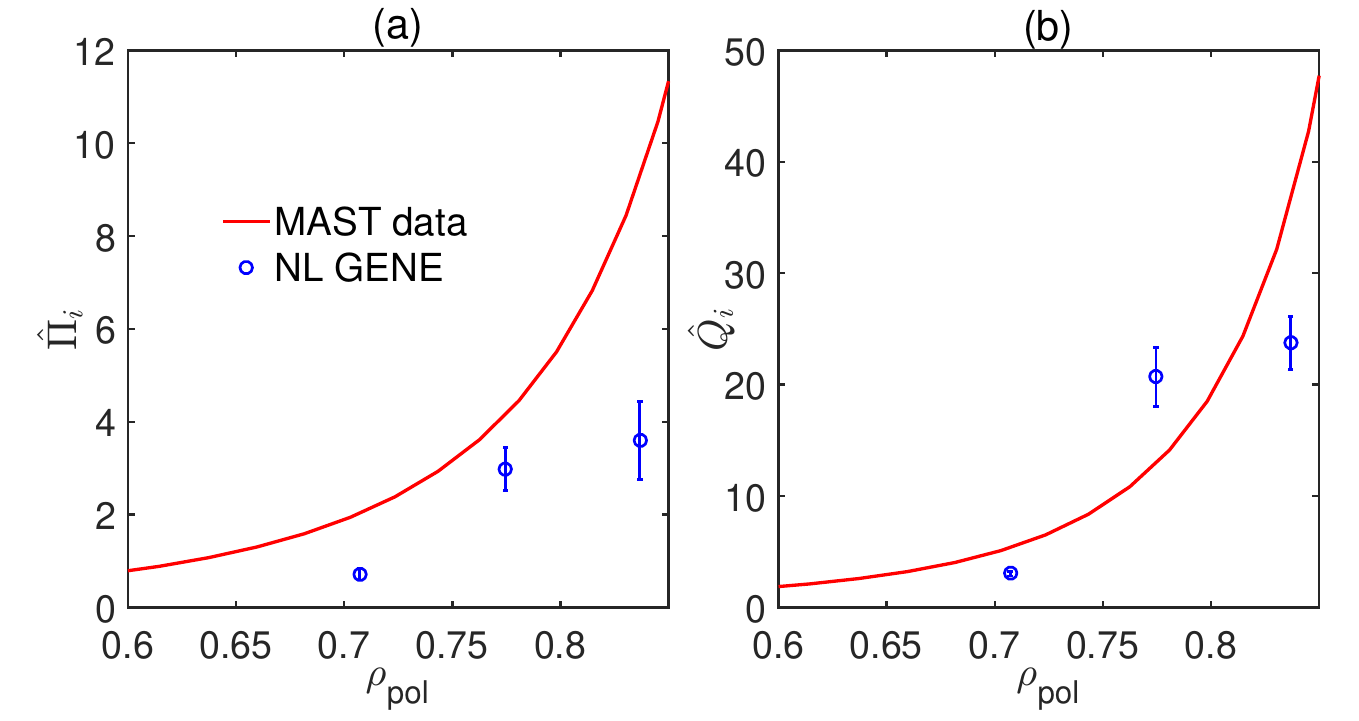}
    \caption{A comparison between fluxes calculated with the TRANSP interpretative code (red lines) and NL GENE simulations (blue points) for (a) the momentum and (b) the heat fluxes in local gyroBohm units.}
    \label{fig10heatmomwithMAST}
\end{figure}

Figure \ref{fig9MASTgeo} (a) shows the shapes of the MAST magnetic flux surfaces at $\rho_{pol}=\bigl\{0.71, 0.77, 0.84\bigl\}$. In order to validate GENE simulations, we first performed simulations using the experimental parameters, including flow shear, and compared the heat flux and momentum flux with the experimental values. Note that all the simulations using the MAST geometry include kinetic electrons. The simulation grid parameters are $(n_{k_x},n_{k_y},n_z,n_{v_{||}},n_{\mu})=(192,64,64,32,16)$, with box size $L_x=200\rho_i$, $L_y=125\rho_i$, $v_{||}/\sqrt{2T_i/m_i}\in [-3,3]$ and $\sqrt{\mu/(T_i/B)}\in [0,3]$. Figure \ref{fig10heatmomwithMAST} compares the experimental heat flux and momentum flux at different radial locations calculated from TRANSP in local gyroBohm units with our GENE simulations. The deviations are within a factor of $3$, indicating that the gyrokinetic simulations reasonably capture the experimental heat flux and momentum flux levels. Note that the purpose of these simulations is not to use the MAST equilibrium for experimental validation. Rather we use it to provide a self-consistent and realistic equilibrium to demonstrate that the LMD regime can be achieved in real devices.

\begin{figure}
    \centering
    \includegraphics[width=0.62\textwidth]{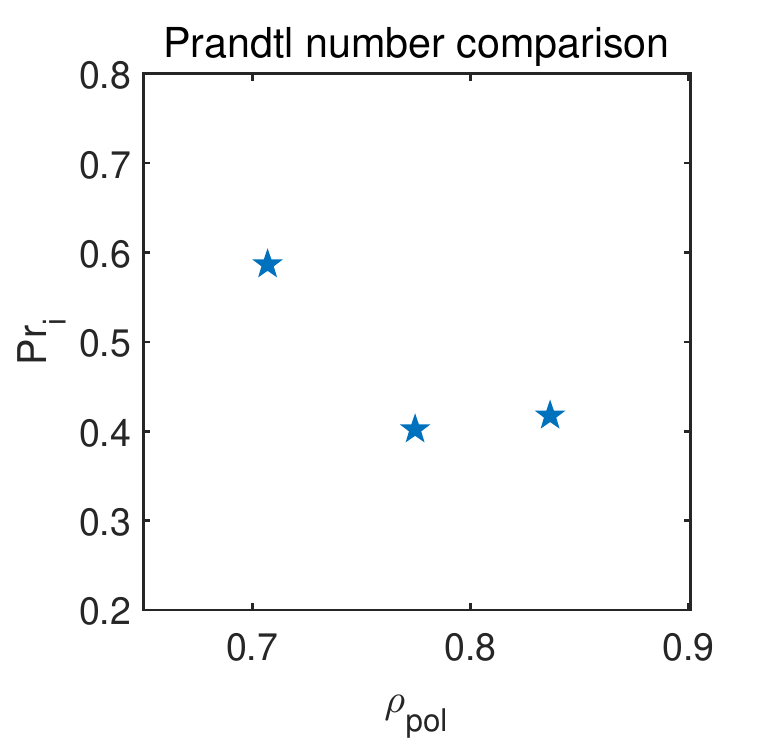}
    \caption{The Prandtl number calculated by NL simulations (blue stars) in the experimental MAST equilibrium at different radial locations.}
    \label{fig15MASTPrandtl}
\end{figure}

For reference, we start by simulating the experimental conditions, setting the background flow shear to zero ($\omega_{\perp}=0$) using both linear and nonlinear simulations. Linear simulations show that the three considered radial locations are dominated by ITG turbulence. Nonlinear simulations checking the grid resolution have been performed. The converged grid parameters for the three radial locations are listed in Table \ref{table2}. Then, with NL simulations, we can calculate the Prandtl number for the three radial locations shown in Fig. \ref{fig15MASTPrandtl}. Importantly, one can see that the Prandtl numbers are low $\text{Pr}_i\sim 0.4$ at two of the radial locations that we consider here. This can help explain why MAST rotates very fast in this discharge. 

\begin{table*}
\caption{\label{table11} The physical parameters for the NL GENE simulations at three radial locations of MAST shot number 24600, where $\delta$ is triangularity and the flow shear values are the experimental values at the corresponding locations.}
\scalebox{0.95}{
\begin{ruledtabular}
\begin{tabular}{cccccccccc}
$\rho_{pol}$ & $R_0/L_{Ti}$&$R_0/L_{Te}$&$R_0/L_n$&$q$& $\hat{s}$&$\epsilon$&$\kappa$&$\delta$&$\omega_{\perp}R_0/c_s$\\ \hline
$0.71$&$7.15$ & $6.29$&$4.05$&$1.28$&$1.17$&$0.42$&$1.45$&$0.18$&$0.59$\\
$0.77$&$5.74$ & $5.84$&$3.65$&$1.48$&$2.09$&$0.46$&$1.45$&$0.19$&$0.54$\\
$0.84$&$5.56$ & $7.94$&$2.43$&$1.86$&$3.20$&$0.51$&$1.47$&$0.19$&$0.47$\\
\end{tabular}
\end{ruledtabular}
}
\end{table*}

\begin{table*}
\caption{\label{table2} The converged grid parameters for the NL GENE simulations without flow shear at three radial locations of MAST shot number 24600.}
\begin{ruledtabular}
\begin{tabular}{ccccccc}
$\rho_{pol}$ & $L_x/\rho_i$&$L_y/\rho_i$&$z$&$v_{||}/\sqrt{2T_i/m_i}$&$\sqrt{\mu/(T_i/B)}$&$(n_{k_x},n_{k_y},n_z,n_{v_{||}},n_{\mu})$\\ \hline
$0.71$ & $234$&$179$&$[-\pi,\pi)$&$[-3,3]$&$[0,3]$&$(256,64,64,32,18)$\\
$0.77$ & $225$&$125$&$[-\pi,\pi)$&$[-3,3]$&$[0,3]$&$(256,64,96,32,18)$\\
$0.84$ & $244$&$125$&$[-\pi,\pi)$&$[-3,3]$&$[0,3]$&$(256,96,64,32,18)$\\
\end{tabular}
\end{ruledtabular}
\end{table*}

\begin{figure*}
    \centering
    \includegraphics[width=0.87\textwidth]{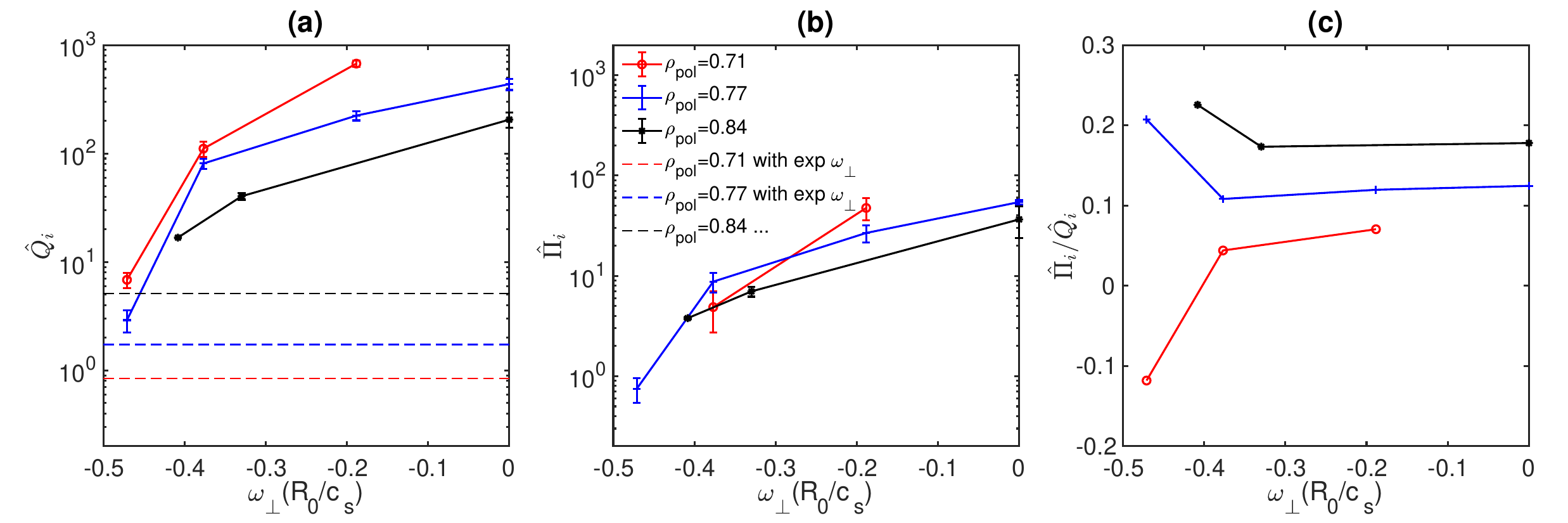}
    \caption{The (a) ion heat flux $\hat{Q}_i$, (b) ion toroidal angular momentum flux $\hat{\Pi}_i$, and (c) their ratio $\hat{\Pi}_i/\hat{Q}_i$ as a function of flow shear from NL GENE simulations of the tilted elliptical flux surfaces in Fig. \ref{fig9MASTgeo} (b). The horizontal dashed lines in (a) denote the heat flux values with experimental flow shear. Note that  red point with a negative value is omitted in (b) due to the logarithmic scale.}
    \label{fig13fluxsummary}
\end{figure*}

\begin{figure}
    \centering
    \includegraphics[width=0.62\textwidth]{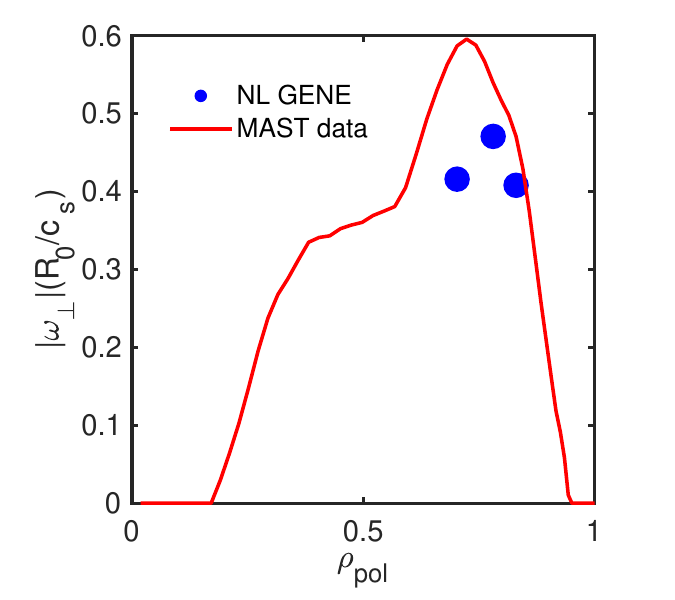}
    \caption{A comparison of the (absolute value of) flow shear profile driven by NBI measured in MAST experiments (red line) and GENE predictions for values of flow shear (blue dots) that could hypothetically be created by intrinsic rotation if MAST could create the tilted elliptical geometries of Fig. \ref{fig9MASTgeo} (b).}
    \label{fig14reconstruct}
\end{figure}

Next, to drive intrinsic flow shear, we artificially modify the magnetic flux surfaces to make them up-down asymmetric. To maximize the drive from up-down asymmetry, the poloidal cross-sectional shapes of the modified flux surfaces are chosen to be ellipses with a tilt angle of $\theta_{\kappa}=\pi/8$ \cite{ball2018}. These cross-sections are shown in Fig. \ref{fig9MASTgeo} (b). At the three radial locations $\rho_{pol}=\bigl\{0.71,0.77,0.84\bigl\}$, we performed NL gyrokinetic simulations to scan the flow shear value (using the grid parameters shown in Table \ref{table2}). Figure \ref{fig13fluxsummary} shows that when we increase the strength of the flow shear (making $\omega_{\perp}$ more and more negative), both the heat flux and momentum flux decrease significantly. The heat flux drops to a low level almost simultaneously with the momentum flux. This implies that for the artificially tilted MAST geometry, the intrinsic flow shear is actually more effective at stabilizing turbulence than in the idealized tilted elliptical geometry considered in Section \ref{tiltellipsegeo} (see Fig. \ref{fig6heatreduction}). For the tilted MAST case, the intrinsic flow shear can thus be expected to significantly reduce the ITG turbulence. 
Therefore, if we perform a linear extrapolation of the toroidal angular momentum flux, we can get an estimate for the flow shear $\omega_{\perp}$ created by the up-down asymmetry. This is presented in Fig. \ref{fig14reconstruct}, which shows that the up-down asymmetry can create flow shear value comparable to the experimental flow shear created by NBI. Note that there is no reason these two values should agree (as the flow shear is driven by different mechanisms). However, this indicates that the already very strong flow shear from NBI could be approximately doubled if the flux surfaces were tilted. Benefiting from the LMD regime, this value of intrinsic flow shear from up-down asymmetry is much greater than what was calculated or measured in conventional tokamaks \cite{Camenen_2010_TCVEXP,ball2018}. We can therefore conclude that strong and experimentally significant flow shear can be created by up-down asymmetry in the LMD regime. Note that one can always flip the flux surfaces upside down to change the direction of the intrinsic rotation. Therefore, in experiments, one can combine NBI injection together with intrinsic rotation to create an even stronger rotation.


\section{Prediction for the SMART tokamak}\label{SMARTgeo}
\begin{figure}
    \centering
    \includegraphics[width=0.3\textwidth]{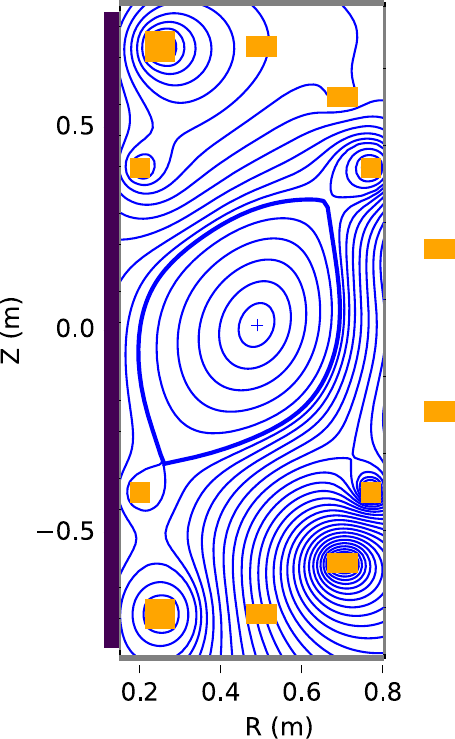}
    \caption{An up-down asymmetric magnetic geometry that can be achieved by the SMART tokamak with its advanced coil set. The orange rectangles represent the shaping coils, the purple rectangle is the central solenoid, the gray rectangles are the vacuum vessel walls and the blue lines are the flux surfaces.}
    \label{fig9SMARTgeo}
\end{figure}

\begin{figure}
    \centering
    \includegraphics[width=0.85\textwidth]{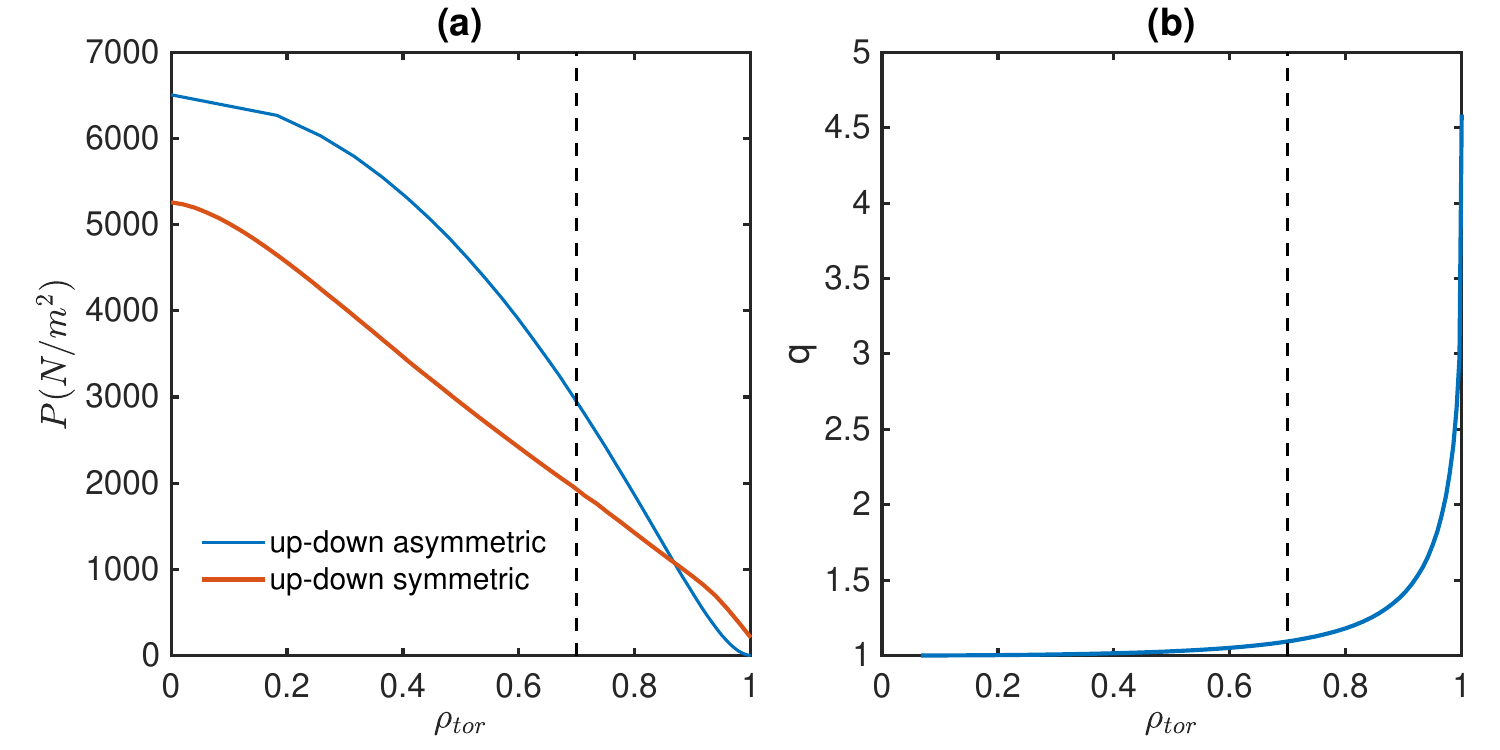}
    \caption{Profiles of (a) pressure and (b) safety factor for the SMART geometry shown in Fig. \ref{fig9SMARTgeo}. Subplot (a) also shows the pressure profile of the up-down symmetric equilibrium from which we took the density and temperature profiles.}
    \label{fig10profileSMART}
\end{figure}

The SMall Aspect Ratio Tokamak (SMART) is a newly built spherical tokamak in Spain \cite{DOYLE_SMART_2021,MANCINI_SMART_2021,AGREDANOTORRES_SMART_2021,SEGADOFERNANDE_SMART_2023,Podesta_SMART_2024}. It has a strong shaping capability and is an ideal tokamak to validate the predictions made in this paper. Here, we show a preliminary calculation for the flow shear that can be created in SMART, using a low $q$, up-down asymmetric geometry that can be achieved on SMART (see Fig. \ref{fig9SMARTgeo}). This up-down asymmetric geometry was designed using the TokaMaker code \cite{HANSEN2024_tokamaker} and relies on a new set of advanced shaping coils. This coil set enables the optimal elongation tilt angle of $\theta_{\kappa}=\pi/8$. While this coil set is not part of the initial operation of SMART, the existing coil set can still create a significantly tilted geometry.
The pressure and safety factor profiles are given in Fig. \ref{fig10profileSMART}. Note that here we use the radial coordinate $\rho_{tor}$ instead of $\rho_{pol}$. Unfortunately, as no experiments have been performed on SMART, this particular equilibrium only has a prediction for the pressure profile (calculated as part of the MHD equilibrium). This is primarily because that the geometry has to be modified to consider the advanced shaping coils, making it difficult to use TRANSP code to create the temperature and density profiles. To get the profiles of temperature and density (which are required for GENE simulations), we take the temperature and density data from a different up-down symmetric geometry expected for SMART, which was created using TRANSP. The purpose of doing this is just to give a preliminary prediction for SMART. Note that only the temperature and density data are taken from an up-down symmetric geometry. The physical and numerical grid parameters are summarized in Tables \ref{tablesmart1} and \ref{tablesmart2}, respectively. We can see from the $q$ and $\hat{s}$ values that this radial location is in the LMD regime and linear simulations indicate that this radial location is dominated by ITG. We have also tried radial locations further out, but they are dominated by the MTM instability and are very difficult to simulate.

\begin{table*}
\caption{\label{tablesmart1} The physical parameters for the NL GENE simulations at $\rho_{tor}=0.7$ for the SMART geometry shown in Fig. \ref{fig9SMARTgeo}.}
\scalebox{0.85}{
\begin{ruledtabular}
\begin{tabular}{ccccccc}
$\rho_{tor}$ & $R_0/L_{Ti}$&$R_0/L_{Te}$&$R_0/L_n$&$q$& $\hat{s}$&$\epsilon$\\ \hline
$0.7$&$6.32$ & $7.86$&$2.81$&$1.34$&$1.25$&$0.39$\\
\end{tabular}
\end{ruledtabular}
}
\end{table*}

\begin{table*}
\caption{\label{tablesmart2} The converged grid parameters for the NL GENE simulations at $\rho_{tor}=0.7$ for the SMART geometry shown in Fig. \ref{fig9SMARTgeo}.}
\begin{ruledtabular}
\begin{tabular}{ccccccc}
$\rho_{pol}$ & $L_x/\rho_i$&$L_y/\rho_i$&$z$&$v_{||}/\sqrt{2T_i/m_i}$&$\sqrt{\mu/(T_i/B)}$&$(n_{k_x},n_{k_y},n_z,n_{v_{||}},n_{\mu})$\\ \hline
$0.7$ & $185$&$125$&$[-\pi,\pi)$&$[-3,3]$&$[0,3]$&$(192,64,96,32,16)$\\
\end{tabular}
\end{ruledtabular}
\end{table*}

\begin{figure*}
    \centering
    \includegraphics[width=0.87\textwidth]{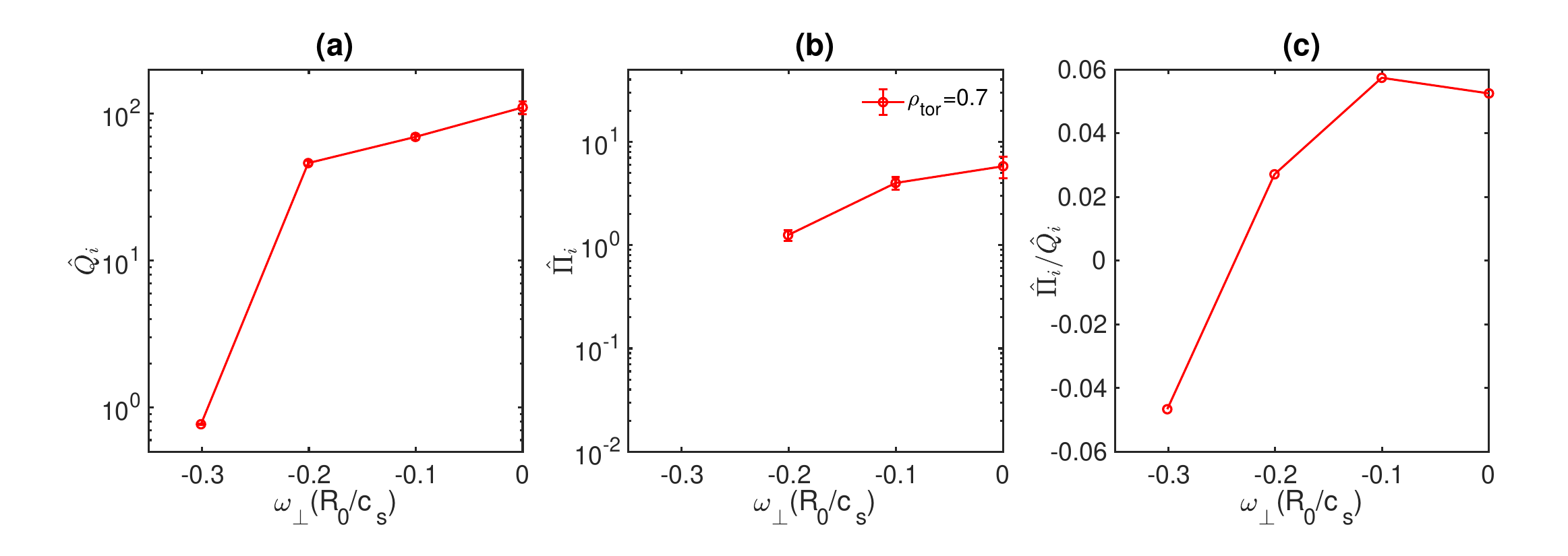}
    \caption{The (a) ion heat flux $\hat{Q}_i$, (b) ion toroidal angular momentum flux $\hat{\Pi}_i$, and (c) their ratio $\hat{\Pi}_i/\hat{Q}_i$ as a function of flow shear from NL GENE simulations of the SMART geometry shown in Fig. \ref{fig9SMARTgeo}. Note that a point with a negative value is omitted from (b) due to the logarithmic scale.}
    \label{fig13fluxsummarySMART}
\end{figure*}

The NL simulation results are summarized in Fig. \ref{fig13fluxsummarySMART}. Based on Fig. \ref{fig13fluxsummarySMART} (c), we can see that a flow shear of $\omega_{\perp}\approx -0.25c_s/R_0$ is predicted. The heat flux reduction due to this intrinsic flow shear is also significant. As can be seen in Fig. \ref{fig13fluxsummarySMART} (a), the heat flux reduces from $\hat{Q}_i\sim 100$ to $\hat{Q}_i\sim 10$ in gyroBohm units. Experimentally, this could be compared with a similar up-down symmetric equilibrium (for which no flow shear from up-down asymmetry would be present). The Prandtl number calculated based on the second and third points in Fig. \ref{fig13fluxsummarySMART} (c) and Eq. \eqref{eq_Prandtlnumber22} is $\text{Pr}_i=0.55$, indicating that the case is very close to the LMD regime. 

\section{Conclusions}\label{conclusion}
In this paper, we studied the Low Momentum Diffusivity (LMD) regime using a large number of Non-Linear (NL) and Quasi-Linear (QL) gyrokinetic simulations. We first considered an ideal circular geometry and showed that the ion Prandtl number can be much smaller than 1 at tight aspect ratio, low safety factor $q$, normal to high magnetic shear $\hat{s}$ and low temperature gradient drive $R_0/L_{T}$. This occurs for both ITG and TEM driven turbulence. Including kinetic electrons tends to somewhat increase the Prandtl number by making the turbulence further away from marginal stability. However, the parameter dependence of the Prandtl number remains essentially the same and a very low Prandtl number can also be achieved using kinetic electrons. 

Combining the LMD regime with up-down asymmetric flux surfaces creates strong intrinsic flow shear, which can significantly reduce the heat flux. This method to create flow shear does not depend on external momentum injection, and is expected to scale well to large devices. To verify the practicality of creating flow shear with up-down asymmetry, we considered a particular experimental MAST equilibrium. Using NL gyrokinetic simulations, we first showed that it is indeed in the LMD regime. We then artificially modified the flux surface shapes to be up-down asymmetric to drive intrinsic momentum flux. This illustrated that combining up-down asymmetry and the LMD regime can create experimentally significant flow shear, much larger than what has been achieved in previous theoretical \cite{ball2018} and experimental \cite{Camenen_2010_TCVEXP} studies. A prediction for the SMART tokamak \cite{DOYLE_SMART_2021} is also made, indicating that an up-down asymmetric spherical tokamak could exhibit a fast intrinsic rotation. Moreover, Appendix \ref{AppendixA} shows that considering the contribution of the pinch term to the momentum flux, which was ignored in our analysis, will make the intrinsic flow shear even larger. Therefore, the flow shear values calculated in this paper represent a lower bound of the flow shear that can be created by combining the LMD regime and up-down asymmetry. Thus, this represents a new way of creating strong rotation in future tokamaks. 

However, creating strongly up-down asymmetric flux surfaces with tight aspect ratio is difficult in existing devices. Fortunately, SMART has strong shaping capabilities and should be able to test the predictions of this paper in the near future. 

\section{Acknowledgement}
The authors thank Prof. Ben McMillan, Dr. Antoine Hoffmann, Dr. Francis Casson, Arnas Volcokas for the fruitful discussions. This work has been carried out within the framework of the EUROfusion Consortium, partially funded by the European Union via the Euratom Research and Training Programme (Grant Agreement No. 101052200 - EUROfusion). The Swiss contribution to this work has been funded by the Swiss State Secretariat for Education, Research and Innovation (SERI). Views and opinions expressed are however those of the author(s) only and do not necessarily reflect those of the European Union, the European Commission or SERI. Neither the European Union nor the European Commission nor SERI can be held responsible for them. This work was supported in part by the Swiss National Science Foundation. This work has been partly funded by the EPSRC Energy Programme (Grant Number EP/W006839/1).


\appendix
\section{Effect of the pinch term on momentum transport}\label{AppendixA}
In this Appendix, we will discuss the effect of the pinch term, which has been neglected in the main text of this paper. We start from Eq. \eqref{eq_momentum}, the momentum transport equation. It is important to note that both $D_{\Pi_i}$ and $P_{\Pi_i}$ are positive as explained in Ref. \cite{Peeters2007PRLpinchterm}. This means that the diffusive term (proportional to $d\Omega_i/dx$) contributes positively to the momentum flux when the gradient is negative and the Coriolis pinch term (proportional to $\Omega_i$) contributes negatively to the momentum flux when the rotation is positive. Considering the steady state operation of a tokamak ($\Pi_i=0$), based on Eq. \eqref{eq_momentum}, one obtains 
\begin{equation}\label{momentumwithpinch}
\Pi_{i,int}=n_im_iR^2_0(D_{\Pi i}\frac{d\Omega_i}{dx}+P_{\Pi i}\Omega_i).
\end{equation}
Equation \eqref{momentumwithpinch} is a simple first order differential equation. Consider the boundary condition $\Omega_i(a)=\Omega_{edge}$, where $a$ is the minor radius of the last closed flux surface and $\Omega_{edge}$ is the rotation at the edge, the solution at a given radial location $x$ is
\begin{equation}\label{rotationfinal}
    \Omega_i(x)=-e^{\int^{a}_x\frac{P_{\Pi i}}{D_{\Pi i}}dx'}\int^{a}_x\frac{\Pi_{i,int}}{n_im_iR^2_0D_{\Pi i}}e^{-\int^{a}_{x''}\frac{P_{\Pi i}}{D_{\Pi i}}dx'}dx''+\Omega_{edge}e^{\int^{a}_x\frac{P_{\Pi i}}{D_{\Pi i}}dx'}.
\end{equation}
Thus, the flow shear is 
\begin{equation}\label{flowshearfinal}
    \frac{d\Omega_i}{dx}(x)=\frac{P_{\Pi i}}{D_{\Pi i}}e^{\int^{a}_x\frac{P_{\Pi i}}{D_{\Pi i}}dx'}\int^{a}_x\frac{\Pi_{i,int}}{n_im_iR^2_0D_{\Pi i}}e^{-\int^{a}_{x''}\frac{P_{\Pi i}}{D_{\Pi i}}dx'}dx''+\frac{\Pi_{i,int}}{n_im_iR^2_0D_{\Pi i}}-\Omega_{edge}\frac{P_{\Pi i}}{D_{\Pi i}}e^{\int^{a}_x \frac{P_{\Pi i}}{D_{\Pi i}}dx'},
\end{equation}
where $P_{\Pi i}$, $D_{\Pi i}$, $D_{\Pi i}$ and $n_i$ are functions of $x$. From Eq. \eqref{flowshearfinal}, one can easily see that, no matter what the sign of $\Omega_{edge}$ is, one can always choose the appropriate sign of $\Pi_{i,int}$ to make the absolute value of flow shear larger. This can be accomplished by flipping the shape of the flux surface across the midplane (which flips the sign of $\Pi_{i,int}$). Additionally, since $P_{\Pi i}$ and $D_{\Pi i}$ are both positive, including the pinch effect (a positive $P_{\Pi i}$) will always make the flow shear generated by the intrinsic rotation larger. Since this paper is ignoring the pinch term, it provides a lower bound on the flow shear created by the intrinsic momentum flux. For typical experimental values of the pinch term, the value of $P_{\Pi i}$ is comparable to $D_{\Pi i}$ \cite{Zimmermann_2022,Zimmermann_2023}. We thus expect a significant increase of the flow shear due to the pinch term.

\bibliographystyle{apsrev4-2} 
\bibliography{apssamp}

\begin{thebibliography}{92}%
\makeatletter
\providecommand \@ifxundefined [1]{%
 \@ifx{#1\undefined}
}%
\providecommand \@ifnum [1]{%
 \ifnum #1\expandafter \@firstoftwo
 \else \expandafter \@secondoftwo
 \fi
}%
\providecommand \@ifx [1]{%
 \ifx #1\expandafter \@firstoftwo
 \else \expandafter \@secondoftwo
 \fi
}%
\providecommand \natexlab [1]{#1}%
\providecommand \enquote  [1]{``#1''}%
\providecommand \bibnamefont  [1]{#1}%
\providecommand \bibfnamefont [1]{#1}%
\providecommand \citenamefont [1]{#1}%
\providecommand \href@noop [0]{\@secondoftwo}%
\providecommand \href [0]{\begingroup \@sanitize@url \@href}%
\providecommand \@href[1]{\@@startlink{#1}\@@href}%
\providecommand \@@href[1]{\endgroup#1\@@endlink}%
\providecommand \@sanitize@url [0]{\catcode `\\12\catcode `\$12\catcode `\&12\catcode `\#12\catcode `\^12\catcode `\_12\catcode `\%12\relax}%
\providecommand \@@startlink[1]{}%
\providecommand \@@endlink[0]{}%
\providecommand \url  [0]{\begingroup\@sanitize@url \@url }%
\providecommand \@url [1]{\endgroup\@href {#1}{\urlprefix }}%
\providecommand \urlprefix  [0]{URL }%
\providecommand \Eprint [0]{\href }%
\providecommand \doibase [0]{https://doi.org/}%
\providecommand \selectlanguage [0]{\@gobble}%
\providecommand \bibinfo  [0]{\@secondoftwo}%
\providecommand \bibfield  [0]{\@secondoftwo}%
\providecommand \translation [1]{[#1]}%
\providecommand \BibitemOpen [0]{}%
\providecommand \bibitemStop [0]{}%
\providecommand \bibitemNoStop [0]{.\EOS\space}%
\providecommand \EOS [0]{\spacefactor3000\relax}%
\providecommand \BibitemShut  [1]{\csname bibitem#1\endcsname}%
\let\auto@bib@innerbib\@empty
\bibitem [{\citenamefont {Dimits}\ \emph {et~al.}(2000)\citenamefont {Dimits}, \citenamefont {Bateman}, \citenamefont {Beer}, \citenamefont {Cohen}, \citenamefont {Dorland}, \citenamefont {Hammett}, \citenamefont {Kim}, \citenamefont {Kinsey}, \citenamefont {Kotschenreuther}, \citenamefont {Kritz} \emph {et~al.}}]{DimitsShift2000}%
  \BibitemOpen
  \bibfield  {author} {\bibinfo {author} {\bibfnamefont {A.}~\bibnamefont {Dimits}}, \bibinfo {author} {\bibfnamefont {G.}~\bibnamefont {Bateman}}, \bibinfo {author} {\bibfnamefont {M.}~\bibnamefont {Beer}}, \bibinfo {author} {\bibfnamefont {B.}~\bibnamefont {Cohen}}, \bibinfo {author} {\bibfnamefont {W.}~\bibnamefont {Dorland}}, \bibinfo {author} {\bibfnamefont {G.}~\bibnamefont {Hammett}}, \bibinfo {author} {\bibfnamefont {C.}~\bibnamefont {Kim}}, \bibinfo {author} {\bibfnamefont {J.}~\bibnamefont {Kinsey}}, \bibinfo {author} {\bibfnamefont {M.}~\bibnamefont {Kotschenreuther}}, \bibinfo {author} {\bibfnamefont {A.}~\bibnamefont {Kritz}}, \emph {et~al.},\ }\href@noop {} {\bibfield  {journal} {\bibinfo  {journal} {Physics of Plasmas}\ }\textbf {\bibinfo {volume} {7}},\ \bibinfo {pages} {969} (\bibinfo {year} {2000})}\BibitemShut {NoStop}%
\bibitem [{\citenamefont {Biglari}\ \emph {et~al.}(1990)\citenamefont {Biglari}, \citenamefont {Diamond},\ and\ \citenamefont {Terry}}]{Biglari1990}%
  \BibitemOpen
  \bibfield  {author} {\bibinfo {author} {\bibfnamefont {H.}~\bibnamefont {Biglari}}, \bibinfo {author} {\bibfnamefont {P.~H.}\ \bibnamefont {Diamond}},\ and\ \bibinfo {author} {\bibfnamefont {P.~W.}\ \bibnamefont {Terry}},\ }\href@noop {} {\bibfield  {journal} {\bibinfo  {journal} {Physics of Fluids B: Plasma Physics}\ }\textbf {\bibinfo {volume} {2}},\ \bibinfo {pages} {1} (\bibinfo {year} {1990})}\BibitemShut {NoStop}%
\bibitem [{\citenamefont {Stambaugh}\ \emph {et~al.}(1990)\citenamefont {Stambaugh}, \citenamefont {Wolfe}, \citenamefont {Hawryluk}, \citenamefont {Harris}, \citenamefont {Biglari}, \citenamefont {Prager}, \citenamefont {Goldston}, \citenamefont {Fonck}, \citenamefont {Ohkawa}, \citenamefont {Logan},\ and\ \citenamefont {Oktay}}]{Stambaugh1990enhanced}%
  \BibitemOpen
  \bibfield  {author} {\bibinfo {author} {\bibfnamefont {R.~D.}\ \bibnamefont {Stambaugh}}, \bibinfo {author} {\bibfnamefont {S.~M.}\ \bibnamefont {Wolfe}}, \bibinfo {author} {\bibfnamefont {R.~J.}\ \bibnamefont {Hawryluk}}, \bibinfo {author} {\bibfnamefont {J.~H.}\ \bibnamefont {Harris}}, \bibinfo {author} {\bibfnamefont {H.}~\bibnamefont {Biglari}}, \bibinfo {author} {\bibfnamefont {S.~C.}\ \bibnamefont {Prager}}, \bibinfo {author} {\bibfnamefont {R.~J.}\ \bibnamefont {Goldston}}, \bibinfo {author} {\bibfnamefont {R.~J.}\ \bibnamefont {Fonck}}, \bibinfo {author} {\bibfnamefont {T.}~\bibnamefont {Ohkawa}}, \bibinfo {author} {\bibfnamefont {B.~G.}\ \bibnamefont {Logan}},\ and\ \bibinfo {author} {\bibfnamefont {E.}~\bibnamefont {Oktay}},\ }\href@noop {} {\bibfield  {journal} {\bibinfo  {journal} {Physics of Fluids B: Plasma Physics}\ }\textbf {\bibinfo {volume} {2}},\ \bibinfo {pages} {2941} (\bibinfo {year} {1990})}\BibitemShut {NoStop}%
\bibitem [{\citenamefont {Eriksson}\ \emph {et~al.}(1997)\citenamefont {Eriksson}, \citenamefont {Righi},\ and\ \citenamefont {Zastrow}}]{JETL-GEriksson_1997}%
  \BibitemOpen
  \bibfield  {author} {\bibinfo {author} {\bibfnamefont {L.-G.}\ \bibnamefont {Eriksson}}, \bibinfo {author} {\bibfnamefont {E.}~\bibnamefont {Righi}},\ and\ \bibinfo {author} {\bibfnamefont {K.-D.}\ \bibnamefont {Zastrow}},\ }\href@noop {} {\bibfield  {journal} {\bibinfo  {journal} {Plasma Physics and Controlled Fusion}\ }\textbf {\bibinfo {volume} {39}},\ \bibinfo {pages} {27} (\bibinfo {year} {1997})}\BibitemShut {NoStop}%
\bibitem [{\citenamefont {Chu}\ \emph {et~al.}(1999)\citenamefont {Chu}, \citenamefont {Chen}, \citenamefont {Zheng}, \citenamefont {Ren},\ and\ \citenamefont {Bondeson}}]{RotationMHDChu1999NF}%
  \BibitemOpen
  \bibfield  {author} {\bibinfo {author} {\bibfnamefont {M.}~\bibnamefont {Chu}}, \bibinfo {author} {\bibfnamefont {L.}~\bibnamefont {Chen}}, \bibinfo {author} {\bibfnamefont {L.-J.}\ \bibnamefont {Zheng}}, \bibinfo {author} {\bibfnamefont {C.}~\bibnamefont {Ren}},\ and\ \bibinfo {author} {\bibfnamefont {A.}~\bibnamefont {Bondeson}},\ }\href@noop {} {\bibfield  {journal} {\bibinfo  {journal} {Nuclear Fusion}\ }\textbf {\bibinfo {volume} {39}},\ \bibinfo {pages} {2107} (\bibinfo {year} {1999})}\BibitemShut {NoStop}%
\bibitem [{\citenamefont {Angioni}\ \emph {et~al.}(2011)\citenamefont {Angioni}, \citenamefont {McDermott}, \citenamefont {Casson}, \citenamefont {Fable}, \citenamefont {Bottino}, \citenamefont {Dux}, \citenamefont {Fischer}, \citenamefont {Podoba}, \citenamefont {P\"utterich}, \citenamefont {Ryter},\ and\ \citenamefont {Viezzer}}]{Angioni2001IntrinsicRotation}%
  \BibitemOpen
  \bibfield  {author} {\bibinfo {author} {\bibfnamefont {C.}~\bibnamefont {Angioni}}, \bibinfo {author} {\bibfnamefont {R.~M.}\ \bibnamefont {McDermott}}, \bibinfo {author} {\bibfnamefont {F.~J.}\ \bibnamefont {Casson}}, \bibinfo {author} {\bibfnamefont {E.}~\bibnamefont {Fable}}, \bibinfo {author} {\bibfnamefont {A.}~\bibnamefont {Bottino}}, \bibinfo {author} {\bibfnamefont {R.}~\bibnamefont {Dux}}, \bibinfo {author} {\bibfnamefont {R.}~\bibnamefont {Fischer}}, \bibinfo {author} {\bibfnamefont {Y.}~\bibnamefont {Podoba}}, \bibinfo {author} {\bibfnamefont {T.}~\bibnamefont {P\"utterich}}, \bibinfo {author} {\bibfnamefont {F.}~\bibnamefont {Ryter}},\ and\ \bibinfo {author} {\bibfnamefont {E.}~\bibnamefont {Viezzer}} (\bibinfo {collaboration} {ASDEX Upgrade Team}),\ }\href@noop {} {\bibfield  {journal} {\bibinfo  {journal} {Phys. Rev. Lett.}\ }\textbf {\bibinfo {volume} {107}},\ \bibinfo {pages} {215003} (\bibinfo {year} {2011})}\BibitemShut {NoStop}%
\bibitem [{\citenamefont {Wahlberg}\ and\ \citenamefont {Bondeson}(2000)}]{RotationMHDWahlberg2000POP}%
  \BibitemOpen
  \bibfield  {author} {\bibinfo {author} {\bibfnamefont {C.}~\bibnamefont {Wahlberg}}\ and\ \bibinfo {author} {\bibfnamefont {A.}~\bibnamefont {Bondeson}},\ }\href@noop {} {\bibfield  {journal} {\bibinfo  {journal} {Physics of Plasmas}\ }\textbf {\bibinfo {volume} {7}},\ \bibinfo {pages} {923} (\bibinfo {year} {2000})}\BibitemShut {NoStop}%
\bibitem [{\citenamefont {Peeters}\ \emph {et~al.}(2005)\citenamefont {Peeters}, \citenamefont {Angioni},\ and\ \citenamefont {the ASDEX Upgrade~Team}}]{Peeters2005LinearToroidal}%
  \BibitemOpen
  \bibfield  {author} {\bibinfo {author} {\bibfnamefont {A.~G.}\ \bibnamefont {Peeters}}, \bibinfo {author} {\bibfnamefont {C.}~\bibnamefont {Angioni}},\ and\ \bibinfo {author} {\bibnamefont {the ASDEX Upgrade~Team}},\ }\href@noop {} {\bibfield  {journal} {\bibinfo  {journal} {Physics of Plasmas}\ }\textbf {\bibinfo {volume} {12}} (\bibinfo {year} {2005})},\ \bibinfo {note} {072515}\BibitemShut {NoStop}%
\bibitem [{\citenamefont {Schekochihin}\ \emph {et~al.}(2008)\citenamefont {Schekochihin}, \citenamefont {Cowley},\ and\ \citenamefont {et~al.}}]{schekochihin2008}%
  \BibitemOpen
  \bibfield  {author} {\bibinfo {author} {\bibfnamefont {A.~A.}\ \bibnamefont {Schekochihin}}, \bibinfo {author} {\bibfnamefont {S.~C.}\ \bibnamefont {Cowley}},\ and\ \bibinfo {author} {\bibfnamefont {W.~D.}\ \bibnamefont {et~al.}},\ }\href@noop {} {\bibfield  {journal} {\bibinfo  {journal} {Plasma Physics and Controlled Fusion}\ }\textbf {\bibinfo {volume} {50}},\ \bibinfo {pages} {124024} (\bibinfo {year} {2008})}\BibitemShut {NoStop}%
\bibitem [{\citenamefont {de~Vries}\ \emph {et~al.}(2008)\citenamefont {de~Vries}, \citenamefont {Hua}, \citenamefont {McDonald}, \citenamefont {Giroud}, \citenamefont {Janvier}, \citenamefont {Johnson}, \citenamefont {Tala}, \citenamefont {Zastrow},\ and\ \citenamefont {Contributors}}]{deVries_2008}%
  \BibitemOpen
  \bibfield  {author} {\bibinfo {author} {\bibfnamefont {P.}~\bibnamefont {de~Vries}}, \bibinfo {author} {\bibfnamefont {M.-D.}\ \bibnamefont {Hua}}, \bibinfo {author} {\bibfnamefont {D.}~\bibnamefont {McDonald}}, \bibinfo {author} {\bibfnamefont {C.}~\bibnamefont {Giroud}}, \bibinfo {author} {\bibfnamefont {M.}~\bibnamefont {Janvier}}, \bibinfo {author} {\bibfnamefont {M.}~\bibnamefont {Johnson}}, \bibinfo {author} {\bibfnamefont {T.}~\bibnamefont {Tala}}, \bibinfo {author} {\bibfnamefont {K.-D.}\ \bibnamefont {Zastrow}},\ and\ \bibinfo {author} {\bibfnamefont {J.~E.}\ \bibnamefont {Contributors}},\ }\href@noop {} {\bibfield  {journal} {\bibinfo  {journal} {Nuclear Fusion}\ }\textbf {\bibinfo {volume} {48}},\ \bibinfo {pages} {065006} (\bibinfo {year} {2008})}\BibitemShut {NoStop}%
\bibitem [{\citenamefont {Mantica}\ \emph {et~al.}(2009)\citenamefont {Mantica}, \citenamefont {Strintzi}, \citenamefont {Tala}, \citenamefont {Giroud}, \citenamefont {Johnson}, \citenamefont {Leggate}, \citenamefont {Lerche}, \citenamefont {Loarer}, \citenamefont {Peeters}, \citenamefont {Salmi}, \citenamefont {Sharapov}, \citenamefont {Van~Eester}, \citenamefont {de~Vries}, \citenamefont {Zabeo},\ and\ \citenamefont {Zastrow}}]{JETMantica2009PRL}%
  \BibitemOpen
  \bibfield  {author} {\bibinfo {author} {\bibfnamefont {P.}~\bibnamefont {Mantica}}, \bibinfo {author} {\bibfnamefont {D.}~\bibnamefont {Strintzi}}, \bibinfo {author} {\bibfnamefont {T.}~\bibnamefont {Tala}}, \bibinfo {author} {\bibfnamefont {C.}~\bibnamefont {Giroud}}, \bibinfo {author} {\bibfnamefont {T.}~\bibnamefont {Johnson}}, \bibinfo {author} {\bibfnamefont {H.}~\bibnamefont {Leggate}}, \bibinfo {author} {\bibfnamefont {E.}~\bibnamefont {Lerche}}, \bibinfo {author} {\bibfnamefont {T.}~\bibnamefont {Loarer}}, \bibinfo {author} {\bibfnamefont {A.~G.}\ \bibnamefont {Peeters}}, \bibinfo {author} {\bibfnamefont {A.}~\bibnamefont {Salmi}}, \bibinfo {author} {\bibfnamefont {S.}~\bibnamefont {Sharapov}}, \bibinfo {author} {\bibfnamefont {D.}~\bibnamefont {Van~Eester}}, \bibinfo {author} {\bibfnamefont {P.~C.}\ \bibnamefont {de~Vries}}, \bibinfo {author} {\bibfnamefont {L.}~\bibnamefont {Zabeo}},\ and\ \bibinfo {author} {\bibfnamefont {K.-D.}\ \bibnamefont {Zastrow}},\ }\href@noop {} {\bibfield  {journal}
  {\bibinfo  {journal} {Phys. Rev. Lett.}\ }\textbf {\bibinfo {volume} {102}},\ \bibinfo {pages} {175002} (\bibinfo {year} {2009})}\BibitemShut {NoStop}%
\bibitem [{\citenamefont {Ida}\ \emph {et~al.}(1990)\citenamefont {Ida}, \citenamefont {Hidekuma}, \citenamefont {Miura}, \citenamefont {Fujita}, \citenamefont {Mori}, \citenamefont {Hoshino}, \citenamefont {Suzuki},\ and\ \citenamefont {Yamauchi}}]{Ida2009PRL}%
  \BibitemOpen
  \bibfield  {author} {\bibinfo {author} {\bibfnamefont {K.}~\bibnamefont {Ida}}, \bibinfo {author} {\bibfnamefont {S.}~\bibnamefont {Hidekuma}}, \bibinfo {author} {\bibfnamefont {Y.}~\bibnamefont {Miura}}, \bibinfo {author} {\bibfnamefont {T.}~\bibnamefont {Fujita}}, \bibinfo {author} {\bibfnamefont {M.}~\bibnamefont {Mori}}, \bibinfo {author} {\bibfnamefont {K.}~\bibnamefont {Hoshino}}, \bibinfo {author} {\bibfnamefont {N.}~\bibnamefont {Suzuki}},\ and\ \bibinfo {author} {\bibfnamefont {T.}~\bibnamefont {Yamauchi}} (\bibinfo {collaboration} {JFT-2M Group}),\ }\href@noop {} {\bibfield  {journal} {\bibinfo  {journal} {Phys. Rev. Lett.}\ }\textbf {\bibinfo {volume} {65}},\ \bibinfo {pages} {1364} (\bibinfo {year} {1990})}\BibitemShut {NoStop}%
\bibitem [{\citenamefont {Aiba}\ \emph {et~al.}(2009)\citenamefont {Aiba}, \citenamefont {Tokuda}, \citenamefont {Furukawa}, \citenamefont {Oyama},\ and\ \citenamefont {Ozeki}}]{RotationMHDAiba2009NF}%
  \BibitemOpen
  \bibfield  {author} {\bibinfo {author} {\bibfnamefont {N.}~\bibnamefont {Aiba}}, \bibinfo {author} {\bibfnamefont {S.}~\bibnamefont {Tokuda}}, \bibinfo {author} {\bibfnamefont {M.}~\bibnamefont {Furukawa}}, \bibinfo {author} {\bibfnamefont {N.}~\bibnamefont {Oyama}},\ and\ \bibinfo {author} {\bibfnamefont {T.}~\bibnamefont {Ozeki}},\ }\href@noop {} {\bibfield  {journal} {\bibinfo  {journal} {Nuclear Fusion}\ }\textbf {\bibinfo {volume} {49}},\ \bibinfo {pages} {065015} (\bibinfo {year} {2009})}\BibitemShut {NoStop}%
\bibitem [{\citenamefont {Newton}\ \emph {et~al.}(2010)\citenamefont {Newton}, \citenamefont {Cowley},\ and\ \citenamefont {Loureiro}}]{NewtonFlowShearUnderstanding2010}%
  \BibitemOpen
  \bibfield  {author} {\bibinfo {author} {\bibfnamefont {S.}~\bibnamefont {Newton}}, \bibinfo {author} {\bibfnamefont {S.}~\bibnamefont {Cowley}},\ and\ \bibinfo {author} {\bibfnamefont {N.}~\bibnamefont {Loureiro}},\ }\href@noop {} {\bibfield  {journal} {\bibinfo  {journal} {Plasma Physics and Controlled Fusion}\ }\textbf {\bibinfo {volume} {52}},\ \bibinfo {pages} {125001} (\bibinfo {year} {2010})}\BibitemShut {NoStop}%
\bibitem [{\citenamefont {Aiba}\ \emph {et~al.}(2011)\citenamefont {Aiba}, \citenamefont {Furukawa}, \citenamefont {Hirota}, \citenamefont {Oyama}, \citenamefont {Kojima}, \citenamefont {Tokuda},\ and\ \citenamefont {Yagi}}]{RotationMHDAiba2011NF}%
  \BibitemOpen
  \bibfield  {author} {\bibinfo {author} {\bibfnamefont {N.}~\bibnamefont {Aiba}}, \bibinfo {author} {\bibfnamefont {M.}~\bibnamefont {Furukawa}}, \bibinfo {author} {\bibfnamefont {M.}~\bibnamefont {Hirota}}, \bibinfo {author} {\bibfnamefont {N.}~\bibnamefont {Oyama}}, \bibinfo {author} {\bibfnamefont {A.}~\bibnamefont {Kojima}}, \bibinfo {author} {\bibfnamefont {S.}~\bibnamefont {Tokuda}},\ and\ \bibinfo {author} {\bibfnamefont {M.}~\bibnamefont {Yagi}},\ }\href@noop {} {\bibfield  {journal} {\bibinfo  {journal} {Nuclear Fusion}\ }\textbf {\bibinfo {volume} {51}},\ \bibinfo {pages} {073012} (\bibinfo {year} {2011})}\BibitemShut {NoStop}%
\bibitem [{\citenamefont {Barnes}\ \emph {et~al.}(2011)\citenamefont {Barnes}, \citenamefont {Parra}, \citenamefont {Highcock}, \citenamefont {Schekochihin}, \citenamefont {Cowley},\ and\ \citenamefont {Roach}}]{BarnesFlowShear2011}%
  \BibitemOpen
  \bibfield  {author} {\bibinfo {author} {\bibfnamefont {M.}~\bibnamefont {Barnes}}, \bibinfo {author} {\bibfnamefont {F.}~\bibnamefont {Parra}}, \bibinfo {author} {\bibfnamefont {E.}~\bibnamefont {Highcock}}, \bibinfo {author} {\bibfnamefont {A.}~\bibnamefont {Schekochihin}}, \bibinfo {author} {\bibfnamefont {S.}~\bibnamefont {Cowley}},\ and\ \bibinfo {author} {\bibfnamefont {C.}~\bibnamefont {Roach}},\ }\href@noop {} {\bibfield  {journal} {\bibinfo  {journal} {Phys. Rev. Lett.}\ }\textbf {\bibinfo {volume} {106}},\ \bibinfo {pages} {175004} (\bibinfo {year} {2011})}\BibitemShut {NoStop}%
\bibitem [{\citenamefont {Schekochihin}\ \emph {et~al.}(2012)\citenamefont {Schekochihin}, \citenamefont {Highcock},\ and\ \citenamefont {Cowley}}]{schekochihin2012}%
  \BibitemOpen
  \bibfield  {author} {\bibinfo {author} {\bibfnamefont {A.~A.}\ \bibnamefont {Schekochihin}}, \bibinfo {author} {\bibfnamefont {E.~G.}\ \bibnamefont {Highcock}},\ and\ \bibinfo {author} {\bibfnamefont {S.~C.}\ \bibnamefont {Cowley}},\ }\href@noop {} {\bibfield  {journal} {\bibinfo  {journal} {Plasma Physics and Controlled Fusion}\ }\textbf {\bibinfo {volume} {54}},\ \bibinfo {pages} {055011} (\bibinfo {year} {2012})}\BibitemShut {NoStop}%
\bibitem [{\citenamefont {Highcock}\ \emph {et~al.}(2011)\citenamefont {Highcock}, \citenamefont {Barnes}, \citenamefont {Parra}, \citenamefont {Schekochihin}, \citenamefont {Roach},\ and\ \citenamefont {Cowley}}]{Highcock2011POP}%
  \BibitemOpen
  \bibfield  {author} {\bibinfo {author} {\bibfnamefont {E.~G.}\ \bibnamefont {Highcock}}, \bibinfo {author} {\bibfnamefont {M.}~\bibnamefont {Barnes}}, \bibinfo {author} {\bibfnamefont {F.~I.}\ \bibnamefont {Parra}}, \bibinfo {author} {\bibfnamefont {A.~A.}\ \bibnamefont {Schekochihin}}, \bibinfo {author} {\bibfnamefont {C.~M.}\ \bibnamefont {Roach}},\ and\ \bibinfo {author} {\bibfnamefont {S.~C.}\ \bibnamefont {Cowley}},\ }\href@noop {} {\bibfield  {journal} {\bibinfo  {journal} {Physics of Plasmas}\ }\textbf {\bibinfo {volume} {18}},\ \bibinfo {pages} {102304} (\bibinfo {year} {2011})}\BibitemShut {NoStop}%
\bibitem [{\citenamefont {Highcock}\ \emph {et~al.}(2012)\citenamefont {Highcock}, \citenamefont {Schekochihin},\ and\ \citenamefont {et~al.}}]{highcock2012}%
  \BibitemOpen
  \bibfield  {author} {\bibinfo {author} {\bibfnamefont {E.~G.}\ \bibnamefont {Highcock}}, \bibinfo {author} {\bibfnamefont {A.~A.}\ \bibnamefont {Schekochihin}},\ and\ \bibinfo {author} {\bibfnamefont {S.~C.~C.}\ \bibnamefont {et~al.}},\ }\href@noop {} {\bibfield  {journal} {\bibinfo  {journal} {Phys. Rev. Lett.}\ }\textbf {\bibinfo {volume} {109}},\ \bibinfo {pages} {265001} (\bibinfo {year} {2012})}\BibitemShut {NoStop}%
\bibitem [{\citenamefont {Christen}\ \emph {et~al.}(2021)\citenamefont {Christen}, \citenamefont {Barnes},\ and\ \citenamefont {Parra}}]{ChristenFlowShear2018}%
  \BibitemOpen
  \bibfield  {author} {\bibinfo {author} {\bibfnamefont {N.}~\bibnamefont {Christen}}, \bibinfo {author} {\bibfnamefont {M.}~\bibnamefont {Barnes}},\ and\ \bibinfo {author} {\bibfnamefont {F.~I.}\ \bibnamefont {Parra}},\ }\href {https://doi.org/10.1017/S0022377821000453} {\bibfield  {journal} {\bibinfo  {journal} {Journal of Plasma Physics}\ }\textbf {\bibinfo {volume} {87}},\ \bibinfo {pages} {905870230} (\bibinfo {year} {2021})}\BibitemShut {NoStop}%
\bibitem [{\citenamefont {McMillan}\ \emph {et~al.}(2019)\citenamefont {McMillan}, \citenamefont {Ball},\ and\ \citenamefont {Brunner}}]{ben2019}%
  \BibitemOpen
  \bibfield  {author} {\bibinfo {author} {\bibfnamefont {B.~F.}\ \bibnamefont {McMillan}}, \bibinfo {author} {\bibfnamefont {J.}~\bibnamefont {Ball}},\ and\ \bibinfo {author} {\bibfnamefont {S.}~\bibnamefont {Brunner}},\ }\href@noop {} {\bibfield  {journal} {\bibinfo  {journal} {Plasma Phys. Control. Fusion}\ }\textbf {\bibinfo {volume} {61}},\ \bibinfo {pages} {055006} (\bibinfo {year} {2019})}\BibitemShut {NoStop}%
\bibitem [{\citenamefont {Ball}\ \emph {et~al.}(2019)\citenamefont {Ball}, \citenamefont {Brunner},\ and\ \citenamefont {McMillan}}]{ball2019}%
  \BibitemOpen
  \bibfield  {author} {\bibinfo {author} {\bibfnamefont {J.}~\bibnamefont {Ball}}, \bibinfo {author} {\bibfnamefont {S.}~\bibnamefont {Brunner}},\ and\ \bibinfo {author} {\bibfnamefont {B.~F.}\ \bibnamefont {McMillan}},\ }\href@noop {} {\bibfield  {journal} {\bibinfo  {journal} {Plasma Phys. Control. Fusion}\ }\textbf {\bibinfo {volume} {61}},\ \bibinfo {pages} {064004} (\bibinfo {year} {2019})}\BibitemShut {NoStop}%
\bibitem [{\citenamefont {Noterdaeme}\ \emph {et~al.}(2003)\citenamefont {Noterdaeme}, \citenamefont {Righi}, \citenamefont {Chan}, \citenamefont {deGrassie}, \citenamefont {Kirov}, \citenamefont {Mantsinen}, \citenamefont {Nave}, \citenamefont {Testa}, \citenamefont {Zastrow}, \citenamefont {Budny}, \citenamefont {Cesario}, \citenamefont {Gondhalekar}, \citenamefont {Hawkes}, \citenamefont {Hellsten}, \citenamefont {Lamalle}, \citenamefont {Meo}, \citenamefont {Nguyen},\ and\ \citenamefont {contributors}}]{JETJ.M.Noterdaeme_2003}%
  \BibitemOpen
  \bibfield  {author} {\bibinfo {author} {\bibfnamefont {J.-M.}\ \bibnamefont {Noterdaeme}}, \bibinfo {author} {\bibfnamefont {E.}~\bibnamefont {Righi}}, \bibinfo {author} {\bibfnamefont {V.}~\bibnamefont {Chan}}, \bibinfo {author} {\bibfnamefont {J.}~\bibnamefont {deGrassie}}, \bibinfo {author} {\bibfnamefont {K.}~\bibnamefont {Kirov}}, \bibinfo {author} {\bibfnamefont {M.}~\bibnamefont {Mantsinen}}, \bibinfo {author} {\bibfnamefont {M.}~\bibnamefont {Nave}}, \bibinfo {author} {\bibfnamefont {D.}~\bibnamefont {Testa}}, \bibinfo {author} {\bibfnamefont {K.-D.}\ \bibnamefont {Zastrow}}, \bibinfo {author} {\bibfnamefont {R.}~\bibnamefont {Budny}}, \bibinfo {author} {\bibfnamefont {R.}~\bibnamefont {Cesario}}, \bibinfo {author} {\bibfnamefont {A.}~\bibnamefont {Gondhalekar}}, \bibinfo {author} {\bibfnamefont {N.}~\bibnamefont {Hawkes}}, \bibinfo {author} {\bibfnamefont {T.}~\bibnamefont {Hellsten}}, \bibinfo {author} {\bibfnamefont {P.}~\bibnamefont {Lamalle}}, \bibinfo {author} {\bibfnamefont {F.}~\bibnamefont
  {Meo}}, \bibinfo {author} {\bibfnamefont {F.}~\bibnamefont {Nguyen}},\ and\ \bibinfo {author} {\bibfnamefont {E.-J.-E.}\ \bibnamefont {contributors}},\ }\href@noop {} {\bibfield  {journal} {\bibinfo  {journal} {Nuclear Fusion}\ }\textbf {\bibinfo {volume} {43}},\ \bibinfo {pages} {274} (\bibinfo {year} {2003})}\BibitemShut {NoStop}%
\bibitem [{\citenamefont {de~Vries}\ \emph {et~al.}(2006)\citenamefont {de~Vries}, \citenamefont {Rantam\"aki}, \citenamefont {Giroud}, \citenamefont {Asp}, \citenamefont {Corrigan}, \citenamefont {Eriksson}, \citenamefont {de~Greef}, \citenamefont {Jenkins}, \citenamefont {Knoops}, \citenamefont {Mantica}, \citenamefont {Nordman}, \citenamefont {Strand}, \citenamefont {Tala}, \citenamefont {Weiland}, \citenamefont {Zastrow},\ and\ \citenamefont {Contributors}}]{JETdeVries_2006}%
  \BibitemOpen
  \bibfield  {author} {\bibinfo {author} {\bibfnamefont {P.~C.}\ \bibnamefont {de~Vries}}, \bibinfo {author} {\bibfnamefont {K.~M.}\ \bibnamefont {Rantam\"aki}}, \bibinfo {author} {\bibfnamefont {C.}~\bibnamefont {Giroud}}, \bibinfo {author} {\bibfnamefont {E.}~\bibnamefont {Asp}}, \bibinfo {author} {\bibfnamefont {G.}~\bibnamefont {Corrigan}}, \bibinfo {author} {\bibfnamefont {A.}~\bibnamefont {Eriksson}}, \bibinfo {author} {\bibfnamefont {M.}~\bibnamefont {de~Greef}}, \bibinfo {author} {\bibfnamefont {I.}~\bibnamefont {Jenkins}}, \bibinfo {author} {\bibfnamefont {H.~C.~M.}\ \bibnamefont {Knoops}}, \bibinfo {author} {\bibfnamefont {P.}~\bibnamefont {Mantica}}, \bibinfo {author} {\bibfnamefont {H.}~\bibnamefont {Nordman}}, \bibinfo {author} {\bibfnamefont {P.}~\bibnamefont {Strand}}, \bibinfo {author} {\bibfnamefont {T.}~\bibnamefont {Tala}}, \bibinfo {author} {\bibfnamefont {J.}~\bibnamefont {Weiland}}, \bibinfo {author} {\bibfnamefont {K.-D.}\ \bibnamefont {Zastrow}},\ and\ \bibinfo {author} {\bibfnamefont
  {J.~E.}\ \bibnamefont {Contributors}},\ }\href@noop {} {\bibfield  {journal} {\bibinfo  {journal} {Plasma Physics and Controlled Fusion}\ }\textbf {\bibinfo {volume} {48}},\ \bibinfo {pages} {1693} (\bibinfo {year} {2006})}\BibitemShut {NoStop}%
\bibitem [{\citenamefont {Groebner}\ \emph {et~al.}(1990)\citenamefont {Groebner}, \citenamefont {Burrell},\ and\ \citenamefont {Seraydarian}}]{Groebner1990NBIrotation}%
  \BibitemOpen
  \bibfield  {author} {\bibinfo {author} {\bibfnamefont {R.~J.}\ \bibnamefont {Groebner}}, \bibinfo {author} {\bibfnamefont {K.~H.}\ \bibnamefont {Burrell}},\ and\ \bibinfo {author} {\bibfnamefont {R.~P.}\ \bibnamefont {Seraydarian}},\ }\href@noop {} {\bibfield  {journal} {\bibinfo  {journal} {Phys. Rev. Lett.}\ }\textbf {\bibinfo {volume} {64}},\ \bibinfo {pages} {3015} (\bibinfo {year} {1990})}\BibitemShut {NoStop}%
\bibitem [{\citenamefont {Suckewer}\ \emph {et~al.}(1981)\citenamefont {Suckewer}, \citenamefont {Eubank}, \citenamefont {Goldston}, \citenamefont {McEnerney}, \citenamefont {Sauthoff},\ and\ \citenamefont {Towner}}]{Suckewer1981NBIrotation}%
  \BibitemOpen
  \bibfield  {author} {\bibinfo {author} {\bibfnamefont {S.}~\bibnamefont {Suckewer}}, \bibinfo {author} {\bibfnamefont {H.}~\bibnamefont {Eubank}}, \bibinfo {author} {\bibfnamefont {R.}~\bibnamefont {Goldston}}, \bibinfo {author} {\bibfnamefont {J.}~\bibnamefont {McEnerney}}, \bibinfo {author} {\bibfnamefont {N.}~\bibnamefont {Sauthoff}},\ and\ \bibinfo {author} {\bibfnamefont {H.}~\bibnamefont {Towner}},\ }\href@noop {} {\bibfield  {journal} {\bibinfo  {journal} {Nuclear Fusion}\ }\textbf {\bibinfo {volume} {21}},\ \bibinfo {pages} {1301} (\bibinfo {year} {1981})}\BibitemShut {NoStop}%
\bibitem [{\citenamefont {Goumiri}\ \emph {et~al.}(2016)\citenamefont {Goumiri}, \citenamefont {Rowley}, \citenamefont {Sabbagh}, \citenamefont {Gates}, \citenamefont {Gerhardt}, \citenamefont {Boyer}, \citenamefont {Andre}, \citenamefont {Kolemen},\ and\ \citenamefont {Taira}}]{Goumiri2016NSTXNBI}%
  \BibitemOpen
  \bibfield  {author} {\bibinfo {author} {\bibfnamefont {I.}~\bibnamefont {Goumiri}}, \bibinfo {author} {\bibfnamefont {C.}~\bibnamefont {Rowley}}, \bibinfo {author} {\bibfnamefont {S.}~\bibnamefont {Sabbagh}}, \bibinfo {author} {\bibfnamefont {D.}~\bibnamefont {Gates}}, \bibinfo {author} {\bibfnamefont {S.}~\bibnamefont {Gerhardt}}, \bibinfo {author} {\bibfnamefont {M.}~\bibnamefont {Boyer}}, \bibinfo {author} {\bibfnamefont {R.}~\bibnamefont {Andre}}, \bibinfo {author} {\bibfnamefont {E.}~\bibnamefont {Kolemen}},\ and\ \bibinfo {author} {\bibfnamefont {K.}~\bibnamefont {Taira}},\ }\href@noop {} {\bibfield  {journal} {\bibinfo  {journal} {Nuclear Fusion}\ }\textbf {\bibinfo {volume} {56}},\ \bibinfo {pages} {036023} (\bibinfo {year} {2016})}\BibitemShut {NoStop}%
\bibitem [{\citenamefont {Hsuan}\ \emph {et~al.}(1996)\citenamefont {Hsuan}, \citenamefont {Bitter}, \citenamefont {Phillips}, \citenamefont {Wilson}, \citenamefont {Bush}, \citenamefont {Duong}, \citenamefont {Darrow}, \citenamefont {Hammett}, \citenamefont {Hill}, \citenamefont {Majeski}, \citenamefont {Medley}, \citenamefont {Petrov}, \citenamefont {Synakowski}, \citenamefont {Zarnstorff},\ and\ \citenamefont {Zweben}}]{Hsuan1996ICHrotation}%
  \BibitemOpen
  \bibfield  {author} {\bibinfo {author} {\bibfnamefont {H.}~\bibnamefont {Hsuan}}, \bibinfo {author} {\bibfnamefont {M.}~\bibnamefont {Bitter}}, \bibinfo {author} {\bibfnamefont {C.~K.}\ \bibnamefont {Phillips}}, \bibinfo {author} {\bibfnamefont {J.~R.}\ \bibnamefont {Wilson}}, \bibinfo {author} {\bibfnamefont {C.}~\bibnamefont {Bush}}, \bibinfo {author} {\bibfnamefont {H.~H.}\ \bibnamefont {Duong}}, \bibinfo {author} {\bibfnamefont {D.}~\bibnamefont {Darrow}}, \bibinfo {author} {\bibfnamefont {G.~W.}\ \bibnamefont {Hammett}}, \bibinfo {author} {\bibfnamefont {K.~W.}\ \bibnamefont {Hill}}, \bibinfo {author} {\bibfnamefont {R.~P.}\ \bibnamefont {Majeski}}, \bibinfo {author} {\bibfnamefont {S.}~\bibnamefont {Medley}}, \bibinfo {author} {\bibfnamefont {M.}~\bibnamefont {Petrov}}, \bibinfo {author} {\bibfnamefont {E.}~\bibnamefont {Synakowski}}, \bibinfo {author} {\bibfnamefont {M.}~\bibnamefont {Zarnstorff}},\ and\ \bibinfo {author} {\bibfnamefont {S.}~\bibnamefont {Zweben}},\ }\href@noop {} {\bibfield
  {journal} {\bibinfo  {journal} {AIP Conference Proceedings}\ }\textbf {\bibinfo {volume} {355}},\ \bibinfo {pages} {39} (\bibinfo {year} {1996})}\BibitemShut {NoStop}%
\bibitem [{\citenamefont {Chang}\ \emph {et~al.}(1999)\citenamefont {Chang}, \citenamefont {Phillips}, \citenamefont {White}, \citenamefont {Zweben}, \citenamefont {Bonoli}, \citenamefont {Rice}, \citenamefont {Greenwald},\ and\ \citenamefont {deGrassie}}]{Chang1999ICRH}%
  \BibitemOpen
  \bibfield  {author} {\bibinfo {author} {\bibfnamefont {C.~S.}\ \bibnamefont {Chang}}, \bibinfo {author} {\bibfnamefont {C.~K.}\ \bibnamefont {Phillips}}, \bibinfo {author} {\bibfnamefont {R.}~\bibnamefont {White}}, \bibinfo {author} {\bibfnamefont {S.}~\bibnamefont {Zweben}}, \bibinfo {author} {\bibfnamefont {P.~T.}\ \bibnamefont {Bonoli}}, \bibinfo {author} {\bibfnamefont {J.~E.}\ \bibnamefont {Rice}}, \bibinfo {author} {\bibfnamefont {M.~J.}\ \bibnamefont {Greenwald}},\ and\ \bibinfo {author} {\bibfnamefont {J.}~\bibnamefont {deGrassie}},\ }\href@noop {} {\bibfield  {journal} {\bibinfo  {journal} {Physics of Plasmas}\ }\textbf {\bibinfo {volume} {6}},\ \bibinfo {pages} {1969} (\bibinfo {year} {1999})}\BibitemShut {NoStop}%
\bibitem [{\citenamefont {Chan}\ \emph {et~al.}(2002)\citenamefont {Chan}, \citenamefont {Chiu},\ and\ \citenamefont {Omelchenko}}]{Chan2002RFrotation}%
  \BibitemOpen
  \bibfield  {author} {\bibinfo {author} {\bibfnamefont {V.~S.}\ \bibnamefont {Chan}}, \bibinfo {author} {\bibfnamefont {S.~C.}\ \bibnamefont {Chiu}},\ and\ \bibinfo {author} {\bibfnamefont {Y.~A.}\ \bibnamefont {Omelchenko}},\ }\href@noop {} {\bibfield  {journal} {\bibinfo  {journal} {Physics of Plasmas}\ }\textbf {\bibinfo {volume} {9}},\ \bibinfo {pages} {501} (\bibinfo {year} {2002})}\BibitemShut {NoStop}%
\bibitem [{\citenamefont {Li}\ and\ \citenamefont {Wan}(2011)}]{Li_2011EASTRF}%
  \BibitemOpen
  \bibfield  {author} {\bibinfo {author} {\bibfnamefont {J.}~\bibnamefont {Li}}\ and\ \bibinfo {author} {\bibfnamefont {B.}~\bibnamefont {Wan}},\ }\href@noop {} {\bibfield  {journal} {\bibinfo  {journal} {Nuclear Fusion}\ }\textbf {\bibinfo {volume} {51}},\ \bibinfo {pages} {094007} (\bibinfo {year} {2011})}\BibitemShut {NoStop}%
\bibitem [{\citenamefont {Lyu}\ \emph {et~al.}(2020)\citenamefont {Lyu}, \citenamefont {Wang}, \citenamefont {Chen}, \citenamefont {Hu}, \citenamefont {Li}, \citenamefont {Fu}, \citenamefont {Zhang}, \citenamefont {Bitter}, \citenamefont {Hill}, \citenamefont {Shi}, \citenamefont {Ye},\ and\ \citenamefont {Wan}}]{Lyu2020RFrotation}%
  \BibitemOpen
  \bibfield  {author} {\bibinfo {author} {\bibfnamefont {B.}~\bibnamefont {Lyu}}, \bibinfo {author} {\bibfnamefont {F.~D.}\ \bibnamefont {Wang}}, \bibinfo {author} {\bibfnamefont {J.}~\bibnamefont {Chen}}, \bibinfo {author} {\bibfnamefont {R.~J.}\ \bibnamefont {Hu}}, \bibinfo {author} {\bibfnamefont {Y.~Y.}\ \bibnamefont {Li}}, \bibinfo {author} {\bibfnamefont {J.}~\bibnamefont {Fu}}, \bibinfo {author} {\bibfnamefont {H.~M.}\ \bibnamefont {Zhang}}, \bibinfo {author} {\bibfnamefont {M.}~\bibnamefont {Bitter}}, \bibinfo {author} {\bibfnamefont {K.~W.}\ \bibnamefont {Hill}}, \bibinfo {author} {\bibfnamefont {Y.~J.}\ \bibnamefont {Shi}}, \bibinfo {author} {\bibfnamefont {M.~Y.}\ \bibnamefont {Ye}},\ and\ \bibinfo {author} {\bibfnamefont {B.~N.}\ \bibnamefont {Wan}},\ }\href@noop {} {\bibfield  {journal} {\bibinfo  {journal} {Physics of Plasmas}\ }\textbf {\bibinfo {volume} {27}},\ \bibinfo {pages} {022511} (\bibinfo {year} {2020})}\BibitemShut {NoStop}%
\bibitem [{\citenamefont {Liu}\ \emph {et~al.}(2004)\citenamefont {Liu}, \citenamefont {Bondeson}, \citenamefont {Gribov},\ and\ \citenamefont {Polevoi}}]{YueqiangLiu_2004}%
  \BibitemOpen
  \bibfield  {author} {\bibinfo {author} {\bibfnamefont {Y.}~\bibnamefont {Liu}}, \bibinfo {author} {\bibfnamefont {A.}~\bibnamefont {Bondeson}}, \bibinfo {author} {\bibfnamefont {Y.}~\bibnamefont {Gribov}},\ and\ \bibinfo {author} {\bibfnamefont {A.}~\bibnamefont {Polevoi}},\ }\href@noop {} {\bibfield  {journal} {\bibinfo  {journal} {Nuclear Fusion}\ }\textbf {\bibinfo {volume} {44}},\ \bibinfo {pages} {232} (\bibinfo {year} {2004})}\BibitemShut {NoStop}%
\bibitem [{\citenamefont {Parra}\ \emph {et~al.}(2011{\natexlab{a}})\citenamefont {Parra}, \citenamefont {Barnes}, \citenamefont {Highcock}, \citenamefont {Schekochihin},\ and\ \citenamefont {Cowley}}]{Parra_2011_PRL_Momentum_optimum}%
  \BibitemOpen
  \bibfield  {author} {\bibinfo {author} {\bibfnamefont {F.~I.}\ \bibnamefont {Parra}}, \bibinfo {author} {\bibfnamefont {M.}~\bibnamefont {Barnes}}, \bibinfo {author} {\bibfnamefont {E.~G.}\ \bibnamefont {Highcock}}, \bibinfo {author} {\bibfnamefont {A.~A.}\ \bibnamefont {Schekochihin}},\ and\ \bibinfo {author} {\bibfnamefont {S.~C.}\ \bibnamefont {Cowley}},\ }\href {https://doi.org/10.1103/PhysRevLett.106.115004} {\bibfield  {journal} {\bibinfo  {journal} {Phys. Rev. Lett.}\ }\textbf {\bibinfo {volume} {106}},\ \bibinfo {pages} {115004} (\bibinfo {year} {2011}{\natexlab{a}})}\BibitemShut {NoStop}%
\bibitem [{\citenamefont {Camenen}\ \emph {et~al.}(2009)\citenamefont {Camenen}, \citenamefont {Peeters}, \citenamefont {Angioni}, \citenamefont {Casson}, \citenamefont {Hornsby}, \citenamefont {Snodin},\ and\ \citenamefont {Strintzi}}]{Camenen2009TUBTrans}%
  \BibitemOpen
  \bibfield  {author} {\bibinfo {author} {\bibfnamefont {Y.}~\bibnamefont {Camenen}}, \bibinfo {author} {\bibfnamefont {A.~G.}\ \bibnamefont {Peeters}}, \bibinfo {author} {\bibfnamefont {C.}~\bibnamefont {Angioni}}, \bibinfo {author} {\bibfnamefont {F.~J.}\ \bibnamefont {Casson}}, \bibinfo {author} {\bibfnamefont {W.~A.}\ \bibnamefont {Hornsby}}, \bibinfo {author} {\bibfnamefont {A.~P.}\ \bibnamefont {Snodin}},\ and\ \bibinfo {author} {\bibfnamefont {D.}~\bibnamefont {Strintzi}},\ }\href@noop {} {\bibfield  {journal} {\bibinfo  {journal} {Physics of Plasmas}\ }\textbf {\bibinfo {volume} {16}} (\bibinfo {year} {2009})},\ \bibinfo {note} {012503}\BibitemShut {NoStop}%
\bibitem [{\citenamefont {Camenen}\ \emph {et~al.}(2010{\natexlab{a}})\citenamefont {Camenen}, \citenamefont {Bortolon}, \citenamefont {Duval}, \citenamefont {Federspiel}, \citenamefont {Peeters}, \citenamefont {Casson}, \citenamefont {Hornsby}, \citenamefont {Karpushov}, \citenamefont {Piras}, \citenamefont {Sauter}, \citenamefont {Snodin},\ and\ \citenamefont {Szepesi}}]{Camenen2010MomentumTransport}%
  \BibitemOpen
  \bibfield  {author} {\bibinfo {author} {\bibfnamefont {Y.}~\bibnamefont {Camenen}}, \bibinfo {author} {\bibfnamefont {A.}~\bibnamefont {Bortolon}}, \bibinfo {author} {\bibfnamefont {B.~P.}\ \bibnamefont {Duval}}, \bibinfo {author} {\bibfnamefont {L.}~\bibnamefont {Federspiel}}, \bibinfo {author} {\bibfnamefont {A.~G.}\ \bibnamefont {Peeters}}, \bibinfo {author} {\bibfnamefont {F.~J.}\ \bibnamefont {Casson}}, \bibinfo {author} {\bibfnamefont {W.~A.}\ \bibnamefont {Hornsby}}, \bibinfo {author} {\bibfnamefont {A.~N.}\ \bibnamefont {Karpushov}}, \bibinfo {author} {\bibfnamefont {F.}~\bibnamefont {Piras}}, \bibinfo {author} {\bibfnamefont {O.}~\bibnamefont {Sauter}}, \bibinfo {author} {\bibfnamefont {A.~P.}\ \bibnamefont {Snodin}},\ and\ \bibinfo {author} {\bibfnamefont {G.}~\bibnamefont {Szepesi}},\ }\href@noop {} {\bibfield  {journal} {\bibinfo  {journal} {Phys. Rev. Lett.}\ }\textbf {\bibinfo {volume} {105}},\ \bibinfo {pages} {135003} (\bibinfo {year} {2010}{\natexlab{a}})}\BibitemShut {NoStop}%
\bibitem [{\citenamefont {Zhu}\ \emph {et~al.}(2024)\citenamefont {Zhu}, \citenamefont {Stoltzfus-Dueck}, \citenamefont {Hager}, \citenamefont {Ku},\ and\ \citenamefont {Chang}}]{zhu2024intrinsic}%
  \BibitemOpen
  \bibfield  {author} {\bibinfo {author} {\bibfnamefont {H.}~\bibnamefont {Zhu}}, \bibinfo {author} {\bibfnamefont {T.}~\bibnamefont {Stoltzfus-Dueck}}, \bibinfo {author} {\bibfnamefont {R.}~\bibnamefont {Hager}}, \bibinfo {author} {\bibfnamefont {S.}~\bibnamefont {Ku}},\ and\ \bibinfo {author} {\bibfnamefont {C.~S.}\ \bibnamefont {Chang}},\ }\href@noop {} {\bibfield  {journal} {\bibinfo  {journal} {Phys. Rev. Lett.}\ }\textbf {\bibinfo {volume} {133}},\ \bibinfo {pages} {025101} (\bibinfo {year} {2024})}\BibitemShut {NoStop}%
\bibitem [{\citenamefont {Parra}\ \emph {et~al.}(2011{\natexlab{b}})\citenamefont {Parra}, \citenamefont {Barnes},\ and\ \citenamefont {Peeters}}]{ParraUpDownSym2011}%
  \BibitemOpen
  \bibfield  {author} {\bibinfo {author} {\bibfnamefont {F.}~\bibnamefont {Parra}}, \bibinfo {author} {\bibfnamefont {M.}~\bibnamefont {Barnes}},\ and\ \bibinfo {author} {\bibfnamefont {A.}~\bibnamefont {Peeters}},\ }\href@noop {} {\bibfield  {journal} {\bibinfo  {journal} {Phys. Plasmas}\ }\textbf {\bibinfo {volume} {18}},\ \bibinfo {pages} {062501} (\bibinfo {year} {2011}{\natexlab{b}})}\BibitemShut {NoStop}%
\bibitem [{\citenamefont {Peeters}\ \emph {et~al.}(2011)\citenamefont {Peeters}, \citenamefont {Angioni}, \citenamefont {Bortolon}, \citenamefont {Camenen}, \citenamefont {Casson}, \citenamefont {Duval}, \citenamefont {Fiederspiel}, \citenamefont {Hornsby}, \citenamefont {Idomura}, \citenamefont {Hein}, \citenamefont {Kluy}, \citenamefont {Mantica}, \citenamefont {Parra}, \citenamefont {Snodin}, \citenamefont {Szepesi}, \citenamefont {Strintzi}, \citenamefont {Tala}, \citenamefont {Tardini}, \citenamefont {de~Vries},\ and\ \citenamefont {Weiland}}]{Peeters_2011NFReview}%
  \BibitemOpen
  \bibfield  {author} {\bibinfo {author} {\bibfnamefont {A.}~\bibnamefont {Peeters}}, \bibinfo {author} {\bibfnamefont {C.}~\bibnamefont {Angioni}}, \bibinfo {author} {\bibfnamefont {A.}~\bibnamefont {Bortolon}}, \bibinfo {author} {\bibfnamefont {Y.}~\bibnamefont {Camenen}}, \bibinfo {author} {\bibfnamefont {F.}~\bibnamefont {Casson}}, \bibinfo {author} {\bibfnamefont {B.}~\bibnamefont {Duval}}, \bibinfo {author} {\bibfnamefont {L.}~\bibnamefont {Fiederspiel}}, \bibinfo {author} {\bibfnamefont {W.}~\bibnamefont {Hornsby}}, \bibinfo {author} {\bibfnamefont {Y.}~\bibnamefont {Idomura}}, \bibinfo {author} {\bibfnamefont {T.}~\bibnamefont {Hein}}, \bibinfo {author} {\bibfnamefont {N.}~\bibnamefont {Kluy}}, \bibinfo {author} {\bibfnamefont {P.}~\bibnamefont {Mantica}}, \bibinfo {author} {\bibfnamefont {F.}~\bibnamefont {Parra}}, \bibinfo {author} {\bibfnamefont {A.}~\bibnamefont {Snodin}}, \bibinfo {author} {\bibfnamefont {G.}~\bibnamefont {Szepesi}}, \bibinfo {author} {\bibfnamefont {D.}~\bibnamefont {Strintzi}},
  \bibinfo {author} {\bibfnamefont {T.}~\bibnamefont {Tala}}, \bibinfo {author} {\bibfnamefont {G.}~\bibnamefont {Tardini}}, \bibinfo {author} {\bibfnamefont {P.}~\bibnamefont {de~Vries}},\ and\ \bibinfo {author} {\bibfnamefont {J.}~\bibnamefont {Weiland}},\ }\href@noop {} {\bibfield  {journal} {\bibinfo  {journal} {Nuclear Fusion}\ }\textbf {\bibinfo {volume} {51}},\ \bibinfo {pages} {094027} (\bibinfo {year} {2011})}\BibitemShut {NoStop}%
\bibitem [{\citenamefont {Parra}\ and\ \citenamefont {Barnes}(2015)}]{Parratheory_2015}%
  \BibitemOpen
  \bibfield  {author} {\bibinfo {author} {\bibfnamefont {F.~I.}\ \bibnamefont {Parra}}\ and\ \bibinfo {author} {\bibfnamefont {M.}~\bibnamefont {Barnes}},\ }\href@noop {} {\bibfield  {journal} {\bibinfo  {journal} {Plasma Physics and Controlled Fusion}\ }\textbf {\bibinfo {volume} {57}},\ \bibinfo {pages} {045002} (\bibinfo {year} {2015})}\BibitemShut {NoStop}%
\bibitem [{\citenamefont {Ball}\ \emph {et~al.}(2014)\citenamefont {Ball}, \citenamefont {Parra}, \citenamefont {Barnes},\ and\ \citenamefont {et~al.}}]{ball2014}%
  \BibitemOpen
  \bibfield  {author} {\bibinfo {author} {\bibfnamefont {J.}~\bibnamefont {Ball}}, \bibinfo {author} {\bibfnamefont {F.~I.}\ \bibnamefont {Parra}}, \bibinfo {author} {\bibfnamefont {M.}~\bibnamefont {Barnes}},\ and\ \bibinfo {author} {\bibfnamefont {W.~D.}\ \bibnamefont {et~al.}},\ }\href@noop {} {\bibfield  {journal} {\bibinfo  {journal} {Plasma Physics and Controlled Fusion}\ }\textbf {\bibinfo {volume} {56}},\ \bibinfo {pages} {095014} (\bibinfo {year} {2014})}\BibitemShut {NoStop}%
\bibitem [{\citenamefont {Ball}\ \emph {et~al.}(2018)\citenamefont {Ball}, \citenamefont {Parra}, \citenamefont {Landreman},\ and\ \citenamefont {Barnes}}]{ball2018}%
  \BibitemOpen
  \bibfield  {author} {\bibinfo {author} {\bibfnamefont {J.}~\bibnamefont {Ball}}, \bibinfo {author} {\bibfnamefont {F.~I.}\ \bibnamefont {Parra}}, \bibinfo {author} {\bibfnamefont {M.}~\bibnamefont {Landreman}},\ and\ \bibinfo {author} {\bibfnamefont {M.}~\bibnamefont {Barnes}},\ }\href@noop {} {\bibfield  {journal} {\bibinfo  {journal} {Nuclear Fusion}\ }\textbf {\bibinfo {volume} {58}},\ \bibinfo {pages} {026003} (\bibinfo {year} {2018})}\BibitemShut {NoStop}%
\bibitem [{\citenamefont {Newton}\ and\ \citenamefont {Helander}(2006)}]{Newton2006neoclassical}%
  \BibitemOpen
  \bibfield  {author} {\bibinfo {author} {\bibfnamefont {S.}~\bibnamefont {Newton}}\ and\ \bibinfo {author} {\bibfnamefont {P.}~\bibnamefont {Helander}},\ }\href@noop {} {\bibfield  {journal} {\bibinfo  {journal} {Physics of Plasmas}\ }\textbf {\bibinfo {volume} {13}},\ \bibinfo {pages} {012505} (\bibinfo {year} {2006})}\BibitemShut {NoStop}%
\bibitem [{\citenamefont {Wang}\ \emph {et~al.}(2009)\citenamefont {Wang}, \citenamefont {Hahm}, \citenamefont {Ethier}, \citenamefont {Rewoldt}, \citenamefont {Lee}, \citenamefont {Tang}, \citenamefont {Kaye},\ and\ \citenamefont {Diamond}}]{Wang2009PRL}%
  \BibitemOpen
  \bibfield  {author} {\bibinfo {author} {\bibfnamefont {W.~X.}\ \bibnamefont {Wang}}, \bibinfo {author} {\bibfnamefont {T.~S.}\ \bibnamefont {Hahm}}, \bibinfo {author} {\bibfnamefont {S.}~\bibnamefont {Ethier}}, \bibinfo {author} {\bibfnamefont {G.}~\bibnamefont {Rewoldt}}, \bibinfo {author} {\bibfnamefont {W.~W.}\ \bibnamefont {Lee}}, \bibinfo {author} {\bibfnamefont {W.~M.}\ \bibnamefont {Tang}}, \bibinfo {author} {\bibfnamefont {S.~M.}\ \bibnamefont {Kaye}},\ and\ \bibinfo {author} {\bibfnamefont {P.~H.}\ \bibnamefont {Diamond}},\ }\href@noop {} {\bibfield  {journal} {\bibinfo  {journal} {Phys. Rev. Lett.}\ }\textbf {\bibinfo {volume} {102}},\ \bibinfo {pages} {035005} (\bibinfo {year} {2009})}\BibitemShut {NoStop}%
\bibitem [{\citenamefont {Hahm}\ \emph {et~al.}(2007)\citenamefont {Hahm}, \citenamefont {Diamond}, \citenamefont {Gurcan},\ and\ \citenamefont {Rewoldt}}]{Hahm2007NLtheory}%
  \BibitemOpen
  \bibfield  {author} {\bibinfo {author} {\bibfnamefont {T.~S.}\ \bibnamefont {Hahm}}, \bibinfo {author} {\bibfnamefont {P.~H.}\ \bibnamefont {Diamond}}, \bibinfo {author} {\bibfnamefont {O.~D.}\ \bibnamefont {Gurcan}},\ and\ \bibinfo {author} {\bibfnamefont {G.}~\bibnamefont {Rewoldt}},\ }\href@noop {} {\bibfield  {journal} {\bibinfo  {journal} {Physics of Plasmas}\ }\textbf {\bibinfo {volume} {14}} (\bibinfo {year} {2007})},\ \bibinfo {note} {072302}\BibitemShut {NoStop}%
\bibitem [{\citenamefont {Diamond}\ \emph {et~al.}(2008)\citenamefont {Diamond}, \citenamefont {McDevitt}, \citenamefont {G\"urcan}, \citenamefont {Hahm},\ and\ \citenamefont {Naulin}}]{Diamond2008momentumtransport}%
  \BibitemOpen
  \bibfield  {author} {\bibinfo {author} {\bibfnamefont {P.~H.}\ \bibnamefont {Diamond}}, \bibinfo {author} {\bibfnamefont {C.~J.}\ \bibnamefont {McDevitt}}, \bibinfo {author} {\bibfnamefont {{\"O}.~D.}\ \bibnamefont {G\"urcan}}, \bibinfo {author} {\bibfnamefont {T.~S.}\ \bibnamefont {Hahm}},\ and\ \bibinfo {author} {\bibfnamefont {V.}~\bibnamefont {Naulin}},\ }\href@noop {} {\bibfield  {journal} {\bibinfo  {journal} {Physics of Plasmas}\ }\textbf {\bibinfo {volume} {15}} (\bibinfo {year} {2008})},\ \bibinfo {note} {012303}\BibitemShut {NoStop}%
\bibitem [{\citenamefont {Holod}\ and\ \citenamefont {Lin}(2008)}]{Holod2008GKSim}%
  \BibitemOpen
  \bibfield  {author} {\bibinfo {author} {\bibfnamefont {I.}~\bibnamefont {Holod}}\ and\ \bibinfo {author} {\bibfnamefont {Z.}~\bibnamefont {Lin}},\ }\href@noop {} {\bibfield  {journal} {\bibinfo  {journal} {Physics of Plasmas}\ }\textbf {\bibinfo {volume} {15}} (\bibinfo {year} {2008})},\ \bibinfo {note} {092302}\BibitemShut {NoStop}%
\bibitem [{\citenamefont {Casson}\ \emph {et~al.}(2009)\citenamefont {Casson}, \citenamefont {Peeters}, \citenamefont {Camenen}, \citenamefont {Hornsby}, \citenamefont {Snodin}, \citenamefont {Strintzi},\ and\ \citenamefont {Szepesi}}]{CassonExBshear2009}%
  \BibitemOpen
  \bibfield  {author} {\bibinfo {author} {\bibfnamefont {F.}~\bibnamefont {Casson}}, \bibinfo {author} {\bibfnamefont {A.}~\bibnamefont {Peeters}}, \bibinfo {author} {\bibfnamefont {Y.}~\bibnamefont {Camenen}}, \bibinfo {author} {\bibfnamefont {W.}~\bibnamefont {Hornsby}}, \bibinfo {author} {\bibfnamefont {A.}~\bibnamefont {Snodin}}, \bibinfo {author} {\bibfnamefont {D.}~\bibnamefont {Strintzi}},\ and\ \bibinfo {author} {\bibfnamefont {G.}~\bibnamefont {Szepesi}},\ }\href@noop {} {\bibfield  {journal} {\bibinfo  {journal} {Physics of Plasmas}\ }\textbf {\bibinfo {volume} {16}},\ \bibinfo {pages} {092303} (\bibinfo {year} {2009})}\BibitemShut {NoStop}%
\bibitem [{\citenamefont {Roach}\ \emph {et~al.}(2009)\citenamefont {Roach}, \citenamefont {Abel}, \citenamefont {Akers}, \citenamefont {Arter}, \citenamefont {Barnes}, \citenamefont {Camenen}, \citenamefont {Casson}, \citenamefont {Colyer}, \citenamefont {Connor}, \citenamefont {Cowley}, \citenamefont {Dickinson}, \citenamefont {Dorland}, \citenamefont {Field}, \citenamefont {Guttenfelder}, \citenamefont {Hammett}, \citenamefont {Hastie}, \citenamefont {Highcock}, \citenamefont {Loureiro}, \citenamefont {Peeters}, \citenamefont {Reshko}, \citenamefont {Saarelma}, \citenamefont {Schekochihin}, \citenamefont {Valovic},\ and\ \citenamefont {Wilson}}]{Roach_2009}%
  \BibitemOpen
  \bibfield  {author} {\bibinfo {author} {\bibfnamefont {C.~M.}\ \bibnamefont {Roach}}, \bibinfo {author} {\bibfnamefont {I.~G.}\ \bibnamefont {Abel}}, \bibinfo {author} {\bibfnamefont {R.~J.}\ \bibnamefont {Akers}}, \bibinfo {author} {\bibfnamefont {W.}~\bibnamefont {Arter}}, \bibinfo {author} {\bibfnamefont {M.}~\bibnamefont {Barnes}}, \bibinfo {author} {\bibfnamefont {Y.}~\bibnamefont {Camenen}}, \bibinfo {author} {\bibfnamefont {F.~J.}\ \bibnamefont {Casson}}, \bibinfo {author} {\bibfnamefont {G.}~\bibnamefont {Colyer}}, \bibinfo {author} {\bibfnamefont {J.~W.}\ \bibnamefont {Connor}}, \bibinfo {author} {\bibfnamefont {S.~C.}\ \bibnamefont {Cowley}}, \bibinfo {author} {\bibfnamefont {D.}~\bibnamefont {Dickinson}}, \bibinfo {author} {\bibfnamefont {W.}~\bibnamefont {Dorland}}, \bibinfo {author} {\bibfnamefont {A.~R.}\ \bibnamefont {Field}}, \bibinfo {author} {\bibfnamefont {W.}~\bibnamefont {Guttenfelder}}, \bibinfo {author} {\bibfnamefont {G.~W.}\ \bibnamefont {Hammett}}, \bibinfo {author} {\bibfnamefont
  {R.~J.}\ \bibnamefont {Hastie}}, \bibinfo {author} {\bibfnamefont {E.}~\bibnamefont {Highcock}}, \bibinfo {author} {\bibfnamefont {N.~F.}\ \bibnamefont {Loureiro}}, \bibinfo {author} {\bibfnamefont {A.~G.}\ \bibnamefont {Peeters}}, \bibinfo {author} {\bibfnamefont {M.}~\bibnamefont {Reshko}}, \bibinfo {author} {\bibfnamefont {S.}~\bibnamefont {Saarelma}}, \bibinfo {author} {\bibfnamefont {A.~A.}\ \bibnamefont {Schekochihin}}, \bibinfo {author} {\bibfnamefont {M.}~\bibnamefont {Valovic}},\ and\ \bibinfo {author} {\bibfnamefont {H.~R.}\ \bibnamefont {Wilson}},\ }\href@noop {} {\bibfield  {journal} {\bibinfo  {journal} {Plasma Physics and Controlled Fusion}\ }\textbf {\bibinfo {volume} {51}},\ \bibinfo {pages} {124020} (\bibinfo {year} {2009})}\BibitemShut {NoStop}%
\bibitem [{\citenamefont {Yoon}\ and\ \citenamefont {Hahm}(2010)}]{Yoon_2010momentumtransportITG}%
  \BibitemOpen
  \bibfield  {author} {\bibinfo {author} {\bibfnamefont {E.}~\bibnamefont {Yoon}}\ and\ \bibinfo {author} {\bibfnamefont {T.}~\bibnamefont {Hahm}},\ }\href@noop {} {\bibfield  {journal} {\bibinfo  {journal} {Nuclear Fusion}\ }\textbf {\bibinfo {volume} {50}},\ \bibinfo {pages} {064006} (\bibinfo {year} {2010})}\BibitemShut {NoStop}%
\bibitem [{\citenamefont {Camenen}\ \emph {et~al.}(2011)\citenamefont {Camenen}, \citenamefont {Idomura}, \citenamefont {Jolliet},\ and\ \citenamefont {Peeters}}]{Camenen_2011NFReview}%
  \BibitemOpen
  \bibfield  {author} {\bibinfo {author} {\bibfnamefont {Y.}~\bibnamefont {Camenen}}, \bibinfo {author} {\bibfnamefont {Y.}~\bibnamefont {Idomura}}, \bibinfo {author} {\bibfnamefont {S.}~\bibnamefont {Jolliet}},\ and\ \bibinfo {author} {\bibfnamefont {A.}~\bibnamefont {Peeters}},\ }\href@noop {} {\bibfield  {journal} {\bibinfo  {journal} {Nuclear Fusion}\ }\textbf {\bibinfo {volume} {51}},\ \bibinfo {pages} {073039} (\bibinfo {year} {2011})}\BibitemShut {NoStop}%
\bibitem [{\citenamefont {Angioni}\ \emph {et~al.}(2012)\citenamefont {Angioni}, \citenamefont {Camenen}, \citenamefont {Casson}, \citenamefont {Fable}, \citenamefont {McDermott}, \citenamefont {Peeters},\ and\ \citenamefont {Rice}}]{Angioni_2012NFReview}%
  \BibitemOpen
  \bibfield  {author} {\bibinfo {author} {\bibfnamefont {C.}~\bibnamefont {Angioni}}, \bibinfo {author} {\bibfnamefont {Y.}~\bibnamefont {Camenen}}, \bibinfo {author} {\bibfnamefont {F.}~\bibnamefont {Casson}}, \bibinfo {author} {\bibfnamefont {E.}~\bibnamefont {Fable}}, \bibinfo {author} {\bibfnamefont {R.}~\bibnamefont {McDermott}}, \bibinfo {author} {\bibfnamefont {A.}~\bibnamefont {Peeters}},\ and\ \bibinfo {author} {\bibfnamefont {J.}~\bibnamefont {Rice}},\ }\href@noop {} {\bibfield  {journal} {\bibinfo  {journal} {Nuclear Fusion}\ }\textbf {\bibinfo {volume} {52}},\ \bibinfo {pages} {114003} (\bibinfo {year} {2012})}\BibitemShut {NoStop}%
\bibitem [{\citenamefont {Diamond}\ \emph {et~al.}(2013)\citenamefont {Diamond}, \citenamefont {Kosuga}, \citenamefont {G\"urcan}, \citenamefont {McDevitt}, \citenamefont {Hahm}, \citenamefont {Fedorczak}, \citenamefont {Rice}, \citenamefont {Wang}, \citenamefont {Ku}, \citenamefont {Kwon}, \citenamefont {Dif-Pradalier}, \citenamefont {Abiteboul}, \citenamefont {Wang}, \citenamefont {Ko}, \citenamefont {Shi}, \citenamefont {Ida}, \citenamefont {Solomon}, \citenamefont {Jhang}, \citenamefont {Kim}, \citenamefont {Yi}, \citenamefont {Ko}, \citenamefont {Sarazin}, \citenamefont {Singh},\ and\ \citenamefont {Chang}}]{Diamond_2013NFreview}%
  \BibitemOpen
  \bibfield  {author} {\bibinfo {author} {\bibfnamefont {P.}~\bibnamefont {Diamond}}, \bibinfo {author} {\bibfnamefont {Y.}~\bibnamefont {Kosuga}}, \bibinfo {author} {\bibfnamefont {{\"O}.}~\bibnamefont {G\"urcan}}, \bibinfo {author} {\bibfnamefont {C.}~\bibnamefont {McDevitt}}, \bibinfo {author} {\bibfnamefont {T.}~\bibnamefont {Hahm}}, \bibinfo {author} {\bibfnamefont {N.}~\bibnamefont {Fedorczak}}, \bibinfo {author} {\bibfnamefont {J.}~\bibnamefont {Rice}}, \bibinfo {author} {\bibfnamefont {W.}~\bibnamefont {Wang}}, \bibinfo {author} {\bibfnamefont {S.}~\bibnamefont {Ku}}, \bibinfo {author} {\bibfnamefont {J.}~\bibnamefont {Kwon}}, \bibinfo {author} {\bibfnamefont {G.}~\bibnamefont {Dif-Pradalier}}, \bibinfo {author} {\bibfnamefont {J.}~\bibnamefont {Abiteboul}}, \bibinfo {author} {\bibfnamefont {L.}~\bibnamefont {Wang}}, \bibinfo {author} {\bibfnamefont {W.}~\bibnamefont {Ko}}, \bibinfo {author} {\bibfnamefont {Y.}~\bibnamefont {Shi}}, \bibinfo {author} {\bibfnamefont {K.}~\bibnamefont {Ida}}, \bibinfo
  {author} {\bibfnamefont {W.}~\bibnamefont {Solomon}}, \bibinfo {author} {\bibfnamefont {H.}~\bibnamefont {Jhang}}, \bibinfo {author} {\bibfnamefont {S.}~\bibnamefont {Kim}}, \bibinfo {author} {\bibfnamefont {S.}~\bibnamefont {Yi}}, \bibinfo {author} {\bibfnamefont {S.}~\bibnamefont {Ko}}, \bibinfo {author} {\bibfnamefont {Y.}~\bibnamefont {Sarazin}}, \bibinfo {author} {\bibfnamefont {R.}~\bibnamefont {Singh}},\ and\ \bibinfo {author} {\bibfnamefont {C.}~\bibnamefont {Chang}},\ }\href@noop {} {\bibfield  {journal} {\bibinfo  {journal} {Nuclear Fusion}\ }\textbf {\bibinfo {volume} {53}},\ \bibinfo {pages} {104019} (\bibinfo {year} {2013})}\BibitemShut {NoStop}%
\bibitem [{\citenamefont {Peeters}\ \emph {et~al.}(2007)\citenamefont {Peeters}, \citenamefont {Angioni},\ and\ \citenamefont {Strintzi}}]{Peeters2007PRLpinchterm}%
  \BibitemOpen
  \bibfield  {author} {\bibinfo {author} {\bibfnamefont {A.~G.}\ \bibnamefont {Peeters}}, \bibinfo {author} {\bibfnamefont {C.}~\bibnamefont {Angioni}},\ and\ \bibinfo {author} {\bibfnamefont {D.}~\bibnamefont {Strintzi}},\ }\href@noop {} {\bibfield  {journal} {\bibinfo  {journal} {Phys. Rev. Lett.}\ }\textbf {\bibinfo {volume} {98}},\ \bibinfo {pages} {265003} (\bibinfo {year} {2007})}\BibitemShut {NoStop}%
\bibitem [{\citenamefont {Guttenfelder}\ \emph {et~al.}(2017)\citenamefont {Guttenfelder}, \citenamefont {Field}, \citenamefont {Lupelli}, \citenamefont {Tala}, \citenamefont {Kaye}, \citenamefont {Ren},\ and\ \citenamefont {Solomon}}]{Guttenfelder_2017}%
  \BibitemOpen
  \bibfield  {author} {\bibinfo {author} {\bibfnamefont {W.}~\bibnamefont {Guttenfelder}}, \bibinfo {author} {\bibfnamefont {A.}~\bibnamefont {Field}}, \bibinfo {author} {\bibfnamefont {I.}~\bibnamefont {Lupelli}}, \bibinfo {author} {\bibfnamefont {T.}~\bibnamefont {Tala}}, \bibinfo {author} {\bibfnamefont {S.}~\bibnamefont {Kaye}}, \bibinfo {author} {\bibfnamefont {Y.}~\bibnamefont {Ren}},\ and\ \bibinfo {author} {\bibfnamefont {W.}~\bibnamefont {Solomon}},\ }\href@noop {} {\bibfield  {journal} {\bibinfo  {journal} {Nuclear Fusion}\ }\textbf {\bibinfo {volume} {57}},\ \bibinfo {pages} {056022} (\bibinfo {year} {2017})}\BibitemShut {NoStop}%
\bibitem [{\citenamefont {Zimmermann}\ \emph {et~al.}(2022)\citenamefont {Zimmermann}, \citenamefont {McDermott}, \citenamefont {Fable}, \citenamefont {Angioni}, \citenamefont {Duval}, \citenamefont {Dux}, \citenamefont {Salmi}, \citenamefont {Stroth}, \citenamefont {Tala}, \citenamefont {Tardini}, \citenamefont {P\"utterich}, \citenamefont {the ASDEX~Upgrade},\ and\ \citenamefont {Teams}}]{Zimmermann_2022}%
  \BibitemOpen
  \bibfield  {author} {\bibinfo {author} {\bibfnamefont {C.~F.~B.}\ \bibnamefont {Zimmermann}}, \bibinfo {author} {\bibfnamefont {R.~M.}\ \bibnamefont {McDermott}}, \bibinfo {author} {\bibfnamefont {E.}~\bibnamefont {Fable}}, \bibinfo {author} {\bibfnamefont {C.}~\bibnamefont {Angioni}}, \bibinfo {author} {\bibfnamefont {B.~P.}\ \bibnamefont {Duval}}, \bibinfo {author} {\bibfnamefont {R.}~\bibnamefont {Dux}}, \bibinfo {author} {\bibfnamefont {A.}~\bibnamefont {Salmi}}, \bibinfo {author} {\bibfnamefont {U.}~\bibnamefont {Stroth}}, \bibinfo {author} {\bibfnamefont {T.}~\bibnamefont {Tala}}, \bibinfo {author} {\bibfnamefont {G.}~\bibnamefont {Tardini}}, \bibinfo {author} {\bibfnamefont {T.}~\bibnamefont {P\"utterich}}, \bibinfo {author} {\bibnamefont {the ASDEX~Upgrade}},\ and\ \bibinfo {author} {\bibfnamefont {E.~M.}\ \bibnamefont {Teams}},\ }\href@noop {} {\bibfield  {journal} {\bibinfo  {journal} {Plasma Physics and Controlled Fusion}\ }\textbf {\bibinfo {volume} {64}},\ \bibinfo {pages} {055020} (\bibinfo
  {year} {2022})}\BibitemShut {NoStop}%
\bibitem [{\citenamefont {Zimmermann}\ \emph {et~al.}(2023)\citenamefont {Zimmermann}, \citenamefont {McDermott}, \citenamefont {Angioni}, \citenamefont {Duval}, \citenamefont {Dux}, \citenamefont {Fable}, \citenamefont {Salmi}, \citenamefont {Stroth}, \citenamefont {Tala}, \citenamefont {Tardini}, \citenamefont {P\"utterich},\ and\ \citenamefont {the ASDEX Upgrade~Team}}]{Zimmermann_2023}%
  \BibitemOpen
  \bibfield  {author} {\bibinfo {author} {\bibfnamefont {C.}~\bibnamefont {Zimmermann}}, \bibinfo {author} {\bibfnamefont {R.}~\bibnamefont {McDermott}}, \bibinfo {author} {\bibfnamefont {C.}~\bibnamefont {Angioni}}, \bibinfo {author} {\bibfnamefont {B.}~\bibnamefont {Duval}}, \bibinfo {author} {\bibfnamefont {R.}~\bibnamefont {Dux}}, \bibinfo {author} {\bibfnamefont {E.}~\bibnamefont {Fable}}, \bibinfo {author} {\bibfnamefont {A.}~\bibnamefont {Salmi}}, \bibinfo {author} {\bibfnamefont {U.}~\bibnamefont {Stroth}}, \bibinfo {author} {\bibfnamefont {T.}~\bibnamefont {Tala}}, \bibinfo {author} {\bibfnamefont {G.}~\bibnamefont {Tardini}}, \bibinfo {author} {\bibfnamefont {T.}~\bibnamefont {P\"utterich}},\ and\ \bibinfo {author} {\bibnamefont {the ASDEX Upgrade~Team}},\ }\href@noop {} {\bibfield  {journal} {\bibinfo  {journal} {Nuclear Fusion}\ }\textbf {\bibinfo {volume} {63}},\ \bibinfo {pages} {126006} (\bibinfo {year} {2023})}\BibitemShut {NoStop}%
\bibitem [{\citenamefont {Diamond}\ \emph {et~al.}(2009)\citenamefont {Diamond}, \citenamefont {McDevitt}, \citenamefont {G\"urcan}, \citenamefont {Hahm}, \citenamefont {Wang}, \citenamefont {Yoon}, \citenamefont {Holod}, \citenamefont {Lin}, \citenamefont {Naulin},\ and\ \citenamefont {Singh}}]{Diamond_2009}%
  \BibitemOpen
  \bibfield  {author} {\bibinfo {author} {\bibfnamefont {P.}~\bibnamefont {Diamond}}, \bibinfo {author} {\bibfnamefont {C.}~\bibnamefont {McDevitt}}, \bibinfo {author} {\bibfnamefont {Ã.}~\bibnamefont {G\"urcan}}, \bibinfo {author} {\bibfnamefont {T.}~\bibnamefont {Hahm}}, \bibinfo {author} {\bibfnamefont {W.~X.}\ \bibnamefont {Wang}}, \bibinfo {author} {\bibfnamefont {E.}~\bibnamefont {Yoon}}, \bibinfo {author} {\bibfnamefont {I.}~\bibnamefont {Holod}}, \bibinfo {author} {\bibfnamefont {Z.}~\bibnamefont {Lin}}, \bibinfo {author} {\bibfnamefont {V.}~\bibnamefont {Naulin}},\ and\ \bibinfo {author} {\bibfnamefont {R.}~\bibnamefont {Singh}},\ }\href@noop {} {\bibfield  {journal} {\bibinfo  {journal} {Nuclear Fusion}\ }\textbf {\bibinfo {volume} {49}},\ \bibinfo {pages} {045002} (\bibinfo {year} {2009})}\BibitemShut {NoStop}%
\bibitem [{\citenamefont {Tala}\ \emph {et~al.}(2011)\citenamefont {Tala}, \citenamefont {Salmi}, \citenamefont {Angioni}, \citenamefont {Casson}, \citenamefont {Corrigan}, \citenamefont {Ferreira}, \citenamefont {Giroud}, \citenamefont {Mantica}, \citenamefont {Naulin}, \citenamefont {Peeters}, \citenamefont {Solomon}, \citenamefont {Strintzi}, \citenamefont {Tsalas}, \citenamefont {Versloot}, \citenamefont {de~Vries}, \citenamefont {Zastrow},\ and\ \citenamefont {{JET-EFDA contributors}}}]{Tala_2011}%
  \BibitemOpen
  \bibfield  {author} {\bibinfo {author} {\bibfnamefont {T.}~\bibnamefont {Tala}}, \bibinfo {author} {\bibfnamefont {A.}~\bibnamefont {Salmi}}, \bibinfo {author} {\bibfnamefont {C.}~\bibnamefont {Angioni}}, \bibinfo {author} {\bibfnamefont {F.}~\bibnamefont {Casson}}, \bibinfo {author} {\bibfnamefont {G.}~\bibnamefont {Corrigan}}, \bibinfo {author} {\bibfnamefont {J.}~\bibnamefont {Ferreira}}, \bibinfo {author} {\bibfnamefont {C.}~\bibnamefont {Giroud}}, \bibinfo {author} {\bibfnamefont {P.}~\bibnamefont {Mantica}}, \bibinfo {author} {\bibfnamefont {V.}~\bibnamefont {Naulin}}, \bibinfo {author} {\bibfnamefont {A.}~\bibnamefont {Peeters}}, \bibinfo {author} {\bibfnamefont {W.}~\bibnamefont {Solomon}}, \bibinfo {author} {\bibfnamefont {D.}~\bibnamefont {Strintzi}}, \bibinfo {author} {\bibfnamefont {M.}~\bibnamefont {Tsalas}}, \bibinfo {author} {\bibfnamefont {T.}~\bibnamefont {Versloot}}, \bibinfo {author} {\bibfnamefont {P.}~\bibnamefont {de~Vries}}, \bibinfo {author} {\bibfnamefont {K.-D.}\ \bibnamefont
  {Zastrow}},\ and\ \bibinfo {author} {\bibnamefont {{JET-EFDA contributors}}},\ }\href@noop {} {\bibfield  {journal} {\bibinfo  {journal} {Nuclear Fusion}\ }\textbf {\bibinfo {volume} {51}},\ \bibinfo {pages} {123002} (\bibinfo {year} {2011})}\BibitemShut {NoStop}%
\bibitem [{\citenamefont {Weisen}\ \emph {et~al.}(2012)\citenamefont {Weisen}, \citenamefont {Camenen}, \citenamefont {Salmi}, \citenamefont {Versloot}, \citenamefont {de~Vries}, \citenamefont {Maslov}, \citenamefont {Tala}, \citenamefont {Beurskens}, \citenamefont {Giroud},\ and\ \citenamefont {{JET-EFDA contributors}}}]{Weisen_2012}%
  \BibitemOpen
  \bibfield  {author} {\bibinfo {author} {\bibfnamefont {H.}~\bibnamefont {Weisen}}, \bibinfo {author} {\bibfnamefont {Y.}~\bibnamefont {Camenen}}, \bibinfo {author} {\bibfnamefont {A.}~\bibnamefont {Salmi}}, \bibinfo {author} {\bibfnamefont {T.}~\bibnamefont {Versloot}}, \bibinfo {author} {\bibfnamefont {P.}~\bibnamefont {de~Vries}}, \bibinfo {author} {\bibfnamefont {M.}~\bibnamefont {Maslov}}, \bibinfo {author} {\bibfnamefont {T.}~\bibnamefont {Tala}}, \bibinfo {author} {\bibfnamefont {M.}~\bibnamefont {Beurskens}}, \bibinfo {author} {\bibfnamefont {C.}~\bibnamefont {Giroud}},\ and\ \bibinfo {author} {\bibnamefont {{JET-EFDA contributors}}},\ }\href@noop {} {\bibfield  {journal} {\bibinfo  {journal} {Nuclear Fusion}\ }\textbf {\bibinfo {volume} {52}},\ \bibinfo {pages} {042001} (\bibinfo {year} {2012})}\BibitemShut {NoStop}%
\bibitem [{\citenamefont {Highcock}\ \emph {et~al.}(2010)\citenamefont {Highcock}, \citenamefont {Barnes}, \citenamefont {Schekochihin}, \citenamefont {Parra}, \citenamefont {Roach},\ and\ \citenamefont {Cowley}}]{HighcockRotationBifurcation2010}%
  \BibitemOpen
  \bibfield  {author} {\bibinfo {author} {\bibfnamefont {E.}~\bibnamefont {Highcock}}, \bibinfo {author} {\bibfnamefont {M.}~\bibnamefont {Barnes}}, \bibinfo {author} {\bibfnamefont {A.}~\bibnamefont {Schekochihin}}, \bibinfo {author} {\bibfnamefont {F.}~\bibnamefont {Parra}}, \bibinfo {author} {\bibfnamefont {C.}~\bibnamefont {Roach}},\ and\ \bibinfo {author} {\bibfnamefont {S.}~\bibnamefont {Cowley}},\ }\href@noop {} {\bibfield  {journal} {\bibinfo  {journal} {Phys. Rev. Lett.}\ }\textbf {\bibinfo {volume} {105}},\ \bibinfo {pages} {215003} (\bibinfo {year} {2010})}\BibitemShut {NoStop}%
\bibitem [{\citenamefont {McMillan}\ and\ \citenamefont {Dominski}(2019{\natexlab{a}})}]{Ben_2019}%
  \BibitemOpen
  \bibfield  {author} {\bibinfo {author} {\bibfnamefont {B.~F.}\ \bibnamefont {McMillan}}\ and\ \bibinfo {author} {\bibfnamefont {J.}~\bibnamefont {Dominski}},\ }\href@noop {} {\bibfield  {journal} {\bibinfo  {journal} {Journal of Plasma Physics}\ }\textbf {\bibinfo {volume} {85}},\ \bibinfo {pages} {175850301} (\bibinfo {year} {2019}{\natexlab{a}})}\BibitemShut {NoStop}%
\bibitem [{\citenamefont {McMillan}(2015)}]{Ben_2015_intrinsic}%
  \BibitemOpen
  \bibfield  {author} {\bibinfo {author} {\bibfnamefont {B.~F.}\ \bibnamefont {McMillan}},\ }\href@noop {} {\bibfield  {journal} {\bibinfo  {journal} {Physics of Plasmas}\ }\textbf {\bibinfo {volume} {22}},\ \bibinfo {pages} {020707} (\bibinfo {year} {2015})}\BibitemShut {NoStop}%
\bibitem [{\citenamefont {Abel}\ \emph {et~al.}(2013)\citenamefont {Abel}, \citenamefont {Plunk}, \citenamefont {Wang}, \citenamefont {Barnes}, \citenamefont {Cowley}, \citenamefont {Dorland},\ and\ \citenamefont {Schekochihin}}]{AbelGyrokineticsDeriv2012}%
  \BibitemOpen
  \bibfield  {author} {\bibinfo {author} {\bibfnamefont {I.}~\bibnamefont {Abel}}, \bibinfo {author} {\bibfnamefont {G.}~\bibnamefont {Plunk}}, \bibinfo {author} {\bibfnamefont {E.}~\bibnamefont {Wang}}, \bibinfo {author} {\bibfnamefont {M.}~\bibnamefont {Barnes}}, \bibinfo {author} {\bibfnamefont {S.}~\bibnamefont {Cowley}}, \bibinfo {author} {\bibfnamefont {W.}~\bibnamefont {Dorland}},\ and\ \bibinfo {author} {\bibfnamefont {A.}~\bibnamefont {Schekochihin}},\ }\href@noop {} {\bibfield  {journal} {\bibinfo  {journal} {Rep. Prog. Phys}\ }\textbf {\bibinfo {volume} {76}},\ \bibinfo {pages} {116201} (\bibinfo {year} {2013})}\BibitemShut {NoStop}%
\bibitem [{\citenamefont {McDermott}\ \emph {et~al.}(2011)\citenamefont {McDermott}, \citenamefont {Angioni}, \citenamefont {Dux}, \citenamefont {Gude}, \citenamefont {PÃ¼tterich}, \citenamefont {Ryter}, \citenamefont {Tardini},\ and\ \citenamefont {the ASDEX Upgrade~Team}}]{McDermott_2011}%
  \BibitemOpen
  \bibfield  {author} {\bibinfo {author} {\bibfnamefont {R.~M.}\ \bibnamefont {McDermott}}, \bibinfo {author} {\bibfnamefont {C.}~\bibnamefont {Angioni}}, \bibinfo {author} {\bibfnamefont {R.}~\bibnamefont {Dux}}, \bibinfo {author} {\bibfnamefont {A.}~\bibnamefont {Gude}}, \bibinfo {author} {\bibfnamefont {T.}~\bibnamefont {PÃ¼tterich}}, \bibinfo {author} {\bibfnamefont {F.}~\bibnamefont {Ryter}}, \bibinfo {author} {\bibfnamefont {G.}~\bibnamefont {Tardini}},\ and\ \bibinfo {author} {\bibnamefont {the ASDEX Upgrade~Team}},\ }\href {https://doi.org/10.1088/0741-3335/53/3/035007} {\bibfield  {journal} {\bibinfo  {journal} {Plasma Physics and Controlled Fusion}\ }\textbf {\bibinfo {volume} {53}},\ \bibinfo {pages} {035007} (\bibinfo {year} {2011})}\BibitemShut {NoStop}%
\bibitem [{\citenamefont {Buchholz}\ \emph {et~al.}(2015)\citenamefont {Buchholz}, \citenamefont {Grosshauser}, \citenamefont {Guttenfelder}, \citenamefont {Hornsby}, \citenamefont {Migliano}, \citenamefont {Peeters},\ and\ \citenamefont {Strintzi}}]{BuchholzPrandtl2015}%
  \BibitemOpen
  \bibfield  {author} {\bibinfo {author} {\bibfnamefont {R.}~\bibnamefont {Buchholz}}, \bibinfo {author} {\bibfnamefont {S.}~\bibnamefont {Grosshauser}}, \bibinfo {author} {\bibfnamefont {W.}~\bibnamefont {Guttenfelder}}, \bibinfo {author} {\bibfnamefont {W.~A.}\ \bibnamefont {Hornsby}}, \bibinfo {author} {\bibfnamefont {P.}~\bibnamefont {Migliano}}, \bibinfo {author} {\bibfnamefont {A.~G.}\ \bibnamefont {Peeters}},\ and\ \bibinfo {author} {\bibfnamefont {D.}~\bibnamefont {Strintzi}},\ }\href@noop {} {\bibfield  {journal} {\bibinfo  {journal} {Physics of Plasmas}\ }\textbf {\bibinfo {volume} {22}},\ \bibinfo {pages} {082307} (\bibinfo {year} {2015})}\BibitemShut {NoStop}%
\bibitem [{\citenamefont {Hornsby}\ \emph {et~al.}(2018)\citenamefont {Hornsby}, \citenamefont {Angioni}, \citenamefont {Lu}, \citenamefont {Fable}, \citenamefont {Erofeev}, \citenamefont {McDermott}, \citenamefont {Medvedeva}, \citenamefont {Lebschy}, \citenamefont {Peeters},\ and\ \citenamefont {Team}}]{Hornsby_2018}%
  \BibitemOpen
  \bibfield  {author} {\bibinfo {author} {\bibfnamefont {W.}~\bibnamefont {Hornsby}}, \bibinfo {author} {\bibfnamefont {C.}~\bibnamefont {Angioni}}, \bibinfo {author} {\bibfnamefont {Z.}~\bibnamefont {Lu}}, \bibinfo {author} {\bibfnamefont {E.}~\bibnamefont {Fable}}, \bibinfo {author} {\bibfnamefont {I.}~\bibnamefont {Erofeev}}, \bibinfo {author} {\bibfnamefont {R.}~\bibnamefont {McDermott}}, \bibinfo {author} {\bibfnamefont {A.}~\bibnamefont {Medvedeva}}, \bibinfo {author} {\bibfnamefont {A.}~\bibnamefont {Lebschy}}, \bibinfo {author} {\bibfnamefont {A.}~\bibnamefont {Peeters}},\ and\ \bibinfo {author} {\bibfnamefont {T.~A.~U.}\ \bibnamefont {Team}},\ }\href@noop {} {\bibfield  {journal} {\bibinfo  {journal} {Nuclear Fusion}\ }\textbf {\bibinfo {volume} {58}},\ \bibinfo {pages} {056008} (\bibinfo {year} {2018})}\BibitemShut {NoStop}%
\bibitem [{\citenamefont {Meyer}\ \emph {et~al.}(2009)\citenamefont {Meyer}, \citenamefont {Akers}, \citenamefont {Alladio}, \citenamefont {Appel}, \citenamefont {Axon}, \citenamefont {Ayed}, \citenamefont {Boerner}, \citenamefont {Buttery}, \citenamefont {Carolan}, \citenamefont {Ciric}, \citenamefont {Challis}, \citenamefont {Chapman}, \citenamefont {Coyler}, \citenamefont {Connor}, \citenamefont {Conway}, \citenamefont {Cowley}, \citenamefont {Cox}, \citenamefont {Counsell}, \citenamefont {Cunningham}, \citenamefont {Darke}, \citenamefont {deBock}, \citenamefont {deTemmerman}, \citenamefont {Dendy}, \citenamefont {Dowling}, \citenamefont {Dnestrovskij}, \citenamefont {Dnestrovskij}, \citenamefont {Dudson}, \citenamefont {Dunai}, \citenamefont {Dunstan}, \citenamefont {Field}, \citenamefont {Foster}, \citenamefont {Garzotti}, \citenamefont {Gibson}, \citenamefont {Gryaznevich}, \citenamefont {Guttenfelder}, \citenamefont {Hawkes}, \citenamefont {Harrison}, \citenamefont {Helander}, \citenamefont {Hender},
  \citenamefont {Hnat}, \citenamefont {Hole}, \citenamefont {Howell}, \citenamefont {Hua}, \citenamefont {Hubbard}, \citenamefont {Istenic}, \citenamefont {Joiner}, \citenamefont {Keeling}, \citenamefont {Kirk}, \citenamefont {Koslowski}, \citenamefont {Liang}, \citenamefont {Lilley}, \citenamefont {Lisgo}, \citenamefont {Lloyd}, \citenamefont {Maddison}, \citenamefont {Maingi}, \citenamefont {Mancuso}, \citenamefont {Manhood}, \citenamefont {Martin}, \citenamefont {McArdle}, \citenamefont {McCone}, \citenamefont {Michael}, \citenamefont {Micozzi}, \citenamefont {Morgan}, \citenamefont {Morris}, \citenamefont {Muir}, \citenamefont {Nardon}, \citenamefont {Naylor}, \citenamefont {O'Brien}, \citenamefont {O'Gorman}, \citenamefont {Patel}, \citenamefont {Pinches}, \citenamefont {Preinhaelter}, \citenamefont {Price}, \citenamefont {Rachlew}, \citenamefont {Reiter}, \citenamefont {Roach}, \citenamefont {Rozhansky}, \citenamefont {Saarelma}, \citenamefont {Saveliev}, \citenamefont {Scannell}, \citenamefont
  {Sharapov}, \citenamefont {Shevchenko}, \citenamefont {Shibaev}, \citenamefont {Smith}, \citenamefont {Staebler}, \citenamefont {Stork}, \citenamefont {Storrs}, \citenamefont {Sykes}, \citenamefont {Tallents}, \citenamefont {Tamain}, \citenamefont {Taylor}, \citenamefont {Temple}, \citenamefont {Thomas-Davies}, \citenamefont {Thornton}, \citenamefont {Thyagaraja}, \citenamefont {Turnyanskiy}, \citenamefont {Urban}, \citenamefont {Valovic}, \citenamefont {Vann}, \citenamefont {Volpe}, \citenamefont {Voss}, \citenamefont {Walsh}, \citenamefont {Warder}, \citenamefont {Watkins}, \citenamefont {Wilson}, \citenamefont {Windridge}, \citenamefont {Wisse}, \citenamefont {Zabolotski}, \citenamefont {Zoletnik}, \citenamefont {Zolotukhin},\ and\ \citenamefont {{the MAST and NBI teams}}}]{Meyer_2009}%
  \BibitemOpen
  \bibfield  {author} {\bibinfo {author} {\bibfnamefont {H.}~\bibnamefont {Meyer}}, \bibinfo {author} {\bibfnamefont {R.}~\bibnamefont {Akers}}, \bibinfo {author} {\bibfnamefont {F.}~\bibnamefont {Alladio}}, \bibinfo {author} {\bibfnamefont {L.}~\bibnamefont {Appel}}, \bibinfo {author} {\bibfnamefont {K.}~\bibnamefont {Axon}}, \bibinfo {author} {\bibfnamefont {N.~B.}\ \bibnamefont {Ayed}}, \bibinfo {author} {\bibfnamefont {P.}~\bibnamefont {Boerner}}, \bibinfo {author} {\bibfnamefont {R.}~\bibnamefont {Buttery}}, \bibinfo {author} {\bibfnamefont {P.}~\bibnamefont {Carolan}}, \bibinfo {author} {\bibfnamefont {D.}~\bibnamefont {Ciric}}, \bibinfo {author} {\bibfnamefont {C.}~\bibnamefont {Challis}}, \bibinfo {author} {\bibfnamefont {I.}~\bibnamefont {Chapman}}, \bibinfo {author} {\bibfnamefont {G.}~\bibnamefont {Coyler}}, \bibinfo {author} {\bibfnamefont {J.}~\bibnamefont {Connor}}, \bibinfo {author} {\bibfnamefont {N.}~\bibnamefont {Conway}}, \bibinfo {author} {\bibfnamefont {S.}~\bibnamefont {Cowley}},
  \bibinfo {author} {\bibfnamefont {M.}~\bibnamefont {Cox}}, \bibinfo {author} {\bibfnamefont {G.}~\bibnamefont {Counsell}}, \bibinfo {author} {\bibfnamefont {G.}~\bibnamefont {Cunningham}}, \bibinfo {author} {\bibfnamefont {A.}~\bibnamefont {Darke}}, \bibinfo {author} {\bibfnamefont {M.}~\bibnamefont {deBock}}, \bibinfo {author} {\bibfnamefont {G.}~\bibnamefont {deTemmerman}}, \bibinfo {author} {\bibfnamefont {R.}~\bibnamefont {Dendy}}, \bibinfo {author} {\bibfnamefont {J.}~\bibnamefont {Dowling}}, \bibinfo {author} {\bibfnamefont {A.~Y.}\ \bibnamefont {Dnestrovskij}}, \bibinfo {author} {\bibfnamefont {Y.}~\bibnamefont {Dnestrovskij}}, \bibinfo {author} {\bibfnamefont {B.}~\bibnamefont {Dudson}}, \bibinfo {author} {\bibfnamefont {D.}~\bibnamefont {Dunai}}, \bibinfo {author} {\bibfnamefont {M.}~\bibnamefont {Dunstan}}, \bibinfo {author} {\bibfnamefont {A.}~\bibnamefont {Field}}, \bibinfo {author} {\bibfnamefont {A.}~\bibnamefont {Foster}}, \bibinfo {author} {\bibfnamefont {L.}~\bibnamefont {Garzotti}},
  \bibinfo {author} {\bibfnamefont {K.}~\bibnamefont {Gibson}}, \bibinfo {author} {\bibfnamefont {M.}~\bibnamefont {Gryaznevich}}, \bibinfo {author} {\bibfnamefont {W.}~\bibnamefont {Guttenfelder}}, \bibinfo {author} {\bibfnamefont {N.}~\bibnamefont {Hawkes}}, \bibinfo {author} {\bibfnamefont {J.}~\bibnamefont {Harrison}}, \bibinfo {author} {\bibfnamefont {P.}~\bibnamefont {Helander}}, \bibinfo {author} {\bibfnamefont {T.}~\bibnamefont {Hender}}, \bibinfo {author} {\bibfnamefont {B.}~\bibnamefont {Hnat}}, \bibinfo {author} {\bibfnamefont {M.}~\bibnamefont {Hole}}, \bibinfo {author} {\bibfnamefont {D.}~\bibnamefont {Howell}}, \bibinfo {author} {\bibfnamefont {M.~D.}\ \bibnamefont {Hua}}, \bibinfo {author} {\bibfnamefont {A.}~\bibnamefont {Hubbard}}, \bibinfo {author} {\bibfnamefont {M.}~\bibnamefont {Istenic}}, \bibinfo {author} {\bibfnamefont {N.}~\bibnamefont {Joiner}}, \bibinfo {author} {\bibfnamefont {D.}~\bibnamefont {Keeling}}, \bibinfo {author} {\bibfnamefont {A.}~\bibnamefont {Kirk}}, \bibinfo {author}
  {\bibfnamefont {H.}~\bibnamefont {Koslowski}}, \bibinfo {author} {\bibfnamefont {Y.}~\bibnamefont {Liang}}, \bibinfo {author} {\bibfnamefont {M.}~\bibnamefont {Lilley}}, \bibinfo {author} {\bibfnamefont {S.}~\bibnamefont {Lisgo}}, \bibinfo {author} {\bibfnamefont {B.}~\bibnamefont {Lloyd}}, \bibinfo {author} {\bibfnamefont {G.}~\bibnamefont {Maddison}}, \bibinfo {author} {\bibfnamefont {R.}~\bibnamefont {Maingi}}, \bibinfo {author} {\bibfnamefont {A.}~\bibnamefont {Mancuso}}, \bibinfo {author} {\bibfnamefont {S.}~\bibnamefont {Manhood}}, \bibinfo {author} {\bibfnamefont {R.}~\bibnamefont {Martin}}, \bibinfo {author} {\bibfnamefont {G.}~\bibnamefont {McArdle}}, \bibinfo {author} {\bibfnamefont {J.}~\bibnamefont {McCone}}, \bibinfo {author} {\bibfnamefont {C.}~\bibnamefont {Michael}}, \bibinfo {author} {\bibfnamefont {P.}~\bibnamefont {Micozzi}}, \bibinfo {author} {\bibfnamefont {T.}~\bibnamefont {Morgan}}, \bibinfo {author} {\bibfnamefont {A.}~\bibnamefont {Morris}}, \bibinfo {author} {\bibfnamefont
  {D.}~\bibnamefont {Muir}}, \bibinfo {author} {\bibfnamefont {E.}~\bibnamefont {Nardon}}, \bibinfo {author} {\bibfnamefont {G.}~\bibnamefont {Naylor}}, \bibinfo {author} {\bibfnamefont {M.}~\bibnamefont {O'Brien}}, \bibinfo {author} {\bibfnamefont {T.}~\bibnamefont {O'Gorman}}, \bibinfo {author} {\bibfnamefont {A.}~\bibnamefont {Patel}}, \bibinfo {author} {\bibfnamefont {S.}~\bibnamefont {Pinches}}, \bibinfo {author} {\bibfnamefont {J.}~\bibnamefont {Preinhaelter}}, \bibinfo {author} {\bibfnamefont {M.}~\bibnamefont {Price}}, \bibinfo {author} {\bibfnamefont {E.}~\bibnamefont {Rachlew}}, \bibinfo {author} {\bibfnamefont {D.}~\bibnamefont {Reiter}}, \bibinfo {author} {\bibfnamefont {C.}~\bibnamefont {Roach}}, \bibinfo {author} {\bibfnamefont {V.}~\bibnamefont {Rozhansky}}, \bibinfo {author} {\bibfnamefont {S.}~\bibnamefont {Saarelma}}, \bibinfo {author} {\bibfnamefont {A.}~\bibnamefont {Saveliev}}, \bibinfo {author} {\bibfnamefont {R.}~\bibnamefont {Scannell}}, \bibinfo {author} {\bibfnamefont
  {S.}~\bibnamefont {Sharapov}}, \bibinfo {author} {\bibfnamefont {V.}~\bibnamefont {Shevchenko}}, \bibinfo {author} {\bibfnamefont {S.}~\bibnamefont {Shibaev}}, \bibinfo {author} {\bibfnamefont {H.}~\bibnamefont {Smith}}, \bibinfo {author} {\bibfnamefont {G.}~\bibnamefont {Staebler}}, \bibinfo {author} {\bibfnamefont {D.}~\bibnamefont {Stork}}, \bibinfo {author} {\bibfnamefont {J.}~\bibnamefont {Storrs}}, \bibinfo {author} {\bibfnamefont {A.}~\bibnamefont {Sykes}}, \bibinfo {author} {\bibfnamefont {S.}~\bibnamefont {Tallents}}, \bibinfo {author} {\bibfnamefont {P.}~\bibnamefont {Tamain}}, \bibinfo {author} {\bibfnamefont {D.}~\bibnamefont {Taylor}}, \bibinfo {author} {\bibfnamefont {D.}~\bibnamefont {Temple}}, \bibinfo {author} {\bibfnamefont {N.}~\bibnamefont {Thomas-Davies}}, \bibinfo {author} {\bibfnamefont {A.}~\bibnamefont {Thornton}}, \bibinfo {author} {\bibfnamefont {A.}~\bibnamefont {Thyagaraja}}, \bibinfo {author} {\bibfnamefont {M.}~\bibnamefont {Turnyanskiy}}, \bibinfo {author} {\bibfnamefont
  {J.}~\bibnamefont {Urban}}, \bibinfo {author} {\bibfnamefont {M.}~\bibnamefont {Valovic}}, \bibinfo {author} {\bibfnamefont {R.}~\bibnamefont {Vann}}, \bibinfo {author} {\bibfnamefont {F.}~\bibnamefont {Volpe}}, \bibinfo {author} {\bibfnamefont {G.}~\bibnamefont {Voss}}, \bibinfo {author} {\bibfnamefont {M.}~\bibnamefont {Walsh}}, \bibinfo {author} {\bibfnamefont {S.}~\bibnamefont {Warder}}, \bibinfo {author} {\bibfnamefont {R.}~\bibnamefont {Watkins}}, \bibinfo {author} {\bibfnamefont {H.}~\bibnamefont {Wilson}}, \bibinfo {author} {\bibfnamefont {M.}~\bibnamefont {Windridge}}, \bibinfo {author} {\bibfnamefont {M.}~\bibnamefont {Wisse}}, \bibinfo {author} {\bibfnamefont {A.}~\bibnamefont {Zabolotski}}, \bibinfo {author} {\bibfnamefont {S.}~\bibnamefont {Zoletnik}}, \bibinfo {author} {\bibfnamefont {O.}~\bibnamefont {Zolotukhin}},\ and\ \bibinfo {author} {\bibnamefont {{the MAST and NBI teams}}},\ }\href@noop {} {\bibfield  {journal} {\bibinfo  {journal} {Nuclear Fusion}\ }\textbf {\bibinfo {volume} {49}},\
  \bibinfo {pages} {104017} (\bibinfo {year} {2009})}\BibitemShut {NoStop}%
\bibitem [{\citenamefont {Kaye}\ \emph {et~al.}(2021)\citenamefont {Kaye}, \citenamefont {Connor},\ and\ \citenamefont {Roach}}]{Kaye_2021_ST_Review}%
  \BibitemOpen
  \bibfield  {author} {\bibinfo {author} {\bibfnamefont {S.~M.}\ \bibnamefont {Kaye}}, \bibinfo {author} {\bibfnamefont {J.~W.}\ \bibnamefont {Connor}},\ and\ \bibinfo {author} {\bibfnamefont {C.~M.}\ \bibnamefont {Roach}},\ }\href {https://doi.org/10.1088/1361-6587/ac2b38} {\bibfield  {journal} {\bibinfo  {journal} {Plasma Physics and Controlled Fusion}\ }\textbf {\bibinfo {volume} {63}},\ \bibinfo {pages} {123001} (\bibinfo {year} {2021})}\BibitemShut {NoStop}%
\bibitem [{\citenamefont {Jenko}\ \emph {et~al.}(2000)\citenamefont {Jenko}, \citenamefont {Dorland}, \citenamefont {Kotschenreuther},\ and\ \citenamefont {Rogers}}]{JenkoGENE2000}%
  \BibitemOpen
  \bibfield  {author} {\bibinfo {author} {\bibfnamefont {F.}~\bibnamefont {Jenko}}, \bibinfo {author} {\bibfnamefont {W.}~\bibnamefont {Dorland}}, \bibinfo {author} {\bibfnamefont {M.}~\bibnamefont {Kotschenreuther}},\ and\ \bibinfo {author} {\bibfnamefont {B.}~\bibnamefont {Rogers}},\ }\href@noop {} {\bibfield  {journal} {\bibinfo  {journal} {Physics of Plasmas}\ }\textbf {\bibinfo {volume} {7}},\ \bibinfo {pages} {1904} (\bibinfo {year} {2000})}\BibitemShut {NoStop}%
\bibitem [{\citenamefont {Goerler}\ \emph {et~al.}(2011)\citenamefont {Goerler}, \citenamefont {Lapillonne}, \citenamefont {Brunner}, \citenamefont {Dannert}, \citenamefont {Jenko}, \citenamefont {Merz},\ and\ \citenamefont {Told}}]{GoerlerGENE2011}%
  \BibitemOpen
  \bibfield  {author} {\bibinfo {author} {\bibfnamefont {T.}~\bibnamefont {Goerler}}, \bibinfo {author} {\bibfnamefont {X.}~\bibnamefont {Lapillonne}}, \bibinfo {author} {\bibfnamefont {S.}~\bibnamefont {Brunner}}, \bibinfo {author} {\bibfnamefont {T.}~\bibnamefont {Dannert}}, \bibinfo {author} {\bibfnamefont {F.}~\bibnamefont {Jenko}}, \bibinfo {author} {\bibfnamefont {F.}~\bibnamefont {Merz}},\ and\ \bibinfo {author} {\bibfnamefont {D.}~\bibnamefont {Told}},\ }\href@noop {} {\bibfield  {journal} {\bibinfo  {journal} {Journal of Computational Physics}\ }\textbf {\bibinfo {volume} {230}},\ \bibinfo {pages} {7053} (\bibinfo {year} {2011})}\BibitemShut {NoStop}%
\bibitem [{\citenamefont {Miller}\ \emph {et~al.}(1998)\citenamefont {Miller}, \citenamefont {Chu}, \citenamefont {Greene}, \citenamefont {Lin-Liu},\ and\ \citenamefont {Waltz}}]{Millergeometry1998}%
  \BibitemOpen
  \bibfield  {author} {\bibinfo {author} {\bibfnamefont {R.~L.}\ \bibnamefont {Miller}}, \bibinfo {author} {\bibfnamefont {M.~S.}\ \bibnamefont {Chu}}, \bibinfo {author} {\bibfnamefont {J.~M.}\ \bibnamefont {Greene}}, \bibinfo {author} {\bibfnamefont {Y.~R.}\ \bibnamefont {Lin-Liu}},\ and\ \bibinfo {author} {\bibfnamefont {R.~E.}\ \bibnamefont {Waltz}},\ }\href {https://doi.org/10.1063/1.872666} {\bibfield  {journal} {\bibinfo  {journal} {Physics of Plasmas}\ }\textbf {\bibinfo {volume} {5}},\ \bibinfo {pages} {973} (\bibinfo {year} {1998})}\BibitemShut {NoStop}%
\bibitem [{\citenamefont {McMillan}\ and\ \citenamefont {Dominski}(2019{\natexlab{b}})}]{mcmillan2019}%
  \BibitemOpen
  \bibfield  {author} {\bibinfo {author} {\bibfnamefont {B.~F.}\ \bibnamefont {McMillan}}\ and\ \bibinfo {author} {\bibfnamefont {J.}~\bibnamefont {Dominski}},\ }\href@noop {} {\bibfield  {journal} {\bibinfo  {journal} {Journal of Plasma Physics}\ }\textbf {\bibinfo {volume} {85}},\ \bibinfo {pages} {175850301} (\bibinfo {year} {2019}{\natexlab{b}})}\BibitemShut {NoStop}%
\bibitem [{\citenamefont {Sugama}\ and\ \citenamefont {Horton}(1998)}]{Sugammafluxderiv1998}%
  \BibitemOpen
  \bibfield  {author} {\bibinfo {author} {\bibfnamefont {H.}~\bibnamefont {Sugama}}\ and\ \bibinfo {author} {\bibfnamefont {W.}~\bibnamefont {Horton}},\ }\href@noop {} {\bibfield  {journal} {\bibinfo  {journal} {Physics of Plasmas}\ }\textbf {\bibinfo {volume} {5}},\ \bibinfo {pages} {2560} (\bibinfo {year} {1998})}\BibitemShut {NoStop}%
\bibitem [{\citenamefont {Ball}(2016)}]{ball2016a}%
  \BibitemOpen
  \bibfield  {author} {\bibinfo {author} {\bibfnamefont {J.}~\bibnamefont {Ball}},\ }\emph {\bibinfo {title} {Up-down asymmetric tokamaks}},\ \href@noop {} {Ph.D. thesis},\ \bibinfo  {school} {University of Oxford} (\bibinfo {year} {2016})\BibitemShut {NoStop}%
\bibitem [{\citenamefont {Sun}\ \emph {et~al.}(2024)\citenamefont {Sun}, \citenamefont {Ball}, \citenamefont {Brunner},\ and\ \citenamefont {Volcokas}}]{Sun2024NF}%
  \BibitemOpen
  \bibfield  {author} {\bibinfo {author} {\bibfnamefont {H.}~\bibnamefont {Sun}}, \bibinfo {author} {\bibfnamefont {J.}~\bibnamefont {Ball}}, \bibinfo {author} {\bibfnamefont {S.}~\bibnamefont {Brunner}},\ and\ \bibinfo {author} {\bibfnamefont {A.}~\bibnamefont {Volcokas}},\ }\href {http://iopscience.iop.org/article/10.1088/1741-4326/ad2583} {\bibfield  {journal} {\bibinfo  {journal} {Nuclear Fusion}\ } (\bibinfo {year} {2024})}\BibitemShut {NoStop}%
\bibitem [{\citenamefont {Waltz}\ \emph {et~al.}(1998)\citenamefont {Waltz}, \citenamefont {Dewar},\ and\ \citenamefont {Garbet}}]{WaltzFlowShear1998}%
  \BibitemOpen
  \bibfield  {author} {\bibinfo {author} {\bibfnamefont {R.}~\bibnamefont {Waltz}}, \bibinfo {author} {\bibfnamefont {R.}~\bibnamefont {Dewar}},\ and\ \bibinfo {author} {\bibfnamefont {X.}~\bibnamefont {Garbet}},\ }\href@noop {} {\bibfield  {journal} {\bibinfo  {journal} {Physics of Plasmas}\ }\textbf {\bibinfo {volume} {5}},\ \bibinfo {pages} {1784} (\bibinfo {year} {1998})}\BibitemShut {NoStop}%
\bibitem [{\citenamefont {Antonsen}\ \emph {et~al.}(1996)\citenamefont {Antonsen}, \citenamefont {Drake}, \citenamefont {Guzdar}, \citenamefont {Hassam}, \citenamefont {Lau}, \citenamefont {Liu},\ and\ \citenamefont {Novakovskii}}]{Antonsen_toroidalITG_1996}%
  \BibitemOpen
  \bibfield  {author} {\bibinfo {author} {\bibfnamefont {J.}~\bibnamefont {Antonsen}, \bibfnamefont {T.~M.}}, \bibinfo {author} {\bibfnamefont {J.~F.}\ \bibnamefont {Drake}}, \bibinfo {author} {\bibfnamefont {P.~N.}\ \bibnamefont {Guzdar}}, \bibinfo {author} {\bibfnamefont {A.~B.}\ \bibnamefont {Hassam}}, \bibinfo {author} {\bibfnamefont {Y.~T.}\ \bibnamefont {Lau}}, \bibinfo {author} {\bibfnamefont {C.~S.}\ \bibnamefont {Liu}},\ and\ \bibinfo {author} {\bibfnamefont {S.~V.}\ \bibnamefont {Novakovskii}},\ }\href {https://doi.org/10.1063/1.871928} {\bibfield  {journal} {\bibinfo  {journal} {Physics of Plasmas}\ }\textbf {\bibinfo {volume} {3}},\ \bibinfo {pages} {2221} (\bibinfo {year} {1996})},\ \Eprint {https://arxiv.org/abs/https://pubs.aip.org/aip/pop/article-pdf/3/6/2221/19236868/2221\_1\_online.pdf} {https://pubs.aip.org/aip/pop/article-pdf/3/6/2221/19236868/2221\_1\_online.pdf} \BibitemShut {NoStop}%
\bibitem [{\citenamefont {Dimits}\ \emph {et~al.}(2001)\citenamefont {Dimits}, \citenamefont {Cohen}, \citenamefont {Nevins},\ and\ \citenamefont {Shumaker}}]{Dimits2001ITG}%
  \BibitemOpen
  \bibfield  {author} {\bibinfo {author} {\bibfnamefont {A.}~\bibnamefont {Dimits}}, \bibinfo {author} {\bibfnamefont {B.}~\bibnamefont {Cohen}}, \bibinfo {author} {\bibfnamefont {W.}~\bibnamefont {Nevins}},\ and\ \bibinfo {author} {\bibfnamefont {D.}~\bibnamefont {Shumaker}},\ }\href@noop {} {\bibfield  {journal} {\bibinfo  {journal} {Nuclear Fusion}\ }\textbf {\bibinfo {volume} {41}},\ \bibinfo {pages} {1725} (\bibinfo {year} {2001})}\BibitemShut {NoStop}%
\bibitem [{\citenamefont {Camenen}\ \emph {et~al.}(2016)\citenamefont {Camenen}, \citenamefont {Casson}, \citenamefont {Manas},\ and\ \citenamefont {Peeters}}]{Camenen2016popITG}%
  \BibitemOpen
  \bibfield  {author} {\bibinfo {author} {\bibfnamefont {Y.}~\bibnamefont {Camenen}}, \bibinfo {author} {\bibfnamefont {F.~J.}\ \bibnamefont {Casson}}, \bibinfo {author} {\bibfnamefont {P.}~\bibnamefont {Manas}},\ and\ \bibinfo {author} {\bibfnamefont {A.~G.}\ \bibnamefont {Peeters}},\ }\href@noop {} {\bibfield  {journal} {\bibinfo  {journal} {Physics of Plasmas}\ }\textbf {\bibinfo {volume} {23}},\ \bibinfo {pages} {022507} (\bibinfo {year} {2016})}\BibitemShut {NoStop}%
\bibitem [{\citenamefont {Lloyd}\ \emph {et~al.}(2011)\citenamefont {Lloyd}, \citenamefont {Akers}, \citenamefont {Alladio}, \citenamefont {Allan},\ and\ \citenamefont {{L.C. Appel et al.}}}]{MAST_Lloyd_2011}%
  \BibitemOpen
  \bibfield  {author} {\bibinfo {author} {\bibfnamefont {B.}~\bibnamefont {Lloyd}}, \bibinfo {author} {\bibfnamefont {R.}~\bibnamefont {Akers}}, \bibinfo {author} {\bibfnamefont {F.}~\bibnamefont {Alladio}}, \bibinfo {author} {\bibfnamefont {S.}~\bibnamefont {Allan}},\ and\ \bibinfo {author} {\bibnamefont {{L.C. Appel et al.}}},\ }\href@noop {} {\bibfield  {journal} {\bibinfo  {journal} {Nuclear Fusion}\ }\textbf {\bibinfo {volume} {51}},\ \bibinfo {pages} {094013} (\bibinfo {year} {2011})}\BibitemShut {NoStop}%
\bibitem [{\citenamefont {Harrison}\ \emph {et~al.}(2019)\citenamefont {Harrison}, \citenamefont {Akers}, \citenamefont {Allan}, \citenamefont {Allcock},\ and\ \citenamefont {{J. Allen et al.}}}]{Harrison_2019}%
  \BibitemOpen
  \bibfield  {author} {\bibinfo {author} {\bibfnamefont {J.}~\bibnamefont {Harrison}}, \bibinfo {author} {\bibfnamefont {R.}~\bibnamefont {Akers}}, \bibinfo {author} {\bibfnamefont {S.}~\bibnamefont {Allan}}, \bibinfo {author} {\bibfnamefont {J.}~\bibnamefont {Allcock}},\ and\ \bibinfo {author} {\bibnamefont {{J. Allen et al.}}},\ }\href@noop {} {\bibfield  {journal} {\bibinfo  {journal} {Nuclear Fusion}\ }\textbf {\bibinfo {volume} {59}},\ \bibinfo {pages} {112011} (\bibinfo {year} {2019})}\BibitemShut {NoStop}%
\bibitem [{\citenamefont {Field}\ \emph {et~al.}(2011)\citenamefont {Field}, \citenamefont {Michael}, \citenamefont {Akers}, \citenamefont {Candy}, \citenamefont {Colyer}, \citenamefont {Guttenfelder}, \citenamefont {c.~Ghim}, \citenamefont {Roach}, \citenamefont {Saarelma},\ and\ \citenamefont {the MAST~Team}}]{Field_2011NF}%
  \BibitemOpen
  \bibfield  {author} {\bibinfo {author} {\bibfnamefont {A.}~\bibnamefont {Field}}, \bibinfo {author} {\bibfnamefont {C.}~\bibnamefont {Michael}}, \bibinfo {author} {\bibfnamefont {R.}~\bibnamefont {Akers}}, \bibinfo {author} {\bibfnamefont {J.}~\bibnamefont {Candy}}, \bibinfo {author} {\bibfnamefont {G.}~\bibnamefont {Colyer}}, \bibinfo {author} {\bibfnamefont {W.}~\bibnamefont {Guttenfelder}}, \bibinfo {author} {\bibfnamefont {Y.}~\bibnamefont {c.~Ghim}}, \bibinfo {author} {\bibfnamefont {C.}~\bibnamefont {Roach}}, \bibinfo {author} {\bibfnamefont {S.}~\bibnamefont {Saarelma}},\ and\ \bibinfo {author} {\bibnamefont {the MAST~Team}},\ }\href@noop {} {\bibfield  {journal} {\bibinfo  {journal} {Nuclear Fusion}\ }\textbf {\bibinfo {volume} {51}},\ \bibinfo {pages} {063006} (\bibinfo {year} {2011})}\BibitemShut {NoStop}%
\bibitem [{\citenamefont {Hawryluk}(1981)}]{TRANSPoriginal}%
  \BibitemOpen
  \bibfield  {author} {\bibinfo {author} {\bibfnamefont {R.}~\bibnamefont {Hawryluk}},\ }in\ \href@noop {} {\emph {\bibinfo {booktitle} {Physics of Plasmas Close to Thermonuclear Conditions}}},\ \bibinfo {editor} {edited by\ \bibinfo {editor} {\bibfnamefont {B.}~\bibnamefont {Coppi}}, \bibinfo {editor} {\bibfnamefont {G.}~\bibnamefont {Leotta}}, \bibinfo {editor} {\bibfnamefont {D.}~\bibnamefont {Pfirsch}}, \bibinfo {editor} {\bibfnamefont {R.}~\bibnamefont {Pozzoli}},\ and\ \bibinfo {editor} {\bibfnamefont {E.}~\bibnamefont {Sindoni}}}\ (\bibinfo  {publisher} {Pergamon},\ \bibinfo {year} {1981})\ pp.\ \bibinfo {pages} {19--46}\BibitemShut {NoStop}%
\bibitem [{\citenamefont {Patel}\ \emph {et~al.}(2024)\citenamefont {Patel}, \citenamefont {Hill}, \citenamefont {Pattinson}, \citenamefont {Giacomin}, \citenamefont {Bokshi}, \citenamefont {Kennedy}, \citenamefont {Dudding}, \citenamefont {Parisi}, \citenamefont {Neiser}, \citenamefont {Jayalekshmi}, \citenamefont {Dickinson},\ and\ \citenamefont {Ruiz}}]{Patel_pyrokinetics_2022}%
  \BibitemOpen
  \bibfield  {author} {\bibinfo {author} {\bibfnamefont {B.~S.}\ \bibnamefont {Patel}}, \bibinfo {author} {\bibfnamefont {P.}~\bibnamefont {Hill}}, \bibinfo {author} {\bibfnamefont {L.}~\bibnamefont {Pattinson}}, \bibinfo {author} {\bibfnamefont {M.}~\bibnamefont {Giacomin}}, \bibinfo {author} {\bibfnamefont {A.}~\bibnamefont {Bokshi}}, \bibinfo {author} {\bibfnamefont {D.}~\bibnamefont {Kennedy}}, \bibinfo {author} {\bibfnamefont {H.~G.}\ \bibnamefont {Dudding}}, \bibinfo {author} {\bibfnamefont {J.~F.}\ \bibnamefont {Parisi}}, \bibinfo {author} {\bibfnamefont {T.~F.}\ \bibnamefont {Neiser}}, \bibinfo {author} {\bibfnamefont {A.~C.}\ \bibnamefont {Jayalekshmi}}, \bibinfo {author} {\bibfnamefont {D.}~\bibnamefont {Dickinson}},\ and\ \bibinfo {author} {\bibfnamefont {J.~R.}\ \bibnamefont {Ruiz}},\ }\href {https://doi.org/10.21105/joss.05866} {\bibfield  {journal} {\bibinfo  {journal} {Journal of Open Source Software}\ }\textbf {\bibinfo {volume} {9}},\ \bibinfo {pages} {5866} (\bibinfo {year}
  {2024})}\BibitemShut {NoStop}%
\bibitem [{\citenamefont {Camenen}\ \emph {et~al.}(2010{\natexlab{b}})\citenamefont {Camenen}, \citenamefont {Bortolon}, \citenamefont {Duval}, \citenamefont {Federspiel}, \citenamefont {Peeters}, \citenamefont {Casson}, \citenamefont {Hornsby}, \citenamefont {Karpushov}, \citenamefont {Piras}, \citenamefont {Sauter}, \citenamefont {Snodin},\ and\ \citenamefont {Szepesi}}]{Camenen_2010_TCVEXP}%
  \BibitemOpen
  \bibfield  {author} {\bibinfo {author} {\bibfnamefont {Y.}~\bibnamefont {Camenen}}, \bibinfo {author} {\bibfnamefont {A.}~\bibnamefont {Bortolon}}, \bibinfo {author} {\bibfnamefont {B.~P.}\ \bibnamefont {Duval}}, \bibinfo {author} {\bibfnamefont {L.}~\bibnamefont {Federspiel}}, \bibinfo {author} {\bibfnamefont {A.~G.}\ \bibnamefont {Peeters}}, \bibinfo {author} {\bibfnamefont {F.~J.}\ \bibnamefont {Casson}}, \bibinfo {author} {\bibfnamefont {W.~A.}\ \bibnamefont {Hornsby}}, \bibinfo {author} {\bibfnamefont {A.~N.}\ \bibnamefont {Karpushov}}, \bibinfo {author} {\bibfnamefont {F.}~\bibnamefont {Piras}}, \bibinfo {author} {\bibfnamefont {O.}~\bibnamefont {Sauter}}, \bibinfo {author} {\bibfnamefont {A.~P.}\ \bibnamefont {Snodin}},\ and\ \bibinfo {author} {\bibfnamefont {G.}~\bibnamefont {Szepesi}},\ }\href {https://doi.org/10.1103/PhysRevLett.105.135003} {\bibfield  {journal} {\bibinfo  {journal} {Phys. Rev. Lett.}\ }\textbf {\bibinfo {volume} {105}},\ \bibinfo {pages} {135003} (\bibinfo {year}
  {2010}{\natexlab{b}})}\BibitemShut {NoStop}%
\bibitem [{\citenamefont {Doyle}\ \emph {et~al.}(2021)\citenamefont {Doyle}, \citenamefont {Lopez-Aires}, \citenamefont {Mancini}, \citenamefont {Agredano-Torres}, \citenamefont {Garcia-Sanchez}, \citenamefont {Segado-Fernandez}, \citenamefont {Ayllon-Guerola}, \citenamefont {Garcia-Munoz}, \citenamefont {Viezzer}, \citenamefont {Soria-Hoyo}, \citenamefont {Garcia-Lopez}, \citenamefont {Cunningham}, \citenamefont {Buxton}, \citenamefont {Gryaznevich}, \citenamefont {Hwang},\ and\ \citenamefont {Chung}}]{DOYLE_SMART_2021}%
  \BibitemOpen
  \bibfield  {author} {\bibinfo {author} {\bibfnamefont {S.}~\bibnamefont {Doyle}}, \bibinfo {author} {\bibfnamefont {D.}~\bibnamefont {Lopez-Aires}}, \bibinfo {author} {\bibfnamefont {A.}~\bibnamefont {Mancini}}, \bibinfo {author} {\bibfnamefont {M.}~\bibnamefont {Agredano-Torres}}, \bibinfo {author} {\bibfnamefont {J.}~\bibnamefont {Garcia-Sanchez}}, \bibinfo {author} {\bibfnamefont {J.}~\bibnamefont {Segado-Fernandez}}, \bibinfo {author} {\bibfnamefont {J.}~\bibnamefont {Ayllon-Guerola}}, \bibinfo {author} {\bibfnamefont {M.}~\bibnamefont {Garcia-Munoz}}, \bibinfo {author} {\bibfnamefont {E.}~\bibnamefont {Viezzer}}, \bibinfo {author} {\bibfnamefont {C.}~\bibnamefont {Soria-Hoyo}}, \bibinfo {author} {\bibfnamefont {J.}~\bibnamefont {Garcia-Lopez}}, \bibinfo {author} {\bibfnamefont {G.}~\bibnamefont {Cunningham}}, \bibinfo {author} {\bibfnamefont {P.}~\bibnamefont {Buxton}}, \bibinfo {author} {\bibfnamefont {M.}~\bibnamefont {Gryaznevich}}, \bibinfo {author} {\bibfnamefont {Y.}~\bibnamefont {Hwang}},\ and\
  \bibinfo {author} {\bibfnamefont {K.}~\bibnamefont {Chung}},\ }\href@noop {} {\bibfield  {journal} {\bibinfo  {journal} {Fusion Engineering and Design}\ }\textbf {\bibinfo {volume} {171}},\ \bibinfo {pages} {112706} (\bibinfo {year} {2021})}\BibitemShut {NoStop}%
\bibitem [{\citenamefont {Mancini}\ \emph {et~al.}(2021)\citenamefont {Mancini}, \citenamefont {Ayllon-Guerola}, \citenamefont {Doyle}, \citenamefont {Agredano-Torres}, \citenamefont {Lopez-Aires}, \citenamefont {Toledo-Garrido}, \citenamefont {Viezzer}, \citenamefont {Garcia-Munoz}, \citenamefont {Buxton}, \citenamefont {Chung}, \citenamefont {Garcia-Dominguez}, \citenamefont {Garcia-Lopez}, \citenamefont {Gryaznevich}, \citenamefont {Hidalgo-Salaverri}, \citenamefont {Hwang},\ and\ \citenamefont {Segado-Fern\'andez}}]{MANCINI_SMART_2021}%
  \BibitemOpen
  \bibfield  {author} {\bibinfo {author} {\bibfnamefont {A.}~\bibnamefont {Mancini}}, \bibinfo {author} {\bibfnamefont {J.}~\bibnamefont {Ayllon-Guerola}}, \bibinfo {author} {\bibfnamefont {S.}~\bibnamefont {Doyle}}, \bibinfo {author} {\bibfnamefont {M.}~\bibnamefont {Agredano-Torres}}, \bibinfo {author} {\bibfnamefont {D.}~\bibnamefont {Lopez-Aires}}, \bibinfo {author} {\bibfnamefont {J.}~\bibnamefont {Toledo-Garrido}}, \bibinfo {author} {\bibfnamefont {E.}~\bibnamefont {Viezzer}}, \bibinfo {author} {\bibfnamefont {M.}~\bibnamefont {Garcia-Munoz}}, \bibinfo {author} {\bibfnamefont {P.}~\bibnamefont {Buxton}}, \bibinfo {author} {\bibfnamefont {K.}~\bibnamefont {Chung}}, \bibinfo {author} {\bibfnamefont {J.}~\bibnamefont {Garcia-Dominguez}}, \bibinfo {author} {\bibfnamefont {J.}~\bibnamefont {Garcia-Lopez}}, \bibinfo {author} {\bibfnamefont {M.}~\bibnamefont {Gryaznevich}}, \bibinfo {author} {\bibfnamefont {J.}~\bibnamefont {Hidalgo-Salaverri}}, \bibinfo {author} {\bibfnamefont {Y.}~\bibnamefont {Hwang}},\
  and\ \bibinfo {author} {\bibfnamefont {J.}~\bibnamefont {Segado-Fern\'andez}},\ }\href {https://doi.org/https://doi.org/10.1016/j.fusengdes.2021.112542} {\bibfield  {journal} {\bibinfo  {journal} {Fusion Engineering and Design}\ }\textbf {\bibinfo {volume} {171}},\ \bibinfo {pages} {112542} (\bibinfo {year} {2021})}\BibitemShut {NoStop}%
\bibitem [{\citenamefont {Agredano-Torres}\ \emph {et~al.}(2021)\citenamefont {Agredano-Torres}, \citenamefont {Garcia-Sanchez}, \citenamefont {Mancini}, \citenamefont {Doyle}, \citenamefont {Garcia-Munoz}, \citenamefont {Ayllon-Guerola}, \citenamefont {Barragan-Villarejo}, \citenamefont {Viezzer}, \citenamefont {Segado-Fernandez}, \citenamefont {Lopez-Aires}, \citenamefont {Toledo-Garrido}, \citenamefont {Buxton}, \citenamefont {Chung}, \citenamefont {Garcia-Dominguez}, \citenamefont {Garcia-Franquelo}, \citenamefont {Gryaznevich}, \citenamefont {Hidalgo-Salaverri}, \citenamefont {Hwang}, \citenamefont {Leon-Galvan},\ and\ \citenamefont {Maza-Ortega}}]{AGREDANOTORRES_SMART_2021}%
  \BibitemOpen
  \bibfield  {author} {\bibinfo {author} {\bibfnamefont {M.}~\bibnamefont {Agredano-Torres}}, \bibinfo {author} {\bibfnamefont {J.}~\bibnamefont {Garcia-Sanchez}}, \bibinfo {author} {\bibfnamefont {A.}~\bibnamefont {Mancini}}, \bibinfo {author} {\bibfnamefont {S.}~\bibnamefont {Doyle}}, \bibinfo {author} {\bibfnamefont {M.}~\bibnamefont {Garcia-Munoz}}, \bibinfo {author} {\bibfnamefont {J.}~\bibnamefont {Ayllon-Guerola}}, \bibinfo {author} {\bibfnamefont {M.}~\bibnamefont {Barragan-Villarejo}}, \bibinfo {author} {\bibfnamefont {E.}~\bibnamefont {Viezzer}}, \bibinfo {author} {\bibfnamefont {J.}~\bibnamefont {Segado-Fernandez}}, \bibinfo {author} {\bibfnamefont {D.}~\bibnamefont {Lopez-Aires}}, \bibinfo {author} {\bibfnamefont {J.}~\bibnamefont {Toledo-Garrido}}, \bibinfo {author} {\bibfnamefont {P.}~\bibnamefont {Buxton}}, \bibinfo {author} {\bibfnamefont {K.}~\bibnamefont {Chung}}, \bibinfo {author} {\bibfnamefont {J.}~\bibnamefont {Garcia-Dominguez}}, \bibinfo {author} {\bibfnamefont {L.}~\bibnamefont
  {Garcia-Franquelo}}, \bibinfo {author} {\bibfnamefont {M.}~\bibnamefont {Gryaznevich}}, \bibinfo {author} {\bibfnamefont {J.}~\bibnamefont {Hidalgo-Salaverri}}, \bibinfo {author} {\bibfnamefont {Y.}~\bibnamefont {Hwang}}, \bibinfo {author} {\bibfnamefont {J.}~\bibnamefont {Leon-Galvan}},\ and\ \bibinfo {author} {\bibfnamefont {J.}~\bibnamefont {Maza-Ortega}},\ }\href {https://doi.org/https://doi.org/10.1016/j.fusengdes.2021.112683} {\bibfield  {journal} {\bibinfo  {journal} {Fusion Engineering and Design}\ }\textbf {\bibinfo {volume} {168}},\ \bibinfo {pages} {112683} (\bibinfo {year} {2021})}\BibitemShut {NoStop}%
\bibitem [{\citenamefont {Segado-Fernandez}\ \emph {et~al.}(2023)\citenamefont {Segado-Fernandez}, \citenamefont {Mancini}, \citenamefont {Garcia-Dominguez}, \citenamefont {Ayllon-Guerola}, \citenamefont {Cruz-Zabala}, \citenamefont {Velarde}, \citenamefont {Garcia-Munoz}, \citenamefont {Viezzer}, \citenamefont {Navarro}, \citenamefont {Agredano-Torres},\ and\ \citenamefont {Vicente-Torres}}]{SEGADOFERNANDE_SMART_2023}%
  \BibitemOpen
  \bibfield  {author} {\bibinfo {author} {\bibfnamefont {J.}~\bibnamefont {Segado-Fernandez}}, \bibinfo {author} {\bibfnamefont {A.}~\bibnamefont {Mancini}}, \bibinfo {author} {\bibfnamefont {J.}~\bibnamefont {Garcia-Dominguez}}, \bibinfo {author} {\bibfnamefont {J.}~\bibnamefont {Ayllon-Guerola}}, \bibinfo {author} {\bibfnamefont {D.}~\bibnamefont {Cruz-Zabala}}, \bibinfo {author} {\bibfnamefont {L.}~\bibnamefont {Velarde}}, \bibinfo {author} {\bibfnamefont {M.}~\bibnamefont {Garcia-Munoz}}, \bibinfo {author} {\bibfnamefont {E.}~\bibnamefont {Viezzer}}, \bibinfo {author} {\bibfnamefont {C.}~\bibnamefont {Navarro}}, \bibinfo {author} {\bibfnamefont {M.}~\bibnamefont {Agredano-Torres}},\ and\ \bibinfo {author} {\bibfnamefont {P.}~\bibnamefont {Vicente-Torres}},\ }\href {https://doi.org/https://doi.org/10.1016/j.fusengdes.2023.113832} {\bibfield  {journal} {\bibinfo  {journal} {Fusion Engineering and Design}\ }\textbf {\bibinfo {volume} {193}},\ \bibinfo {pages} {113832} (\bibinfo {year} {2023})}\BibitemShut
  {NoStop}%
\bibitem [{\citenamefont {Podest\'a}\ \emph {et~al.}(2024)\citenamefont {Podest\'a}, \citenamefont {Cruz-Zabala}, \citenamefont {Poli}, \citenamefont {Dominguez-Palacios}, \citenamefont {Berkery}, \citenamefont {Garcia-Munoz}, \citenamefont {Viezzer}, \citenamefont {Mancini}, \citenamefont {Segado}, \citenamefont {Velarde},\ and\ \citenamefont {Kaye}}]{Podesta_SMART_2024}%
  \BibitemOpen
  \bibfield  {author} {\bibinfo {author} {\bibfnamefont {M.}~\bibnamefont {Podest\'a}}, \bibinfo {author} {\bibfnamefont {D.~J.}\ \bibnamefont {Cruz-Zabala}}, \bibinfo {author} {\bibfnamefont {F.~M.}\ \bibnamefont {Poli}}, \bibinfo {author} {\bibfnamefont {J.}~\bibnamefont {Dominguez-Palacios}}, \bibinfo {author} {\bibfnamefont {J.~W.}\ \bibnamefont {Berkery}}, \bibinfo {author} {\bibfnamefont {M.}~\bibnamefont {Garcia-Munoz}}, \bibinfo {author} {\bibfnamefont {E.}~\bibnamefont {Viezzer}}, \bibinfo {author} {\bibfnamefont {A.}~\bibnamefont {Mancini}}, \bibinfo {author} {\bibfnamefont {J.}~\bibnamefont {Segado}}, \bibinfo {author} {\bibfnamefont {L.}~\bibnamefont {Velarde}},\ and\ \bibinfo {author} {\bibfnamefont {S.~M.}\ \bibnamefont {Kaye}},\ }\href@noop {} {\bibfield  {journal} {\bibinfo  {journal} {Plasma Physics and Controlled Fusion}\ }\textbf {\bibinfo {volume} {66}},\ \bibinfo {pages} {045021} (\bibinfo {year} {2024})}\BibitemShut {NoStop}%
\bibitem [{\citenamefont {Hansen}\ \emph {et~al.}(2024)\citenamefont {Hansen}, \citenamefont {Stewart}, \citenamefont {Burgess}, \citenamefont {Pharr}, \citenamefont {Guizzo}, \citenamefont {Logak}, \citenamefont {Nelson},\ and\ \citenamefont {Paz-Soldan}}]{HANSEN2024_tokamaker}%
  \BibitemOpen
  \bibfield  {author} {\bibinfo {author} {\bibfnamefont {C.}~\bibnamefont {Hansen}}, \bibinfo {author} {\bibfnamefont {I.}~\bibnamefont {Stewart}}, \bibinfo {author} {\bibfnamefont {D.}~\bibnamefont {Burgess}}, \bibinfo {author} {\bibfnamefont {M.}~\bibnamefont {Pharr}}, \bibinfo {author} {\bibfnamefont {S.}~\bibnamefont {Guizzo}}, \bibinfo {author} {\bibfnamefont {F.}~\bibnamefont {Logak}}, \bibinfo {author} {\bibfnamefont {A.}~\bibnamefont {Nelson}},\ and\ \bibinfo {author} {\bibfnamefont {C.}~\bibnamefont {Paz-Soldan}},\ }\href {https://doi.org/https://doi.org/10.1016/j.cpc.2024.109111} {\bibfield  {journal} {\bibinfo  {journal} {Computer Physics Communications}\ }\textbf {\bibinfo {volume} {298}},\ \bibinfo {pages} {109111} (\bibinfo {year} {2024})}\BibitemShut {NoStop}%
\end{thebibliography}%

\end{document}